\RequirePackage{fix-cm}
\documentclass[11pt]{article}

\usepackage{fullpage}
\usepackage[small,bf]{caption}
\usepackage{graphicx,amsfonts,amscd,amssymb,bm,cite,epsfig,epsf,url,color,enumerate}
\usepackage{appendix}
\usepackage{subfigure}

\usepackage{algorithm}
\usepackage{algpseudocode}
\usepackage{amsmath}
\usepackage{multicol}
\usepackage{makecell}
\usepackage{dsfont}
\usepackage{ragged2e}

\newtheorem{theorem}{Theorem}
\newtheorem{lemma}{Lemma}
\newtheorem{proposition}{Proposition}
\newtheorem{corollary}{Corollary}
\newtheorem{remark}{Remark}
\newtheorem{assumption}{Assumption}

\allowdisplaybreaks[4]

\title{Quantized Federated Learning under Transmission Delay and Outage Constraints}

\author{
Yanmeng~Wang,
Yanqing~Xu,
Qingjiang~Shi,
and~Tsung-Hui~Chang % stops a space
\thanks{Y. Wang, Y. Xu and T.-H. Chang are with the School of Science and Engineering, The Chinese University of Hong Kong, Shenzhen
518172, China, and also with the Shenzhen Research Institute of Big
Data, Shenzhen 518172, China (e-mail: hiwangym@gmail.com, xuyanqing@cuhk.edu.cn, tsunghui.chang@ieee.org).
Q. Shi is with the School of Software Engineering, Tongji University,
Shanghai 201804, China, and also with the Shenzhen Research Institute of Big
Data, Shenzhen 518172, China (e-mail: shiqj@tongji.edu.cn).
(Corresponding author: Tsung-Hui~Chang.)}}

\date{}

\begin{document}

\maketitle

\begin{abstract}
Federated learning (FL) has been recognized as a viable distributed learning paradigm which trains a machine learning model collaboratively with massive mobile devices in the wireless edge while protecting user privacy.
Although various communication schemes have been proposed to expedite the FL process, most of them have assumed ideal wireless channels which provide reliable and lossless communication links between the server and mobile clients.
Unfortunately,  in practical systems with limited radio resources such as constraint on the training latency and constraints on the transmission  power and bandwidth, transmission of a large number of model parameters  inevitably suffers from quantization errors (QE) and transmission outage (TO).
In this paper, we consider such non-ideal wireless channels, and carry out the first analysis showing that the FL convergence can be severely jeopardized by TO and QE,  but intriguingly can be alleviated if the clients have uniform outage probabilities.
These insightful results motivate us to  propose a robust FL scheme, named \texttt{FedTOE}, which performs joint allocation of wireless resources and quantization bits across the clients to minimize the QE while making the clients have the same TO probability.
Extensive experimental results are presented to show the superior performance of \texttt{FedTOE} for deep learning-based classification tasks with transmission latency constraints.
\\\\
\noindent {\bfseries Keywords}$-$ Federated learning, transmission outage, quantization error, convergence rate, wireless resource allocation.
\\\\
\end{abstract}

\section{Introduction}

With the rapid development of mobile communications and artificial intelligence (AI),  the edge AI,  a system that exploits locally generated data to learn a machine learning (ML) model at the wireless edge, has attracted increasing attentions from both the academia and industries \cite{wang2019edge,zhu2020toward,lim2020federated}.
In particular, federated learning (FL) has been proposed to allow an edge server to coordinate massive mobile clients to collaboratively train a shared ML model without accessing the raw data of clients \cite{mcmahan2017communication}.
However, FL faces several critical challenges.
This includes that the mobile clients have dramatically different data distribution (data heterogeneity) and different computation capabilities (device heterogeneity) \cite{wang2020tackling}.
Moreover, the training is subject to training latency and limited communication resources for serving a large number of clients.
In view of this, the well-known \texttt{FedAvg} algorithm \cite{mcmahan2017communication} with local stochastic gradient descent (local SGD) and partial participation of clients is widely adopted to reduce the training latency and communication overhead \cite{li2019convergence}.
Furthermore, several improved FL algorithms have been proposed to reduce the inter-client variance caused by data heterogeneity \cite{liang2019variance,karimireddy2020scaffold} and device heterogeneity \cite{wang2020tackling,li2018federated}.

\subsection{ Related Works}

Recently,  wireless resource scheduling has been introduced for FL from different perspectives.
Firstly, some works have aimed to reduce the total training latency by
improving the data throughput between the clients and the server under limited resource budget.
For example, \cite{yang2020federated} adopted joint client selection and beamforming design at the server to maximize the number of selected clients while guaranteeing the mean squared error performance of the received data at the server,
while \cite{abad2020hierarchical} introduced a hierarchical FL framework to maximize the transmission rate in the uplink under the bandwidth and transmit power constraints.
With a slight difference, \cite{xu2020client} proposed a 	``later-is-better" principle to jointly optimize the client selection and bandwidth allocation throughout the training process under a total energy budget.
However,  all the above works did not explicitly consider the influence of resource allocation on the FL performance,  and thus cannot directly minimize the training latency.

Secondly,  some works aimed to achieve a high learning performance within a total training latency, through analyzing the theoretical relations between the number of communication rounds and achieved learning accuracy.
For instance, based on the number of communication rounds required to attain a certain model accuracy, \cite{shi2020device} and \cite{yang2020delay} proposed to optimize bandwidth allocation to minimize the total latency of the \texttt{FedAvg} algorithm.
The work \cite{tran2019federated} optimized resource allocation under delay constraints and captured two tradeoffs, including the tradeoff between computation and communication latencies as well as that between training latency and energy consumption of all clients.
While these works can minimize the training latency directly, they have assumed ideal wireless channels with reliable and lossless transmissions.

Some recent works have considered FL and wireless resource allocation under non-ideal wireless environments.
For example, the work \cite{chen2020joint} studied the influence of packet error rate on the convergence of \texttt{FedAvg}, and proposed a joint resource allocation and client selection scheme to improve the convergence speed of \texttt{FedAvg}.
The work \cite{salehi2020federated} attempted to redesign the averaging scheme of local models based on the transmission outage (TO) probabilities.
The work \cite{zhu2020one} exploited the waveform-superposition property of broadband channels to reduce the transmission delay, and also investigated the impacts of channel fading and imperfect channel knowledge on the FL convergence.
The work in \cite{wang2021edge} proposed a unit-modulus over-the-air computation framework for FL to simultaneously upload local model parameters and update global model parameters via analog beamforming, and analyzed the influence of the noise in both uplink and downlink channels on the transmitted model parameters.
On the other hand, some works considered compressed transmission via quantization and analyzed the influence of the quantization error (QE) on the FL performance.
For instance, \cite{reisizadeh2020fedpaq} proposed a communication-efficient FL method, \texttt{FedPAQ}, which sends the quantized global model in the downlink, and then analyzed the effect of QE on the convergence of FL.
Besides, the authors of \cite{ZhenJSAC2020} considered layered quantized transmissions for communication-efficient FL where different quantization levels are assigned to different layers of the trained neural network.
It is noted that in the aforementioned works \cite{chen2020joint, salehi2020federated, zhu2020one, reisizadeh2020fedpaq, ZhenJSAC2020}, the issues of TO and QE have never been considered simultaneously.
An interesting recent work \cite{jin2020design} has considered the distributed \texttt{SignSGD} algorithm (which uses one-bit quantization) with TO in the uplink channels. It analyzed the algorithm convergence properties and studied joint communication and computation resource allocation problems to minimize the device energy consumption and maximize the learning performance, respectively.
However,  \texttt{SignSGD} does not consider local SGD and partial client participation for communication cost reduction, and cannot flexibly adapt different quantization levels.

\subsection{Contributions}

In this paper, we highlight the need of studying the joint impacts of TO and QE on FL,  especially when the transmission latency is constrained.
Specifically, given a transmission delay constraint, a larger number of quantization bits lead to a smaller QE of the transmitted model but demand a higher transmission rate, which however result in a larger TO probability \cite{goldsmith2005wireless}.
Therefore, either when the model size is large or when the latency constraint is stringent, it is essential to take into account both TO and QE in the FL process.
In view of this, unlike the existing works \cite{chen2020joint, salehi2020federated, zhu2020one, reisizadeh2020fedpaq, ZhenJSAC2020},
we generalize \cite{jin2020design} to the celebrated \texttt{FedAvg} algorithm with flexible quantization levels, and study the joint effects of TO and QE. Moreover,  we consider that the clients have non-i.i.d.  data distribution.
To overcome these effects, we propose a new FL scheme,  called \texttt{FedTOE} (\underline{Fed}erated learning with \underline{T}ransmission \underline{O}utage and quantization \underline{E}rror), which performs joint allocation of wireless resources and quantization bits for achieving robust FL performance under such non-ideal learning environment.
In particular,  our main contributions include:

\begin{enumerate}[(1)]
\item
\textbf{FL convergence analysis under both TO and QE:}
We consider a non-convex FL problem, which is more general than the convex problems studied in \cite{chen2020joint, salehi2020federated,ZhenJSAC2020}, and consider non-ideal (uplink) wireless channels with both TO and QE.
To the best of our knowledge, this paper is the first to analyze the influence of both TO and QE on the convergence of \texttt{FedAvg} simultaneously.
The derived theoretical results show that non-uniform TO probabilities not only lead to a biased solution \cite{wang2020tackling} but also amplify the negative effects caused by QE and non-i.i.d.  data distribution (data heterogeneity). Intriguingly,  such undesired property can be alleviated if the clients have the same TO probabilities.

\item
\textbf{\texttt{FedTOE}:}
Inspired by this observation,  we formulate a resource allocation problem to mitigate the impacts of TO and QE.  Specifically,  we propose to carefully allocate the (uplink) transmission bandwidth and quantization bits of clients to
minimize the aggregate QE subject to constraints on the transmission latency and uniform TO probabilities.
We show that a high-quality approximate solution to this problem can be efficiently obtained by a simple gradient projection algorithm.

\item
\textbf{Experiments:}
The proposed \texttt{FedTOE} is implemented for two deep learning-based tasks, including the handwritten-digit recognition on the MNIST dataset and the color image classification on the CIFAR-10 dataset.
The experimental results demonstrate that \texttt{FedTOE} has promising performance over benchmark schemes.
\end{enumerate}

\textbf{Synopsis:} Section \ref{Sec:System model} introduces the proposed system model of FL in the wireless environment.  Section \ref{performance_analysis} presents the convergence rate analysis of FL under both TO and QE.  Based on the results,  the wireless resource allocation scheme (i.e., \texttt{FedTOE}) is formulated in Section \ref{Sec:Resource}. The experiment results are presented in Section \ref{sec: numerical results}.  Section \ref{sec: Conclusion} concludes this paper.

\section{System model}\label{Sec:System model}

\subsection{Federated Learning Algorithm}\label{Sec:System model_FL}

Consider a wireless FL network as shown in Fig. 1 where a central server coordinates $N$ mobile clients to solve the following distributed learning problem
\begin{align}\label{objective function}
\min \limits_{{\mathbf{w}} \in \mathbb{R}^{m} } \; F({\mathbf{w}}) = \sum\limits_{i = 1}^N p_i F_i({\mathbf{w}}) \, \text{,}
\end{align}
where $F_i({\bf{w}})$ is the (possibly) non-convex local loss function, $\mathbf{w} \in \mathbb{R}^{m}$ denotes the ${m}$-dimensional model parameters to be learned,  and $p_i = n_i / \sum_{j = 1}^N n_j$ in which $n_i$ is the number of data samples  stored in client $i$.
Let ${\bm \xi}_{i}$ be the mini-batch samples with size $b$, we denote
$F_i({\mathbf{w}},{\bm \xi}_{i}) = \frac{1}{b} \sum_{j = 1}^{b} f({\mathbf{w}},\xi_{ij})$, where $\xi_{ij}$ is the $j$-th randomly selected sample from the dataset of client $i$, and $f({\mathbf{w}},\xi_{ij})$ is the model loss function with respect to $\xi_{ij}$. When $b = n_i$, ${\bm \xi}_{i}$ refer to the whole local dataset in client $i$ and then $F_i({\mathbf{w}},{\bm \xi}_{i}) = F_i({\mathbf{w}})$.
\begin{figure}[t]
\centerline{\includegraphics[width = 2.85  in]{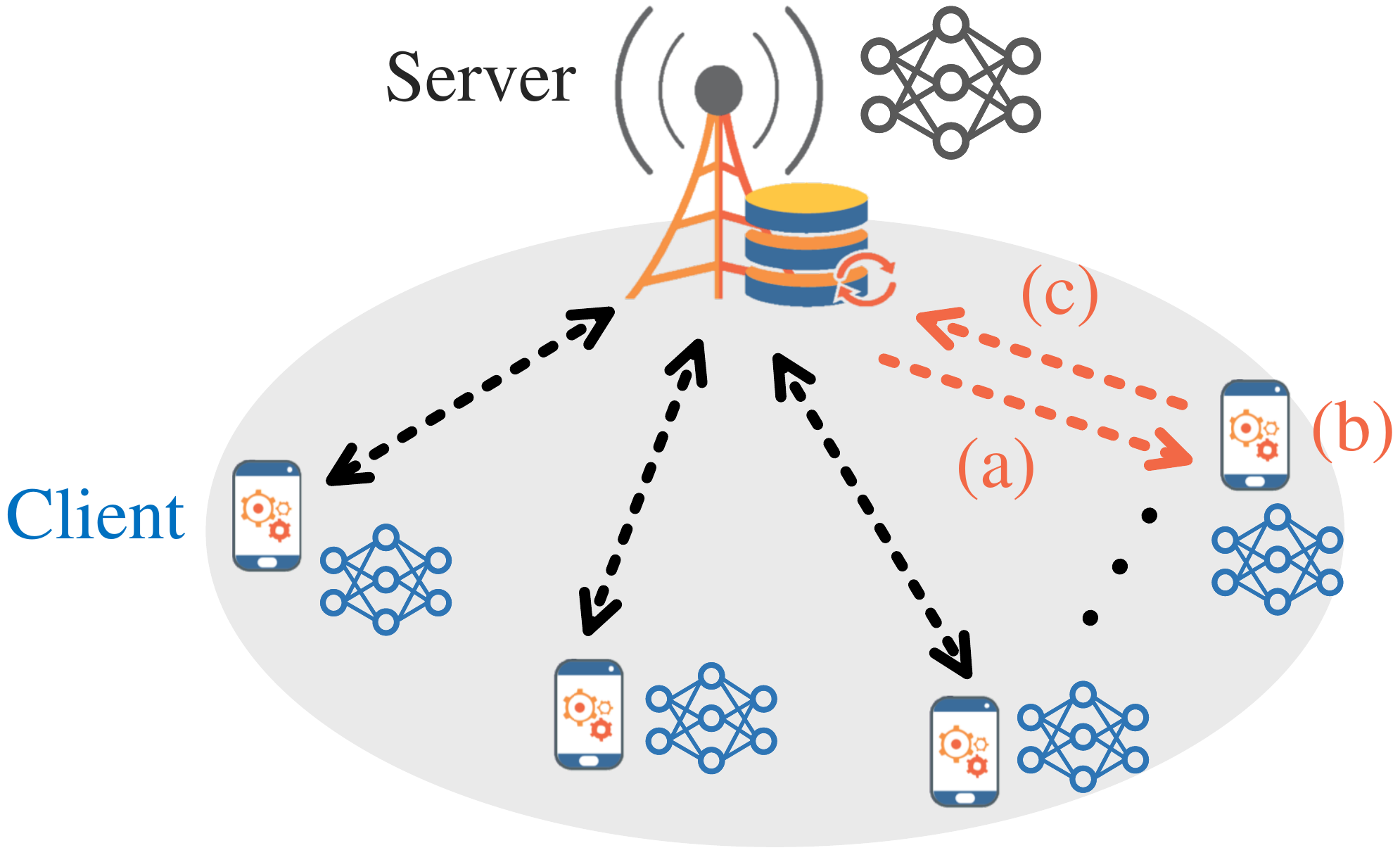}}
\caption{Federated learning in wireless edge.}
\label{fig1}
\end{figure}

We follow the seminal \texttt{FedAvg} algorithm \cite{mcmahan2017communication}.
Specifically, in the ${r}$-th communication round, \texttt{FedAvg} executes the following three steps (see Fig.  1):
\begin{enumerate}[(a)]
\item \textbf{Broadcasting}: The server samples $K$ clients, denoted by the set ${\mathcal{S}}_{r}$ where $|{\mathcal{S}}_{r}|=K$, and then broadcasts the global model ${\bar {\mathbf{w}}}_{r-1}$ in the last communication round to each client $i \in {\mathcal{S}}_{r}$.

\item \textbf{Local model updating}: Each client $i \in {\mathcal{S}}_{r}$ updates local model by local stochastic gradient descent (local SGD) \cite{liang2019variance}.
    It contains $E$ consecutive SGD updates as follows
\begin{align}\label{local model}
\begin{aligned}
& \mathbf{w}^{r,0}_{i}  =  {\bar {\mathbf{w}}}_{r-1} \\
&  \mathbf{w}^{r,\ell}_{i}   =  {\mathbf{w}}^{r,\ell-1}_{i}  -  \gamma \nabla F_{i}({\mathbf{w}}^{r,\ell-1}_{i},{\bm \xi}^{r,\ell}_{i}) , \ell  = 1, \ldots , E \text{,}
\end{aligned}
\end{align}
where $\gamma>0$ is the learning rate.

\item \textbf{Aggregation}: The selected clients upload their local model $\mathbf{w}_i^{r,E}$ to the server for producing a new global model based on certain aggregation principle.
\end{enumerate}
Specifically,  \texttt{FedAvg} considers the following two aggregation schemes, depending on whether all clients participate or not.
\begin{enumerate}[(i)]
\item \textbf{Full participation}:
All clients participate in the aggregation process, i.e., ${\mathcal{S}}_{r} = \{1,\cdots,N\}$ $\forall {r}$, and the global model is updated by
\begin{align}\label{full_participation}
    {\tilde {\mathbf{w}}}_{r} = \sum\limits_{i = 1}^N p_i \mathbf{w}^{r,E}_{i} \text{.}
\end{align}
Considering the massive participates in the network, this scheme would not be feasible under limited communication bandwidth for the uplink channels.
\item \textbf{Partial participation}:
With $|{\mathcal{S}}_{r}| \ll N$, the global model is updated by
\begin{align}\label{partial_participation}
{\bar {\mathbf{w}}}_{r} = \frac{1}{K} \sum\limits_{i \in {\mathcal{S}}_{r}} {\mathbf{w}}^{r,E}_{i} \text{,}
\end{align}
where $K$ clients ($K\ll N$) in ${\mathcal{S}}_{r}$ are selected with replacement according to the probability distribution $\{ p_1,\cdots,p_N\}$.
It should be pointed out that the average scheme in \eqref{partial_participation} leads to an unbiased estimate of ${\bar {\mathbf{w}}}_{r}$ in \eqref{full_participation}, i.e., $\mathbb{E}[{\bar {\mathbf{w}}}_{r}] = {\tilde {\mathbf{w}}}_{r}$ \cite{li2019convergence}.
\end{enumerate}

However, the aforementioned schemes are still far from practice.
In particular, in digital communication systems, the model parameters need to be quantized before being transmitted, which brings QEs to the learned model.  Meanwhile,  channel fadings could cause TO in the delivery of the model parameters from time to time.
Moreover,  given a fixed transmission delay,  QE is strongly coupled with TO.  Specifically,  a larger number of quantization bits lead to a smaller QE of the learned model but require a higher transmission rate,   which however can further elevate the TO probability.
Therefore,  it is essential to consider TO and QE simultaneously in the wireless FL systems.
Such issue has been considered in \cite{jin2020design} for \texttt{SignSGD} with 1-bit quantization and TO in wireless channels,  but neither the partial client participation nor the local SGD is considered.  The influence of different quantization levels on the learning performance cannot be revealed either.
In the next two subsections, we focus on the \texttt{FedAvg} algorithm described above and incorporate QE and TO in the uplink channels\footnote{In the current work,  we only consider the TO and QE in the uplink transmission since the server (i.e., base station) is assumed to be powerful enough to provide reliable and lossless communications for the downlink broadcast channels \cite{reisizadeh2020fedpaq}.}.

\subsection{Quantized Transmission}

For the local model $\mathbf{w}^{r,E}_{i}$, we assume that each parameter $w^{r,E}_{{i}j}$ is bounded satisfying $|w^{r,E}_{{i}j}| \in [{\underline{w} }^{r}_{{i}j}, {\bar w}^{r}_{{i}j} ]$, and is quantized by the stochastic quantization method in  \cite{amiri2020federated}.
In concrete terms, with $B^{r}_{i}$ quantization bits, we denote $\{c_0,c_1,\cdots,c_{2^{B^{r}_{i}} -1} \}$ as the knobs uniformly distributed in $[{\underline{w} }^{r}_{{i}j}, {\bar w}^{r}_{{i}j}]$, where
\begin{align}\label{c_j}
c_{u} = {\underline{w} }^{r}_{{i}j} + u \times \frac{{\bar w}^{r}_{{i}j} - {\underline{w} }^{r}_{{i}j} }{2^{B^{r}_{i}} -1}, \;
u = 0, \cdots, 2^{B^{r}_{i}}-1 \text{.}
\end{align}
Then, the parameter $w^{r,E}_{{i}j} \!$ falling in $[c_u, c_{u+1} )$ is quantized by
\begin{small}
\begin{align}\label{quantization_method}
\mathcal{Q}(w^{r,E}_{{i}j}) =
\left\{
\begin{aligned}
&  {\rm sign}(w^{r,E}_{{i}j}) \cdot c_u , \quad\;  \text{w.p.} \; \frac{c_{u+1} - |w^{r,E}_{{i}j}| }{c_{u+1} - c_u} \text{,} \\
& {\rm sign}(w^{r,E}_{{i}j}) \cdot c_{u+1} , \; \text{w.p.} \; \frac{|w^{r,E}_{{i}j}| - c_u}{c_{u+1} - c_u} \text{,}  \\
\end{aligned}
\right.
\end{align}
\end{small}

\noindent where `w.p.' stands for `with probability'.  In addition,  let $\mu$ be the number of bits used to represent ${\rm sign}(w^{r,E}_{{i}j})$, ${\underline{w} }^{r}_{{i}j}$ and ${\bar w}^{r}_{{i}j}$.
Then,  the quantized local model $\mathcal{Q}(\mathbf{w}^{r,E}_{i}) = [ \mathcal{Q}(w^{r,E}_{{i}1}), \cdots, \mathcal{Q}(w^{r,E}_{im})]$ is expressed by a total number of
\begin{align}\label{quantized model length}
{\hat B}^{r}_{i} = m B^{r}_i + \mu \; \text{bits} \, \text{,}
\end{align}
and is sent to the server.

\begin{lemma}\label{lemma:unbiased_estimation}
With the stochastic quantization method, each local model is unbiasedly estimated as
\begin{align}\label{unbiased_estimation}
\mathbb{E} [ \mathcal{Q}(\mathbf{w}^{r,E}_{i}) ] = \mathbf{w}^{r,E}_{i} \text{,}
\end{align}
and the associated QE is bounded by
\begin{align}\label{quantization error_local_model}
\mathbb{E} [ \| \mathcal{Q}(\mathbf{w}^{r,E}_{i}) - \mathbf{w}^{r,E}_{i} \|^2 ]
\leq { \delta_{ir}^2 }/{( 2^{B^{r}_{i}} - 1 )^2 } \triangleq J_{ir}^2 \, \text{,}
\end{align}
where $\delta_{ir}\triangleq \sqrt{\frac{1}{4} \sum_{j=1}^{m} ({\bar w}^{r}_{{i}j}-{\underline{w} }^{r}_{{i}j})^2} $.
\end{lemma}

\emph{Proof:} Properties like Lemma \ref{lemma:unbiased_estimation} have been discussed in the literature; see \cite{reisizadeh2020fedpaq} and \cite{ZhenJSAC2020}.
For ease of reference,  the proof is presented in Section A of the Supplementary Material.
\hfill  $\blacksquare$

As one can see from (\ref{quantized model length}) and (\ref{quantization error_local_model}) that a higher quantization level $B^{r}_{i}$ leads to a larger number of bits ${\hat B}^{r}_{i} $  for transmission but a smaller QE.

\subsection{Transmission Outage}

The TO can happen in various wireless scenarios.
For example,
1) without channel state information at the transmitter (CSIT), the transmission may suffer from outage due to large-scale fadings such as shadowing \cite{chen2020joint};
2) with imperfect CSIT (e.g., imperfect channel estimation or finite bandwidth feedback), the CSI error could cause transmission outage \cite{wang2014outage};
3) with perfect CSIT, due to finite blocklength transmission, the receiver may fail to decode the message \cite{xu2020transmission}.
In this work, for simplicity, we will consider the case of no CSIT and focus on the impacts of shadowing on the TO of the system.
The system without CSIT removes the need of CSI feedback and power control, which makes the FL system easier to implement especially in the large-scale IoT scenarios \cite{jin2020design,xia2019virtual}.

By assuming that the frequency division multiple access (FDMA) is adopted for uplink transmission, the channel capacity of each client $i \in {\mathcal{S}}_{r}$ is
\begin{align}\label{channel_capacity}
{C}^{r}_{i} = W^{r}_{i} \log_2 \left( 1 + \frac{P^{r}_{i} | h_{i} |^2 }{{W^{r}_{i}N_0}} \right) \; \text{bps,}
\end{align}
where $W^{r}_{i}$ and $P^{r}_{i}$ denote the allocated bandwidth and transmit power of client $i$, respectively, $h_{i}$ is the uplink channel coefficient between the server and client $i$, and $N_0$ represents the power spectrum density (PSD) of the additive noise.
According to the channel coding theorem \cite{goldsmith2005wireless}, if the transmission rate $R^{r}_{i}$ is higher than ${C}^{r}_{i}$, TO occurs and the server fails to decode $\mathcal{Q}(\mathbf{w}_i^{r,E})$ correctly;
that is, the outage probability is given by
\begin{align}\label{outage_probability}
q^{r}_{i} \triangleq {\rm Pr} ( C^{r}_{i} \leq R^{r}_{i} ).
\end{align}
Suppose that the uplink transmission is subject to a delay constraint $\tau_i$, then $R^{r}_{i} = {\hat B}^{r}_{i} / \tau_i$. Thus, either a larger quantization level or a more stringent delay constraint can enlarge the TO.

We model the channel gain in (\ref{channel_capacity}) using the classical path loss model with shadowing \cite{goldsmith2005wireless}, i.e.,
$[ | h_{i} |^2 ]_{\rm dB}
= [\mathcal{K}]_{\rm dB} - \lambda [d_{i}]_{\rm dB} + \psi_{\rm dB}$,
where $[x]_{\rm dB}$ measures $x$ in dB, $\mathcal{K}$ is a constant depending on the antenna characteristics and channel attenuation, $\lambda$ is the path loss exponent, $d_{i}$ (in meter) is the distance between client ${i}$ and the server, and $\psi_{\rm dB} \sim \mathcal{N}(0, \sigma^2_{\rm dB})$ is the shadowing in which $\sigma^2_{\rm dB}$ is the shadowing variance.
Then,  the TO probability in (\ref{outage_probability}) can be computed as
\begin{align}\label{outage_probability_shadow}
q^{r}_{i}
= {\rm Pr} ( \psi_{\rm dB} < \rho _i)
= 1 - Q( \rho_i/\sigma_{\rm dB} )
\, \text{,}
\end{align}
where $Q(x) = \int_{x}^{+\infty} \frac{1}{\sqrt{2 \pi}} \exp ( - \frac{1}{2} z^2 ) {\rm d} z$ is the Q-function and
$\rho_i \triangleq [ (2^{R^{r}_{i}/W^{r}_{i}} -1)W^{r}_{i} N_0 ]_{\rm dB} - [ P^{r}_{i} ]_{\rm dB} - [\mathcal{K}]_{\rm dB} + \lambda [d_{i}]_{\rm dB}$.
As seen,  with $q^{r}_{i} < 0.5$ and $\sigma_{\rm dB} \geq 0$, the TO probability $q^{r}_{i}$ is an increasing function of $\sigma_{\rm dB}$.

\subsection{Federated Learning with QE and TO}

Let us reconsider the \texttt{FedAvg} in Section \ref{Sec:System model_FL} in the presence of both TO and QE in the uplink.
According to \cite{ZhenJSAC2020} and \cite{amiri2020federated}, it is more bit-efficient to transmit the model updates (i.e.,  $\mathbf{w}^{r,E}_{i} - \mathbf{w}^{r,0}_{i}$) than the model $\mathbf{w}^{r,E}_{i}$ itself in the uplink since the dynamic ranges of model updates can decrease with the number of communication rounds.
By adopting this scheme,  each client $i$ sends to the server with
\begin{align}\label{local_model_update}
\mathcal{Q}  \left( \Delta \mathbf{w}^{r}_{i}\right)
 &\triangleq \mathcal{Q} \left( \frac{1}{\gamma} (\mathbf{w}^{r,E}_{i}  -  \mathbf{w}^{r,0}_{i})  \right)\!=\!  \mathcal{Q} \left(\!\sum \limits_{\ell=1}^{E} \nabla F_{i} (\mathbf{w}^{r,\ell-1}_{i}, \bm{\xi}^{r,\ell}_{i} )  \right).
\end{align}
Due to TO, the server may fail to receive the upload messages.
We denote ${\mathds{1}^{r}_{i}}=1$ if the server correctly receives the transmitted local model from client ${i}$, and ${\mathds{1}^{r}_{i}}=0$ otherwise.
Then, with the partial participation scheme in (\ref{partial_participation}), the global model at the server is obtained by
\begin{align}\label{global_model_update}
{\bar {\mathbf{w}}}_{r} = {\bar {\mathbf{w}}}_{r-1} - \gamma \frac{\sum \limits_{{i} \in {\mathcal{S}}_{r}} \mathds{1}^{r}_{i} \mathcal{Q} \left( \Delta \mathbf{w}^{r}_{i} \right)}{\sum \limits_{{i} \in {\mathcal{S}}_{r}} \mathds{1}^{r}_{i}} \text{.}
\end{align}
Note that when the channel is ideal without TO and QE, then \eqref{global_model_update} reduces to the simple averaging scheme in \eqref{partial_participation}.
We assume that the server can use cyclic redundancy check (CRC) to check whether the failure occurs or not \cite{chen2020joint}.
If $\sum_{{i} \in {\mathcal{S}}_{r}} \mathds{1}^{r}_{i} = 0$,
i.e., none of the clients successfully transmit their local updates,
retransmission is carried out until at least one client's message is correctly received by the server

In the downlink transmission, the global model (i.e., ${\bar {\mathbf{w}}}_{r}$) is sent to each client $i \in {\mathcal{S}}_{r}$ (assuming no TO and QE).
Such consideration is based on the following two reasons.
First, the wireless resources of the server for broadcasting transmission are arguably abundant to transmit global model parameters reliably with high precision \cite{reisizadeh2020fedpaq}.
Second, the selected clients differ from round to round,  and thus it requires additional caching mechanism to track the latest global model if the server transmits model difference ${\bar {\mathbf{w}}}_{r} - {\bar {\mathbf{w}}}_{r-1}$; see \cite{ZhenJSAC2020,sattler2019robust} for the details.
The described FL algorithm with uplink TO and QE is summarized in Algorithm 1.

\begin{figure}[t]
\begin{minipage}[t]{1\linewidth}
\centering
\subfigure[\scriptsize{Training loss.}]{
\includegraphics[width= 2.2 in ]{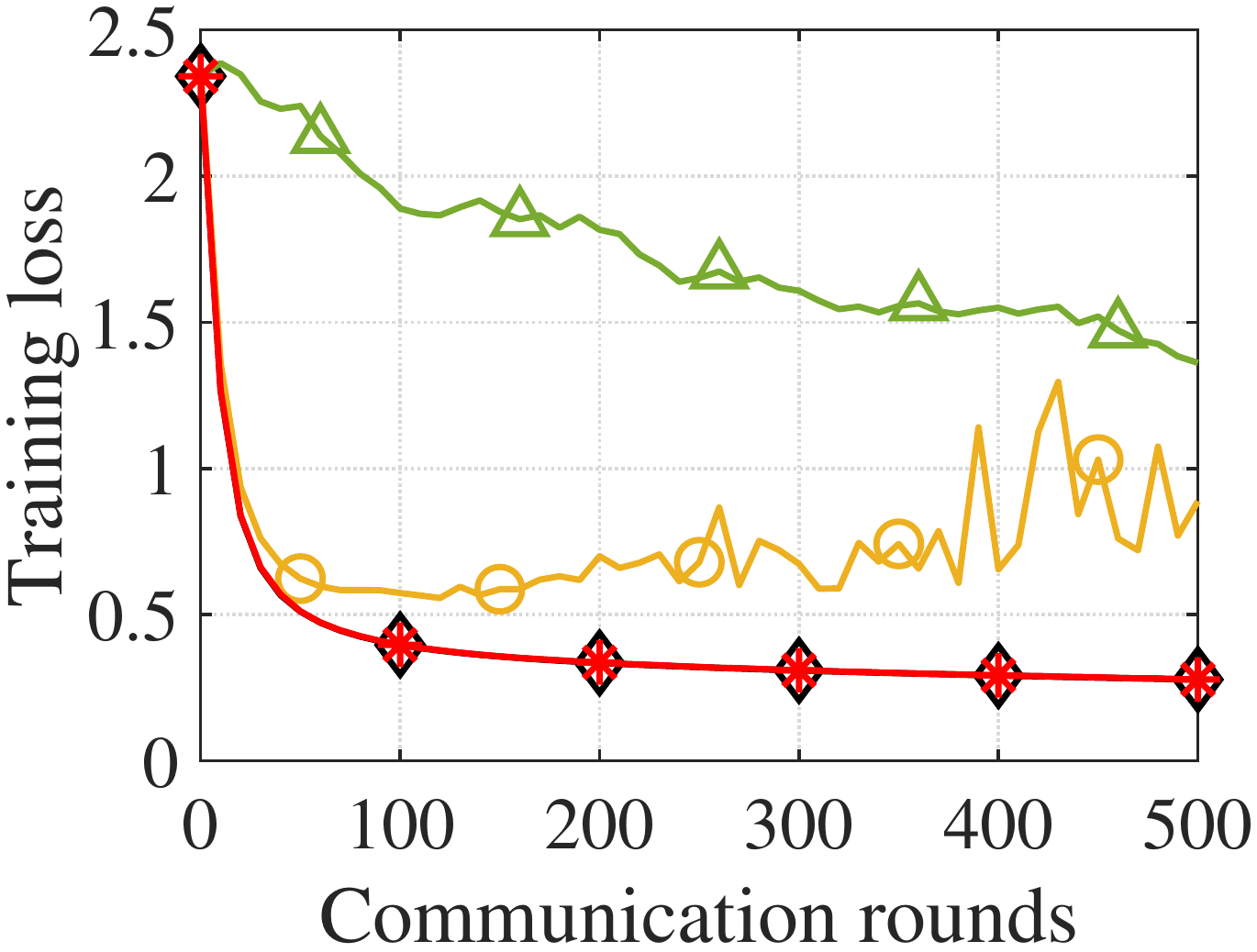}}
\subfigure[\scriptsize{Testing accuracy.}]{
\includegraphics[width= 2.2 in ]{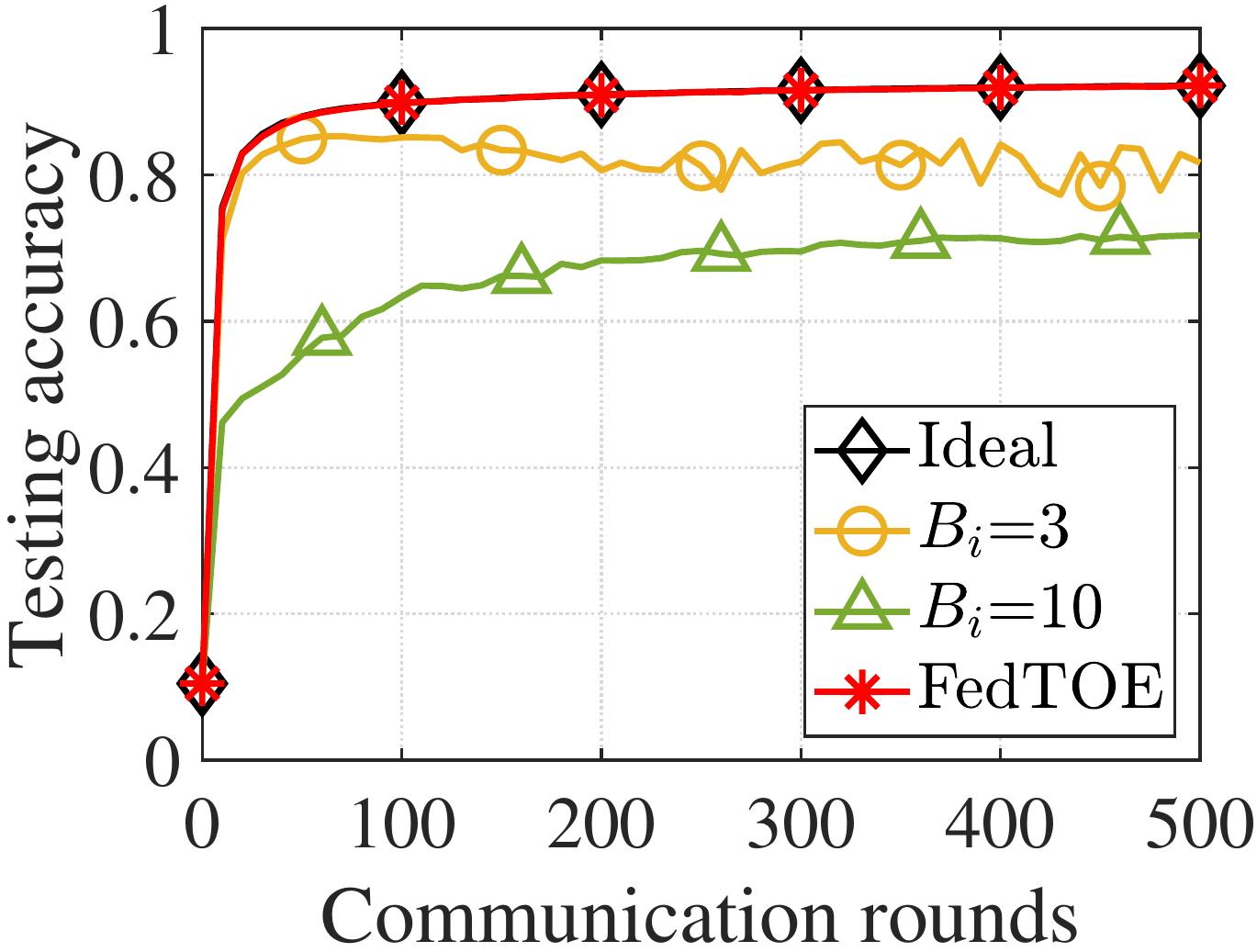}}
\caption{Training loss and testing accuracy comparison of different schemes in wireless environment,
where the uplink transmission delay per communication round is constrained by 100ms.}
\label{fig:Impact_TO_QE}
\end{minipage}
\end{figure}

\begin{remark}\rm
Fig. \ref{fig:Impact_TO_QE} illustrates the influence of TO and QE on the FL with full participation (i.e., $K=N=100$) and the presence of non-i.i.d. data distribution.
The ideal scheme suffers neither TO nor QE, while the curves with $B_i=3$ and $10$ refer to the schemes which allocate uniform bandwidth and same quantization level $B_i$ to all clients.
For a more detailed setting,  refer to Section \ref{parameter_setting}.
One can see from this figure that the scheme with fewer quantization bits (i.e., $B_i = 3$) has an impaired performance due to large QE, whereas the one with more quantization bits (i.e., $B_i = 10$) not only has a slower convergence rate but also does not move to the right solution due to the bias caused by TO (which will be shown in Theorem \ref{Thm_diff_TO_fixed}).
Therefore, the wireless resource and quantization bits need to be carefully allocated.
\end{remark}

In view of this, a robust FL scheme is proposed in this paper,  referred to as \texttt{FedTOE}, which can exhibit robustness in such non-ideal wireless channels with TO and QE as shown in Fig. \ref{fig:Impact_TO_QE}.
We first present a novel theoretical analysis on the convergence of Algorithm 1 in the next section, based on which, a joint wireless resource and quantization bits allocation scheme will be presented to improve the FL performance under TO and QE in Section \ref{Sec:Resource}.

\begin{algorithm}[t]
\caption{\texttt{FedTOE}: FL with uplink TO and QE}
\begin{algorithmic}[1]
\State Initialize global model ${\bar {\mathbf{w}}}_{0}$ by the server.
\For{$r=1,2,\cdots,M$}
\State Server samples $K$ clients $\mathcal{S}_{r}$ with replacement based
\Statex \quad\, on the probabilities $\{p_1,\cdots,p_N\}$;
\State Server broadcasts global model ${\bar {\mathbf{w}}}_{r-1}$ to clients in $\mathcal{S}_{r}$;
\For {client ${i} \in \mathcal{S}_{r}$} (\textbf{in parallel})
\State $\mathbf{w}^{r,0}_{{i}} \leftarrow {\bar {\mathbf{w}}}_{r-1}$
\For {$\ell = 1,2, \cdots, E$}
\State Update local model by mini-batch SGD in (\ref{local model});
\EndFor
\State Send quantized model update in (\ref{local_model_update}) to the server;
\EndFor
\If {$\sum_{{i} \in {\mathcal{S}}_{r}} \mathds{1}^{r}_{{i}} = 0$}
\State Repeat Step 10 for all clients in $\mathcal{S}_{r}$;
\Else
\State {Server updates global model by (\ref{global_model_update})};
\EndIf
\EndFor
\end{algorithmic}
\end{algorithm}

\section{Performance analysis}\label{performance_analysis}

\subsection{Assumptions}\label{assumption}

We consider general smooth non-convex learning problems with the following assumptions.

\begin{assumption}
Each local function $F_i$ is lowered bounded, i.e., $F_i({\mathbf{w}}) \geq \underline{F} > - \infty$,  and
differentiable whose $\nabla F_i$ is Lipschitz continuous with constant $L$: $\forall$${\mathbf{v}}$ and ${\mathbf{w}}$, $F_i({\mathbf{v}}) \leq F_i({\mathbf{w}}) + ({\mathbf{v}} - {\mathbf{w}})^T\nabla F_i({\mathbf{w}}) + \frac{L}{2} \| {\mathbf{v}} - {\mathbf{w}} \|_2^2$.
\end{assumption}

\begin{assumption}
Unbiasedness and bounded variance of SGD: ${\mathbb E}[ \nabla  F_i({\mathbf{w}},{\xi}_{ij}) ] = {\mathbb E}[ \nabla  F_i({\mathbf{w}}) ]$,
${\mathbb E}[\| \nabla  F_i({\mathbf{w}},{\xi}_{ij}) - \nabla  F_i({\mathbf{w}})\|^2]  \leq \sigma^2$.
\end{assumption}

\begin{assumption}
Bounded data variance: ${\mathbb E}[\| \nabla F_i({\mathbf{w}}) - \nabla F ({{\mathbf{w}}})\|^2] \leq D_i^2$, $\forall i =1,\cdots,N$,
which measures the heterogeneity of local datasets \cite{lian2017can}.
\end{assumption}

\subsection{Theoretical results}\label{Theoretical_results}

For ease of presentation,  we consider the fixed quantization level and constant TO probabilities across the training process, i.e.,
$B^{r}_i = B_i$  and  $q^{r}_i = q_i$ for all $r=1,\cdots,M$.
As one will see, such simplification is sufficient to reveal the insight how TO and QE impact on the algorithm convergence.
The extension to the more general case is straightforward and presented in the Supplementary Material.

We first present the following lemma.

\begin{lemma}\label{beta_alpha_K}
Considering the FL algorithm in Algorithm 1, it holds true that
\begin{small}
\begin{align}\label{average_beta_1}
\mathbb{E}
\left[  \left. \frac{ {\sum}_{{i} \in {\mathcal{S}}_{r}} \mathds{1}^{r}_{i}
\Delta \mathbf{w}^{r}_i
 }{ {\sum}_{{i} \in {\mathcal{S}}_{r}} \mathds{1}^{r}_{i}} \right|
\sum \limits_{{i} \in {\mathcal{S}}_{r}} \mathds{1}^{r}_{i} \neq 0  \right]
=
\sum \limits_{i=1}^N {\bar \beta}_i
\Delta \mathbf{w}^{r}_i
\end{align}
\end{small}

\noindent
for some ${\bar \beta}_i\in [0,1]$ with $\sum_{i=1}^N {\bar \beta}_i= 1$, where $\mathbb{E}[\cdot]$ is taken with respect to $\mathcal{S}_{r}$ and $\{ \mathds{1}^{r}_{i} \}$.
Moreover, we also have
\begin{small}
\begin{align}\label{average_alpha_1}
\mathbb{E}
\! \left[ \left. \frac{{\sum}_{{i} \in {\mathcal{S}}_{r}} \mathds{1}^{r}_{i}
\Delta \mathbf{w}^{r}_i }{ \left(\sum_{{i} \in {\mathcal{S}}_{r}} \mathds{1}^{r}_{i} \right)^2 } \right|
\sum \limits_{{i} \in {\mathcal{S}}_{r}} \mathds{1}^{r}_{i} \neq 0 \right]
=
\sum \limits_{i=1}^N {\bar \alpha}_i
\Delta \mathbf{w}^{r}_i
\end{align}
\end{small}

\noindent
for some ${\bar \alpha}_i\geq 0$ $\forall i =1,\ldots,N$, and therefore
\begin{small}
\begin{align}\label{bar_K_1}
& {\mathbb{E}}
\! \left[ \left. \frac{1}{ \sum_{{i} \in {\mathcal{S}}_{r}} \mathds{1}^{r}_{i}} \right| \sum \limits_{{i} \in {\mathcal{S}}_{r}} \mathds{1}^{r}_{i} \neq 0 \right]
= \sum \limits_{i = 1}^N {\bar \alpha}_i
\triangleq \frac{1}{\bar K}
\, \text{.}
\end{align}
\end{small}

When $q_i$ is uniform for all clients, i.e.,  $q_i = q$ $\forall i$, then ${\bar \beta}_i = p_i$  and ${\bar \alpha}_i = p_i /{\bar K}$ $\forall i$ with ${\bar K} = \frac{1 - (q)^K}{\sum_{v = 1}^K \frac{1}{v} \left(\mathbb{C}^v_K (1-q)^{v} (q)^{K-v}\right)}$, where $\mathbb{C}_K^v = \frac{K!}{v!(K-v)!}$.
In addition, if $q_i = 0$ $\forall i$ (no TO), then ${\bar K} = K$.
\end{lemma}

\emph{Proof:} See Appendix A. \hfill $\blacksquare$

From \eqref{average_beta_1}, one can see that $\{{\bar \beta}_i\}$
is the equivalent appearance probabilities of client $i$ in the global aggregation due to client sampling and TO, and they are deviated from $\{p_i\}$
when $\{q_i\}$ are not uniform.
Similarly, $\{{\bar \alpha}_i\}$ defined in \eqref{average_alpha_1} is also related to appearance probabilities of client $i$ but scaled down by the number of active clients (i.e., $\sum_{{i} \in {\mathcal{S}}_{r}} \mathds{1}^{r}_{i}$).
Moreover, in \eqref{bar_K_1}, $\bar K$ represents the average effective number of active clients under TO.
The main convergence result is stated below.

\begin{theorem}\label{Thm_diff_TO_fixed}
Let Assumptions 1 to 3 hold.
If one chooses $\gamma = {\bar K}^{\frac{1}{2}} / (8L{T}^{\frac{1}{2}}) $ and $E \leq T^{\frac{1}{4}}/{\bar K}^{\frac{3}{4}}$  where $T = ME \geq \max \{ {\bar K}^{3}, 1/{\bar K} \}$ is the total number of SGD updates per client,  we have
{\small
\begin{align}\label{theorem_1}
& \frac{1}{M} \sum_{r = 1}^{M} \mathbb{E}\left[ \left\| \nabla  F({\bar {\mathbf{w}}}_{r-1}) \right\|^2 \left| \sum_{{i} \in {\mathcal{S}}_{r}} \mathds{1}^{r}_{i} \neq 0 \right. \right] \notag \\
\leq & \frac{496 L  \left( \mathbb{E}[F({\bar {\mathbf{w}}}_{0})] - \underline{F} \right)}{11  \left( T  {\bar K} \right)^{\frac{1}{2}} }
+  \left(  \frac{ 39 }{ 88  \left( T {\bar K} \right)^{\frac{1}{2}} }
+  \frac{1}{ 88  \left( T  {\bar K} \right)^{\frac{3}{4}} }  \right)  \frac{\sigma^2}{b} 
+  \underbrace{ \frac{31 {\bar K}^{\frac{1}{2}}}{ 88 T^{\frac{3}{2}} } \sum_{r = 1}^{M} \sum_{i = 1}^{N} {\bar \alpha}_i J_{ir}^2}_{ {\rm (a) (caused \ by \ QE)} }
 \notag \\
& +  \underbrace{ \frac{31 }{ 22 \left( T {\bar K} \right)^{\frac{1}{4}} } \sum_{i=1}^N {\bar \alpha}_i D_i^2 }_{ {\rm (b) (caused \ by \ partial \ participation} \atop {\rm  and \ data \ variance)} } 
+ \underbrace{  \left(  \frac{4 }{ 11 \left( T {\bar K} \right)^{\frac{1}{2}} } +  \frac{1}{ 22 \left( T {\bar K} \right)^{\frac{3}{4}} }   \right) \sum \limits_{i = 1}^{N}  {\bar \beta}_i D_i^2 }_{ {\rm (c) (caused \ by \ data \ variance)} }
 + \underbrace{ \frac{62}{11} \chi^2_{\bm{\beta}\|\mathbf{p}}  \sum\limits_{i = 1}^N p_i D_i^2 }_{ {\rm (d) (caused \ by \ TO \ and} \atop {\rm data \ variance)} } \notag \\
& + \underbrace{ \frac{31 }{ 22 \left( T {\bar K} \right)^{\frac{1}{4}} } \sum_{v=2}^K  \frac{(q_{\max})^{K-v}\mathbb{C}^v_K}{1-(q_{\max})^K}
\sum_{i = 1 }^N   p_{i}  (q_i - {\bar q})^2 D_i^2 }_{ {\rm (e) (caused \ by \ TO \ and \ data \ variance)} }
\, \text{,}
\end{align}}

\noindent where $J_{ir}^2$ is given in \eqref{quantization error_local_model},
$\chi^2_{\bm{\beta}\|\mathbf{p}} \triangleq  \sum_{i = 1}^N {({\bar \beta}_i - p_i)^2}/{ p_i}$ is the chi-square divergence \cite{wang2020tackling},  and
$q_{\max} = \max\{q_1,\ldots,q_N\}$ and $\bar q = \sum_{i=1}^N p_i q_i$ are the maximum and average TO probabilities, respectively.
\end{theorem}

\emph{Proof:} Unlike the existing works \cite{chen2020joint, salehi2020federated, zhu2020one, reisizadeh2020fedpaq, ZhenJSAC2020, yu2019parallel, liu2020distributed}, we consider a non-convex FL problem with both TO and QE,  which makes Theorem 1 much more challenging to prove.  In particular,  we adopt the analysis frameworks in \cite{lian2017can,yu2019parallel} and develop several new techniques to deal with the difficulties brought by TO variables  $\mathds{1}^{r}_{i}$ and deviated probabilities ${\bar \beta}_i$ and ${\bar \alpha}_i$.
Details are presented in Appendix B. \hfill $\blacksquare$

It can be found from the right-hand side (RHS) of \eqref{theorem_1} that the convergence of Algorithm 1 can be affected by various parameters, including the quantization error $\{J_{ir}\}$\footnote{
It is worthwhile to remark that the term (a) in the RHS of \eqref{theorem_1} does not depend on specific quantization schemes.
Other quantization,  compression or sparsification methods may also be employed as long as the unbiasedness and bounded error properties in Lemma \ref{lemma:unbiased_estimation} hold.}, the outage probabilities $\{q_i\}$, the local data heterogeneity level $\{D_i\}$, and the effective number of active clients $\bar K$.
As seen from terms (a)-(c), both the quantization error $\{J_{ir}\}$ and local data heterogeneity level $\{D_i\}$ can deteriorate the algorithm convergence.
Besides, when the outage probabilities are non-uniform, i.e., both $(q_i - {\bar q})^2$ and $\chi^2_{\bm{\beta}\|\mathbf{p}}$ are non-zero, it can slow down the convergence by introducing the terms (d) and (e).
Moreover, we have several important insights as follows:
\begin{itemize}
\item Firstly, the upper bound depends on the effective number of clients $\bar K$ instead of $K$, and thus larger TO probabilities directly slow down the algorithm convergence.

\item Secondly,  we observe that,  except for the first two terms,  the terms (a)-(d) are caused by either QE, non-i.i.d. data distribution,  TO or partial client participation.  Therefore,  in ideal wireless channels without QE and TO and with full client participation,  the terms (a),  (b),  (d) and (e) can be removed, whereas the term (c) due to the non-i.i.d. data distribution still impedes the convergence.

\item Thirdly,  the term (d) does not decrease with $T$. Since it is caused by non-uniform TO probabilities and non-i.i.d. data distribution, this implies that the former amplifies the negative effects of the latter and will make the algorithm converge to a biased solution, as observed in Fig. \ref{fig:Impact_TO_QE} and Remark 1.  Intriguingly,  this phenomenon is analogous to the inconsistency issue analyzed in \cite{wang2020tackling} where the clients adopt different numbers of local SGD steps.

\item Last but not the least,  when the clients have an \emph{uniform TO probability, } i.e.,  $q_i=q~\forall i$,   the terms (d) and (e) can vanish,  showing that the algorithm can still converge to a proper stationary solution.
Specifically,  by combining with Lemma 2, we can derive the following result:
\end{itemize}

\begin{corollary}\label{Cor_uniform_TO_TK}
Under the same conditions as Theorem 1, if all clients have a uniform TO probability $q$, we have
\begin{small}
\begin{align}\label{corollary_1}
& \frac{1}{M} \sum \limits_{r = 1}^{M} \mathbb{E}\left[ \| \nabla  F({\bar {\mathbf{w}}}_{r-1}) \|^2 \left| \sum_{{i} \in {\mathcal{S}}_{r}} \mathds{1}^{r}_{i} \neq 0 \right. \right] \notag \\
\leq & \frac{496 L }{11 ( T {\bar K} )^{\frac{1}{2}} }  \left( \mathbb{E}[F({\bar {\mathbf{w}}}_{0})] - \underline{F} \right)
+ \left(  \frac{ 39 }{ 88 ( T {\bar K} )^{\frac{1}{2}} }
 +  \frac{1}{ 88 ( T {\bar K} )^{\frac{3}{4}} }  \right)  \frac{\sigma^2}{b} 
+ \frac{31}{ 88 T^{\frac{3}{2}} {\bar K}^{\frac{1}{2}}} \sum \limits_{r = 1}^{M} \sum \limits_{i = 1}^{N} p_i J_{ir}^2 \notag \\
& + \left( \frac{4 }{ 11 ( T {\bar K} )^{\frac{1}{2}} }
+ \frac{1}{ 22 ( T {\bar K} )^{\frac{3}{4}} } + \frac{31 }{ 22 T^{\frac{1}{4}} {\bar K}^{\frac{5}{4}} } \right) \sum \limits_{i = 1}^{N} p_i D_i^2 \, \text{.}
\end{align}
\end{small}
\end{corollary}

From the last three terms in the RHS of \eqref{corollary_1}, we can observe that with uniform TO probabilities, the impact of the mini-batch SGD variance ${\sigma^2}/{b}$, the quantization error $\{J_{ir}\}$ and the heterogeneity of local datasets $\{D_i\}$ can be reduced with a larger number of effective clients ${\bar K}$, and the FL algorithm can also achieve a \emph{linear speed-up} with respect to ${\bar K}$ even when both TO and QE are present.
This inspiring result implies that balancing the client TO probabilities is crucial for achieving fast and robust FL in non-ideal wireless channels.

\begin{remark}\label{general_case_QE} \rm
To the best of our knowledge,  the claims in Theorem \ref{Thm_diff_TO_fixed} and Corollary \ref{Cor_uniform_TO_TK} and the associated insights have not been discovered in the literature.
Note that these results can readily be extended to the general case where the quantization levels $\{ B^{r}_i \}$ and TO probabilities $\{ q^{r}_i \}$ vary with the communication round $r$.
For example,  the associated upper bound for Corollary \ref{Cor_uniform_TO_TK}
can be obtained by simply replacing
$\sum_{i = 1}^{N} p_i J_{ir}^2$ in the RHS of \eqref{corollary_1} with $\mathbb{E}_{\mathcal{S}_{r}} \left[ \frac{1}{K} \sum_{{i} \in {\mathcal{S}}_{r}} J_{ir}^2 \right]$.
More details are shown in Section B of the Supplementary Material.
\end{remark}

\begin{remark}
\rm One may have noticed that the convergence rate in Corollary \ref{Cor_uniform_TO_TK} is $\mathcal{O}(1/T^{\frac{1}{4}})$ rather than $\mathcal{O}(1/T^{\frac{1}{2}})$ for typical distributed SGD algorithms \cite{yu2019parallel, reisizadeh2020fedpaq}. The cause for such slowdown is the simultaneous presence of partial client participation \eqref{partial_participation},  data heterogeneity (Assumption 3) and TO.  Indeed,  one can verify that when there is no data heterogeneity, i.e.,  $D_i^2 = 0$, and no TO, i.e., $q_i = 0$ $\forall i=1,\ldots, N,$ then the bound in  \eqref{corollary_1} improves to $\mathcal{O}(1/T^{\frac{1}{2}})$. Analogously,  one can show that the same $\mathcal{O}(1/T^{\frac{1}{2}})$ convergence rate can be achieved if all clients are active in each round and no TO.
\end{remark}

\section{Wireless Resource Allocation}\label{Sec:Resource}

Since both TO and QE inevitably occur in the delay constrained wireless communication systems, we aim to minimize their effects on the FL in the wireless edge.
In this section, we formulate a wireless resource allocation problem to minimize the effects due to TO and QE so as to speed up the algorithm convergence.

\subsection{Proposed \texttt{FedTOE}}

Let's first assume an offline scenario, where the bandwidth $W_i$, transmit power $P_i$, quantization level $B_i$ and uplink transmission rate $R_i$ of each client are optimized offline,  and applied to the whole model learning process.
Online scheduling will be considered in Section \ref{online_scheduling}.

\subsubsection{Problem formulation}

According to Theorem \ref{Thm_diff_TO_fixed}, the algorithm convergence is affected by various parameters.
Since the SGD variance $\sigma^2$ and the local data heterogeneity $\{D_i\}$ have nothing to do with the wireless resources, we focus on resource allocation for reducing the impacts of quantization errors $\{J_{ir}\}$ and outage probabilities $\{q_{i}\}$.
As suggested by Corollary 1 that it is crucial to maintain a uniform outage probability across the clients, we enforce the constraint $q_i = q_{\max}$ for all $i = 1,\cdots,N$, where $q_{\max}\in (0,0.5]$ is a preset target outage probability value.
Then, by \eqref{corollary_1}, it remains to reduce the effect of quantization errors.
Therefore, aiming at improving the learning performance, we choose to minimize the accumulative average QE $\sum_{r = 1}^{M} \sum_{i = 1}^{N} p_i J_{ir}^2$ in the RHS of \eqref{corollary_1} under the constraints of uniform outage probability and transmission delay\footnote{
Note that in  \eqref{obj_func_A1_d} the per-round transmission delay constraint $\bar \tau_i\leq \tau_{\max}$ is equivalent to constraining the total transmission delay $\tau_{\rm total} = M \tau_{\max}$ for $M$ communication rounds.
This is because the wireless resource allocation of $\{B_i\}$, $\{P_i\}$, $\{W_i\}$ and $\{R_i\}$ are fixed during the whole training process and applied to each round, and thus the resultant transmission delay $\tau_i$ is the same for all rounds.}.
By \eqref{quantization error_local_model}, this yields the following resource allocation problem.
{\small \begin{subequations}\label{obj_func_A1}
\begin{align}
\min \limits_{W_i, P_i, B_i,R_i \atop i = 1,\cdots, N} &  \sum_{i = 1}^{N} p_i \cdot \frac{ \sum_{r = 1}^{M} \delta_{ir}^2 }{( 2^{B_{i}} - 1 )^2 }  \label{obj_func_A1_a} \\
{\rm s.t.} \quad & \sum_{i=1}^{N} W_i \leq W_{\rm total},
\ W_i \geq 0,
\ i = 1,\cdots, N, \label{obj_func_A1_b} \\
& 0 \leq P_i \leq P_{\max} ,
\ i = 1,\cdots, N, \label{obj_func_A1_c} \\
& 0 \leq \bar \tau_i \leq \tau_{\max} ,
\ i = 1,\cdots, N, \label{obj_func_A1_d} \\
& 0 \leq q_i = q_{\max} ,
\ i = 1,\cdots, N, \label{obj_func_A1_e} \\
& B_i \in \mathbb{Z}_+,\ i = 1,\cdots, N \text{.}
\label{obj_func_A1_f}
\end{align}
\end{subequations}}
\vspace{-0.5cm}

\noindent
where $W_{\rm total}$ is the total bandwidth of the uplink channel, $P_{\max}$ is the maximum transmit power of each client,
$\bar \tau_i$ is the average uplink transmission delay per communication round of client $i$, $\tau_{\max}$ is the constraint on uplink transmission delay, and $\mathbb{Z}_+$ is the positive integer set.

\subsubsection{Uplink delay}

Since retransmission is performed if all selected clients encounter outage in the uplink transmission (i.e., $\sum_{{i} \in {\mathcal{S}}_{r}} \mathds{1}^{r}_{i} = 0$), the average transmission delay of each selected client ${i} \in \mathcal{S}_{r}$ at the ${r}$-th communication round can be shown to be
\begin{align}\label{average_transmission_delay}
{\bar \tau}^{r}_i
= & \frac{1}{1- \prod_{{j} \in \mathcal{S}_{r}} q_{j} } \max \limits_{{j} \in \mathcal{S}_{r}} \frac{{\hat B}_{j}}{R_{j}}
\, \text{,}
\end{align}
where the derivation of \eqref{average_transmission_delay} is presented in Section C of the Supplementary Material.
One can see that $\prod_{{j} \in \mathcal{S}_{r}} q_{j} = (q_{\max})^K \approx 0$ with a large $K$ or smaller $q_{\max}$, and thus ${\bar \tau}^{r}_i \approx \max_{{j} \in \mathcal{S}_{r}} {\hat B}_{j}/{R_{j}}$.
To approximately meet the transmission delay constraint in (\ref{obj_func_A1_d}),  we replace (\ref{obj_func_A1_e}) by $0 \leq {\hat B}_{i}/{R_i} \leq \tau_{\max} \forall i=1,\ldots,N$.

\subsubsection{Optimal condition}

One can prove that the solution to (\ref{obj_func_A1}) satisfies Proposition \ref{optimal_condition}.

\begin{proposition}\label{optimal_condition}
\textbf{(Optimal condition)}
After relaxing $B_i \in \mathbb{Z}_+$ to $B_i \geq 1$ $\forall i=1,\ldots,N$,  for the optimal condition of problem (\ref{obj_func_A1}) it holds that
(a) the transmit power $P_i = P_{\max}$ $\forall i$,
and (b) the uplink delay $\tau_i = {\hat B}_{i}/{R_i} = \tau_{\max}$ $\forall i$.
(c) Moreover,  based on (\ref{quantized model length}) and (\ref{outage_probability_shadow}), the optimal transmission rate $R_i$ satisfies
\begin{align}\label{R_i_optimal}
R_i = {\bar R}_i(W_i)
\triangleq W_i \log_2 \left( 1 + \frac{\theta_i P_{\max}}{W_i N_0} \right)
\text{,}
\end{align}
where $\theta_i \triangleq 10^{\frac{1}{10}\left(\sigma_{\rm dB} \cdot
Q^{-1}\left(1 - q_{\max}\right) +  [ \mathcal{K} ]_{\rm dB} - \lambda [d_i]_{\rm dB} \right)}$, and the optimal quantization level satisfies
\begin{align}\label{quantization_level}
B_i = ({ {\bar R}_i(W_i) \tau_{\max} - \mu})/{m}.
\end{align}
Furthermore,  \eqref{quantization_level} can be equivalently written as $W_i = \overline{W}_i(B_i)$ for some continuously differentiable  and increasing function $\overline{W}_i(\cdot)$.
\end{proposition}

\emph{Proof:}
The conditions (a)-(c) can be easily proved by contradiction and based on the monotonic property of \eqref{obj_func_A1_a} with respect to $B_i$.
The existence of $\overline{W}_i(\cdot)$ and its monotonically increasing property can be obtained by the implicit function theorem \cite{Krantz_Parks02}.
The detailed proof is presented in Section D of the Supplementary Material.
\hfill $\blacksquare$

Combining \eqref{R_i_optimal} with \eqref{outage_probability_shadow}, one can observe that with $q_{\max}<0.5$, a larger shadowing power $\sigma_{\rm dB}^2$ causes the transmission rate in \eqref{R_i_optimal} as well as the quantization level in \eqref{quantization_level} to decline.
It implies a larger quantization error in \eqref{obj_func_A1_a} and consequently deteriorates the learning performing according to Corollary \ref{Cor_uniform_TO_TK}.

\subsubsection{Optimization method}

By Proposition 1,  problem (\ref{obj_func_A1}) after relaxing $B_i \in \mathbb{Z}_+$ to $B_i \geq 0$ $\forall i=1,\ldots,N$, can be reformulated as
{\small
\begin{subequations}\label{obj_func_A2}
\begin{align}
\min \limits_{W_i \atop i =1,\cdots,N} \ &  \sum_{i = 1}^N \frac{p_i \sum_{r = 1}^{M} \delta_{ir}^2}{\left( 2^{\frac{\tau_{\max}}{m} {\bar R}_i(W_i)-  \frac{\mu}{m} } - 1 \right)^2 }  \label{obj_func_A2_a} \\
{\rm s.t.} \ & \sum_{i=1}^{N} W_i \leq W_{\rm total},  \ W_i \geq \overline{W}_i(1), \ i = 1,\ldots, N.  \label{obj_func_A2_b}
\end{align}
\end{subequations}}

One can show that:
\begin{proposition}\label{convexity}
Problem \eqref{obj_func_A2} is convex.
\end{proposition}

\emph{Proof:}
It can be proved by showing that the second-order derivative of each term in the summation of \eqref{obj_func_A2_a} with respect to $W_i$ is non-negative. The details are relegated to Section E of the Supplementary Material.
\hfill $\blacksquare$

Based on Proposition 2,  problem \eqref{obj_func_A2} can be efficiently solved by a simple gradient projection method \cite{figueiredo2007gradient} with an initial point in the feasible region of (\ref{obj_func_A2_b})\footnote{In practice,  the value of $\overline{W}_i(1)$ can be computed by bisection search based on \eqref{quantization_level} and monotonic property of $\overline{W}_i(B_i)$.}.
Since $B_i$ is an positive integer, after each gradient descent step in optimizing \eqref{obj_func_A2}, each $B_i$ obtained by (\ref{quantization_level}) is floored to its nearest integer $\lfloor B_i \rfloor$.
Then, the bandwidth supporting $\lfloor B_i \rfloor$ with the TO probability $q_{\max}$ is given by $\overline{W}_i(\lfloor B_i \rfloor)$, which is further used as the starting point for the next gradient descent step.
Note that such relaxation and rounding strategy is suboptimal since it would underutilize the uplink bandwidth.  Nonetheless,  the experiment results shown in Section \ref{sec: numerical results} show that such a simple strategy is effective.

The details of our proposed wireless resource allocation method for offline scheduling are summarized in Algorithm 2.
We refer to the FL process in Algorithm 1 with the wireless resource allocation solution by Algorithm 2 as \texttt{FedTOE}.
\begin{algorithm}[t]
\caption{\texttt{FedTOE}: Algorithm to solve \eqref{obj_func_A1}}
\begin{algorithmic}[1]
\State $j = 0$
\While {$j <$ maximum iteration number}
\State Update $\{W_i\}$ with one-step gradient descent and
\Statex $\quad\,$ projection on \eqref{obj_func_A2};
\State Compute each $B_i$ $(i = 1,\cdots, N)$ by \eqref{quantization_level};
\State Set each $B_i = \lfloor B_i \rfloor$;
\State Find each $W_i = \overline{W}_i(\lfloor B_i \rfloor)$ by bisection search;
\State $j = j+1$
\EndWhile
\State Compute each $R_i$ by (\ref{R_i_optimal});
\renewcommand{\algorithmicensure}{\textbf{Output:}}
\Ensure Transmit power $P_{\max}$, bandwidth $W_i$, quantization
\Statex $\qquad$ level $B_i$, and transmission rate $R_i$ of each client
\end{algorithmic}
\end{algorithm}

\subsection{Online scheduling}\label{online_scheduling}

In this subsection, let us investigate the online scenario, where the bandwidth $W_i^{r}$, transmit power $P_i^{r}$, quantization level $B_i^{r}$, and uplink transmission rate $R_i^{r}$ of each client are optimized for every communication round $r$.
Since the selected clients in $\mathcal{S}_{r}$ are revealed at each communication round $r$,
such online scheduling can make better use of the wireless resources via dynamically allocating bandwidth and quantization bits.
According to Remark \ref{general_case_QE}, we can consider the following  QE minimization problem at each communication round:
{\small
\begin{subequations}\label{obj_func_B1}
\begin{align}
& \min \limits_{W^{r}_i, P^{r}_i, B^{r}_i,R^{r}_i \atop i \in {\mathcal{S}_{r}} } \frac{1}{K} \sum_{i \in {\mathcal{S}_{r}}} \frac{\delta_{ir}^2 }{( 2^{B^{r}_{i}} - 1 )^2 }  \label{obj_func_B1_a} \\
& \qquad\quad {\rm s.t.} \quad \sum_{i \in {\mathcal{S}_{r}}} W^{r}_i \leq W_{\rm total},
\ W^{r}_i \geq 0,
\ {i} \in {\mathcal{S}_{r}}, \label{obj_func_B1_b} \\
& \qquad\qquad\qquad  0 \leq P^{r}_i \leq P_{\max} ,
\ {i} \in {\mathcal{S}_{r}}, \label{obj_func_B1_c} \\
& \qquad\qquad\qquad 0 \leq {\bar \tau}^{r}_i \leq \tau_{\max} ,
\ {i} \in {\mathcal{S}_{r}}, \label{obj_func_B1_e} \\
& \qquad\qquad\qquad 0 \leq q^{r}_i = q_{\max} ,
\ {i} \in {\mathcal{S}_{r}}, \label{obj_func_B1_d} \\
& \qquad\qquad\qquad B^{r}_i \in \mathbb{Z}_+, \ {i} \in {\mathcal{S}_{r}} \text{.}
\label{obj_func_B1_f}
\end{align}
\end{subequations}}

\vspace{-0.3cm}
Then, following similar derivations as the offline scheme in the previous subsection, (\ref{obj_func_B1}) can be handled by solving
{\small
\begin{subequations}\label{obj_func_B2}
\begin{align}
\min \limits_{W^{r}_i, i \in {\mathcal{S}_{r}}} \ &  \sum \limits_{ {i} \in {\mathcal{S}_{r}} } \frac{\delta_{ir}^2}{\left( 2^{\frac{\tau_{\max}}{m} {\bar R}_i(W^{r}_i)-  \frac{\mu}{m} } - 1 \right)^2 } \\
{\rm s.t.} \ & \sum \limits_{i \in {\mathcal{S}_{r}}} W^{r}_i \leq W_{\rm total}, \ W^{r}_i \geq \overline{W}_i(1), \ {i} \in {\mathcal{S}_{r}}.
\end{align}
\end{subequations}
}

\vspace{-0.3cm}

The procedure of solving \eqref{obj_func_B1} is similar to Algorithm 2, except replacing \eqref{obj_func_A2} in Step 3 with \eqref{obj_func_B2}, replacing $i = 1,\cdots, N,$ in Step 4 with ${i} \in {\mathcal{S}_{r}}$, and replacing $W_i$, $B_i$, and $R_i$ with $W^{r}_i$, $B^{r}_i$, and $R^{r}_i$ respectively.

\section{Numerical results}\label{sec: numerical results}

\subsection{Parameter setting}\label{parameter_setting}

In the simulations, we assume that the server (i.e., base station) is located at the cell center with a cell radius 600m, and $N=100$ clients are uniformly distributed within the cell.
The server employs Algorithm 1 to train neural networks under the following two datasets.

\begin{enumerate}[(a)]
\item
\textbf{MNIST dataset \cite{lecun1998gradient}:}
In the experiments, we consider two types of local datasets, i.e., the i.i.d.  and the non-i.i.d local datasets.
Specifically, in the i.i.d. case, the 60000 training samples in MNIST database are shuffled and then randomly distributed to each client.
In the non-i.i.d. case, the training samples are reordered by their digit labels from 0 to 9 and then partitioned so that each client possesses at most 2 digits of training samples, and the clients farther away from the server have the samples of larger digits.
Besides, each client is assumed to possess the same number of training samples, i.e., $n_i = 600$ $\forall i = 1, \ldots, N$.
We train a 3-layer deep neural network (DNN) with size $784 \times 30 \times 10$ for the classification of digits based on this dataset.

\item
\textbf{CIFAR-10 dataset \cite{krizhevsky2009learning}:}
For the i.i.d. case, the data partition of the 50000 training samples is similar to the MNIST experiment in (a).
In the non-i.i.d. case, we let each client possess at most 5 categories of training samples.
We consider the ResNet-20 \cite{he2016deep} in the experiment.
\end{enumerate}

In the simulations, the size of quantized local model update is represented by
\begin{align}\label{quantized model length_2}
{\hat B}^{r}_{i}
= {m} (1 + B^{r}_{i}) + n_{\min}B_{\min} + n_{\max}B_{\max} \; \text{(bits)}
\, \text{,}
\end{align}
where for the adopted 3-layer DNN, the total number of model parameters is $m = 23860$ which consists of $23820$ $(=784 \times 30 + 30 \times 10)$ weights and $40$ $(=30 + 10)$ bias, while for ResNet-20 with 19 convolution layers and 1 fully-connected layers,  we have $m = 271098$ \cite{he2016deep}.
Meanwhile, 1 bit, $B_{\min}$ bits, and $B_{\max}$ bits in \eqref{quantized model length_2} are used for representing the sign, the lower limit ${\underline{w} }^{r}_{{i}j}$, and the upper limit ${\bar w}^{r}_{{i}j}$ of each parameter update respectively.
In the quantization process as \eqref{quantization_method}, the weight updates belonging to the same layer share the same range $[{\underline{w} }^{r}_{{i}j}, {\bar w}^{r}_{{i}j} ]$, and so do the bias updates.
In this way, with a hidden layer and an output layer in the 3-layer DNN, there are in total $n_{\max} = n_{\min} = 4$ different lower and upper limits respectively adopted by each client to quantize its local model update, while for ResNet-20, $n_{\max} = n_{\min} = 97$.
For simplicity, we assume that the clients in ${\mathcal{S}_{r}}$ have similar constant $\delta_{ir}$ in \eqref{quantization error_local_model}, which leads to a constant $\sum_{r = 1}^{M} \delta_{ir}^2$ for all clients in (\ref{obj_func_A1_a}).
The other simulation parameters are listed in Table \ref{Parameter_value} \cite{chen2020joint,goldsmith2005wireless,misra2013wireless}, and all results were obtained by averaging over 5 independent experiments.
\begin{table}[t]
\centering
\caption{Parameter Setting}\label{Parameter_value}
\begin{tabular}{cc|cc}
\hline
\textbf{Parameter} & \textbf{Value}
& \textbf{Parameter} & \textbf{Value} \\ \hline
$b$ & 128 & $E$ & 5 (MNIST); 10 (CIFAR)  \\
$\gamma$ &  0.05 & $\sigma_{\rm dB}$ & 3.65 (except in Fig. \ref{fig:FedTOE_Performance_different_shadowing}) \\
$q_{\max}$ & 0.1 & $N_0$ & -174 dBm/Hz \\
$W_{\rm total}$ & 20 MHz & $[\mathcal{K}]_{\rm dB}$ & -31.54 \\
$B_{\min}$, $B_{\max}$ & 64 bits & $\lambda$ & 3 \\ \hline
\end{tabular}
\end{table}

Three baselines and the ideal scheme are considered for comparison with \texttt{FedTOE}.
\begin{itemize}
\item \textbf{Baseline 1.}
This scheme performs FL by Algorithm 1 with all clients adopting the maximum transmit power $P_{\max}$, the same quantization level $B_i$, uniform bandwidth $W_i = W_{\rm total}/N$ (offline scheduling) or $W_i = W_{\rm total}/K$ (online scheduling), and date rate $R_i = {\hat B}_{i}/{\tau}_{\max}$.

\item \textbf{Baseline 2.}
We consider the scheme in  \cite{salehi2020federated} where the global model is updated by ${\bar {\mathbf{w}}}_{r} = {\bar {\mathbf{w}}}_{r-1} - \frac{\gamma}{K} \sum_{{i} \in {\mathcal{S}}_{r}} \frac{p_i}{{\hat p}_i \left(1- q_i \right)} \mathds{1}^{r}_{i} \Delta \mathbf{w}^{r}_{i}$,  in which $p_i$ is the weight of client $i$ defined in (\ref{objective function}) and ${\hat p}_i$ is the client selection probability.
For the full-participation case, ${\hat p}_i = 1~ \forall i$, while for the partial participation case, ${\hat p}_i$ is optimized by formulation in \cite[Eqn. (13)]{salehi2020federated}.
Since \cite{salehi2020federated} only considers the influence of TO but not quantization, for fair comparison, we modify the global updating scheme as

\vspace{-0.3cm}
{\small
\begin{align}\label{baseline_2}
{\bar {\mathbf{w}}}_{r} = {\bar {\mathbf{w}}}_{r-1} - \frac{\gamma}{K} \sum \limits_{{i} \in {\mathcal{S}}_{r}} \frac{p_i}{{\hat p}_i \left(1- q_i \right)} \mathds{1}^{r}_{i} \mathcal{Q} \left( \Delta \mathbf{w}^{r}_{i}\right) \text{.}
\end{align}
}

\vspace{-0.3cm} \noindent
Other settings are the same as Baseline 1.

\item \textbf{Baseline 3.}
This scheme considers  (\ref{obj_func_A1}) but with fixed uniform bandwidth $W_i = W_{\rm total}/N$ (offline) or $W_i = W_{\rm total}/K$ (online).
Thus, only the quantization level $B_i$ is optimized and determined by (\ref{quantization_level}).

\item \textbf{Ideal.}
The ideal scheme suffers neither TO nor QE, which acts as the performance upper bound in the simulations.
\end{itemize}

\subsection{Performance Comparison with Offline Resource Allocation}

In this subsection, the performance of the proposed \texttt{FedTOE} with offline scheduling is evaluated unde the MNIST dataset.

\subsubsection{TO versus quantization level}

To examine the effectiveness of $\texttt{FedTOE}$, the performance of different schemes are compared by the MNIST dataset under two different constraints on the total uplink transmission delay $\tau_{\rm total}$, including a tight one with $\tau_{\rm total} = 25$s and a loose one with $\tau_{\rm total} = 100$s.
Then, given the total number of communication rounds $M = 500$, the constraints on the uplink transmission delay per communication round (i.e., $\tau_{\max}$) for the above two cases are 50ms and 200ms respectively.

\begin{figure}[t]
\begin{minipage}[t]{1\linewidth}
\centering
\subfigure[\scriptsize{$\tau_{\max} = 50$ms.}]{
\includegraphics[width= 2.2 in ]{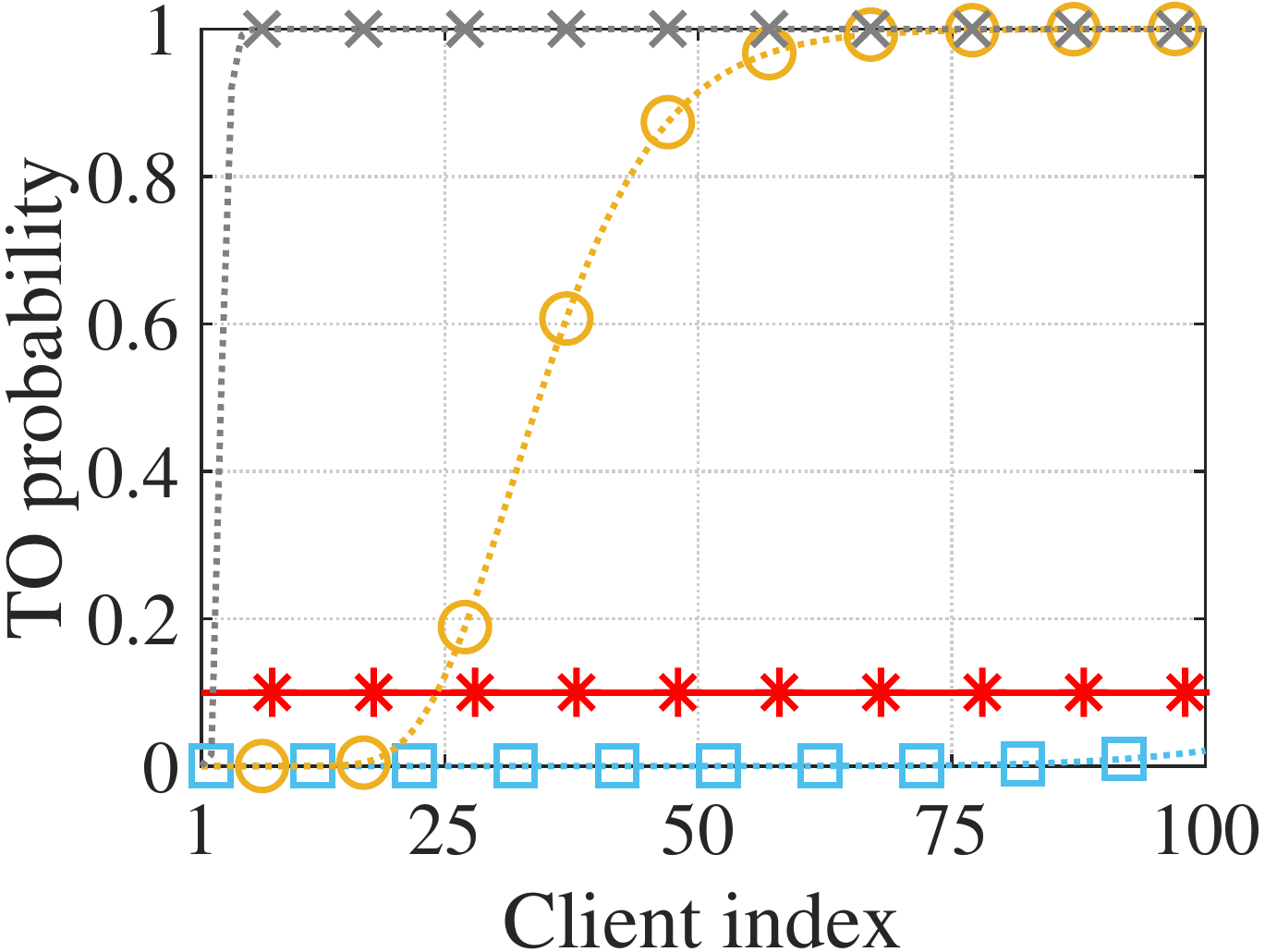}}
\subfigure[\scriptsize{$\tau_{\max} = 200$ms.}]{
\includegraphics[width= 2.2 in ]{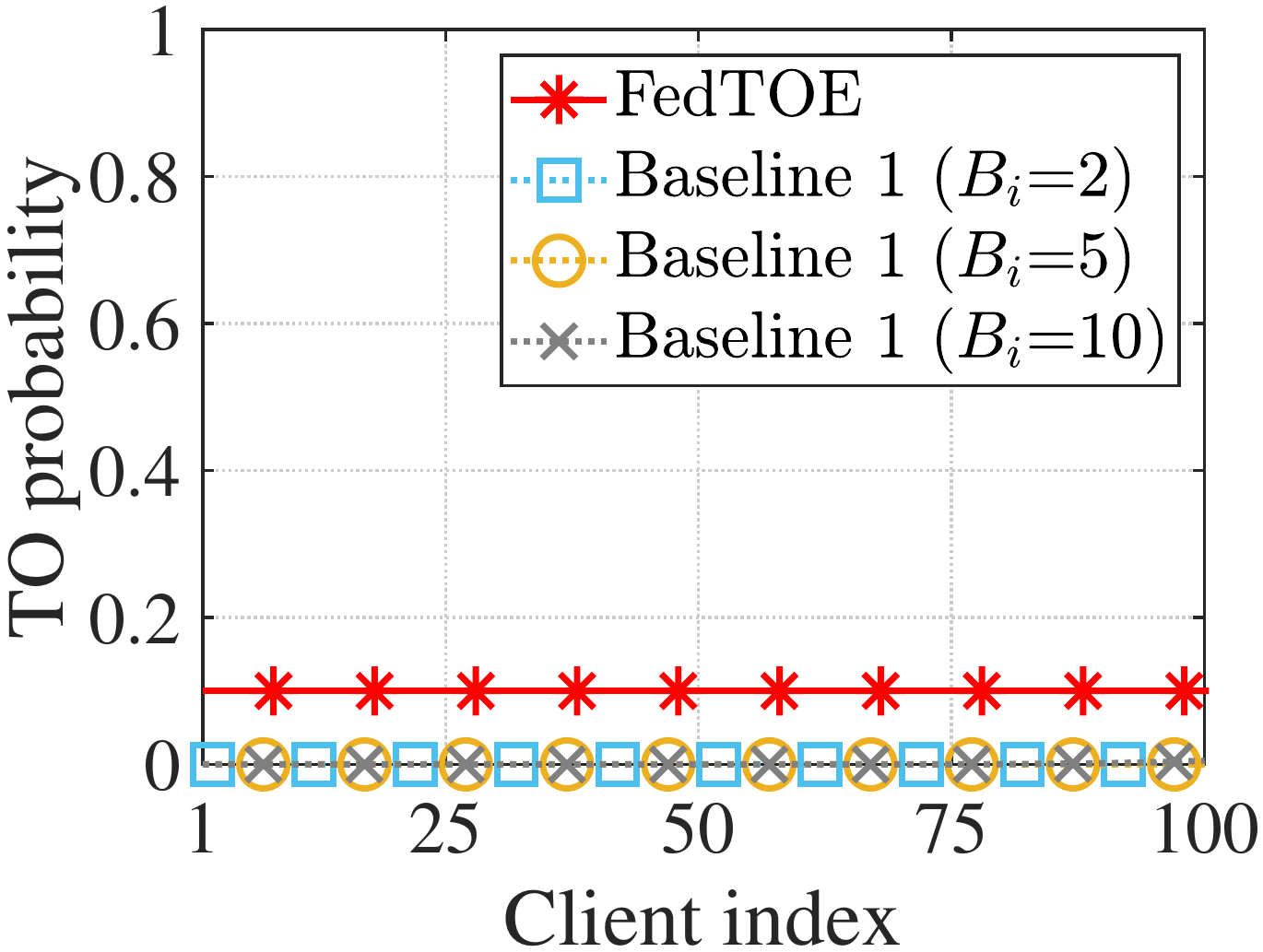}}
\caption{TO probability of each client under different schemes (Client with larger index is farther away from the server).}
\label{fig:PER_uniform_distribution}
\end{minipage}
\end{figure}

Based on the above settings, Fig. \ref{fig:PER_uniform_distribution} compares the TO probabilities of the proposed \texttt{FedTOE} and Baseline 1 (which have different values of $B_i$).
It can be seen from Fig. \ref{fig:PER_uniform_distribution}(a) that all clients in $\texttt{FedTOE}$ have uniform TO probabilities, which is consistent with Proposition \ref{optimal_condition}.
Different from this, for Baseline 1, the clients farther from the server have larger TO probabilities. This is because the data rate $R_i$ for all clients in Baseline 1 is the same, and then the client with longer distance from server has a larger TO probability in (\ref{outage_probability_shadow}).
Meanwhile, as shown in Fig. \ref{fig:PER_uniform_distribution}(a), the Baseline 1 with a larger quantization level $B_i$ leads to a higher TO probability.
The reason is that given a fixed uplink delay, transmitting more bits requires a higher data rate which increases the TO probability.
Further, it can be observed from Fig. \ref{fig:PER_uniform_distribution}(b) that under a relaxed delay constraint ($\tau_{\max} = 200$ms), the TO probabilities in Baseline 1 with all $B_i$ are reduced significantly, since a smaller transmission rate $R_i$ can be used under $\tau_{\max} = 200$ms and then leads to lower TO probabilities.

\begin{figure}[t]
\begin{minipage}[t]{1\linewidth}
\centering
\includegraphics[width= 3.5 in ]{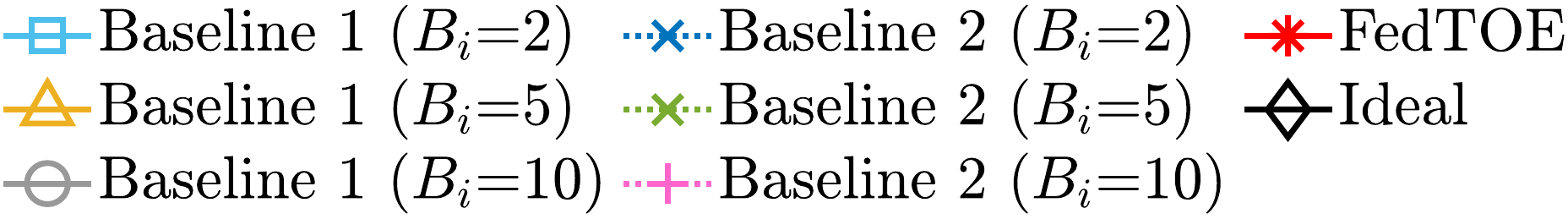}
\end{minipage}
\begin{minipage}[t]{1\linewidth}
\centering
\subfigure[\scriptsize{$K=10$.}]{
\includegraphics[width= 2.2 in ]{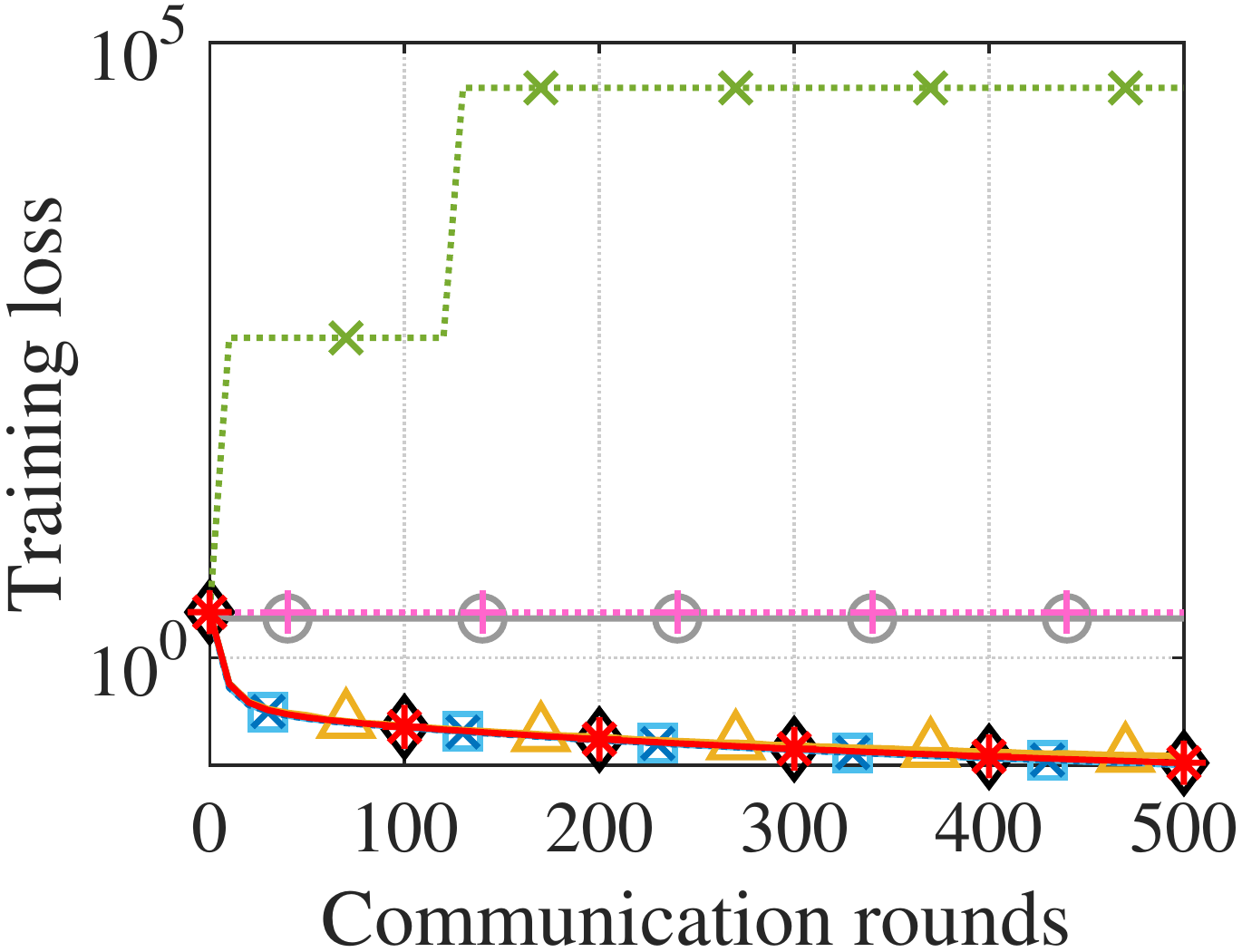}}
\subfigure[\scriptsize{$K=10$.}]{
\includegraphics[width= 2.2 in ]{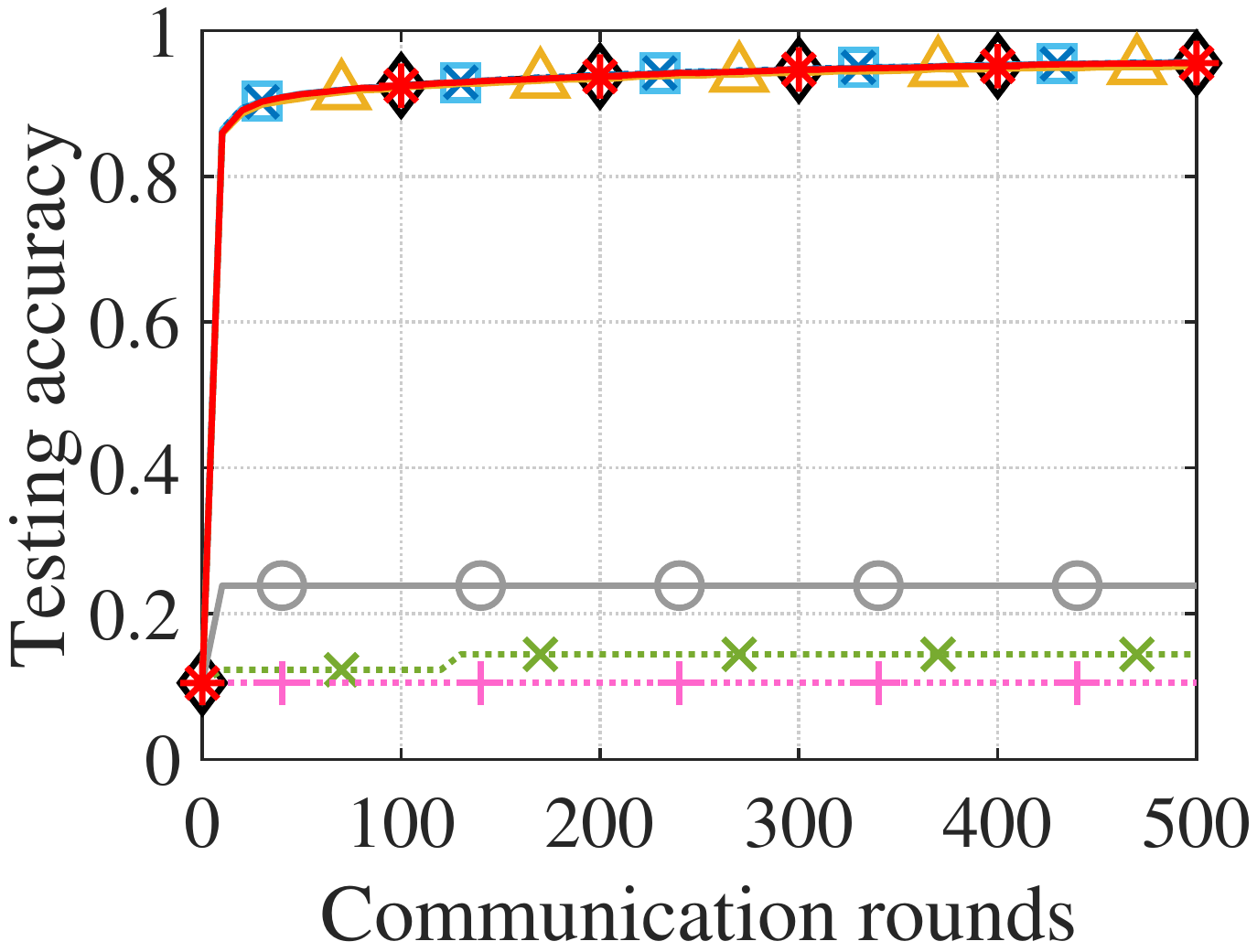}}
\subfigure[\scriptsize{$K=100$.}]{
\includegraphics[width= 2.2 in ]{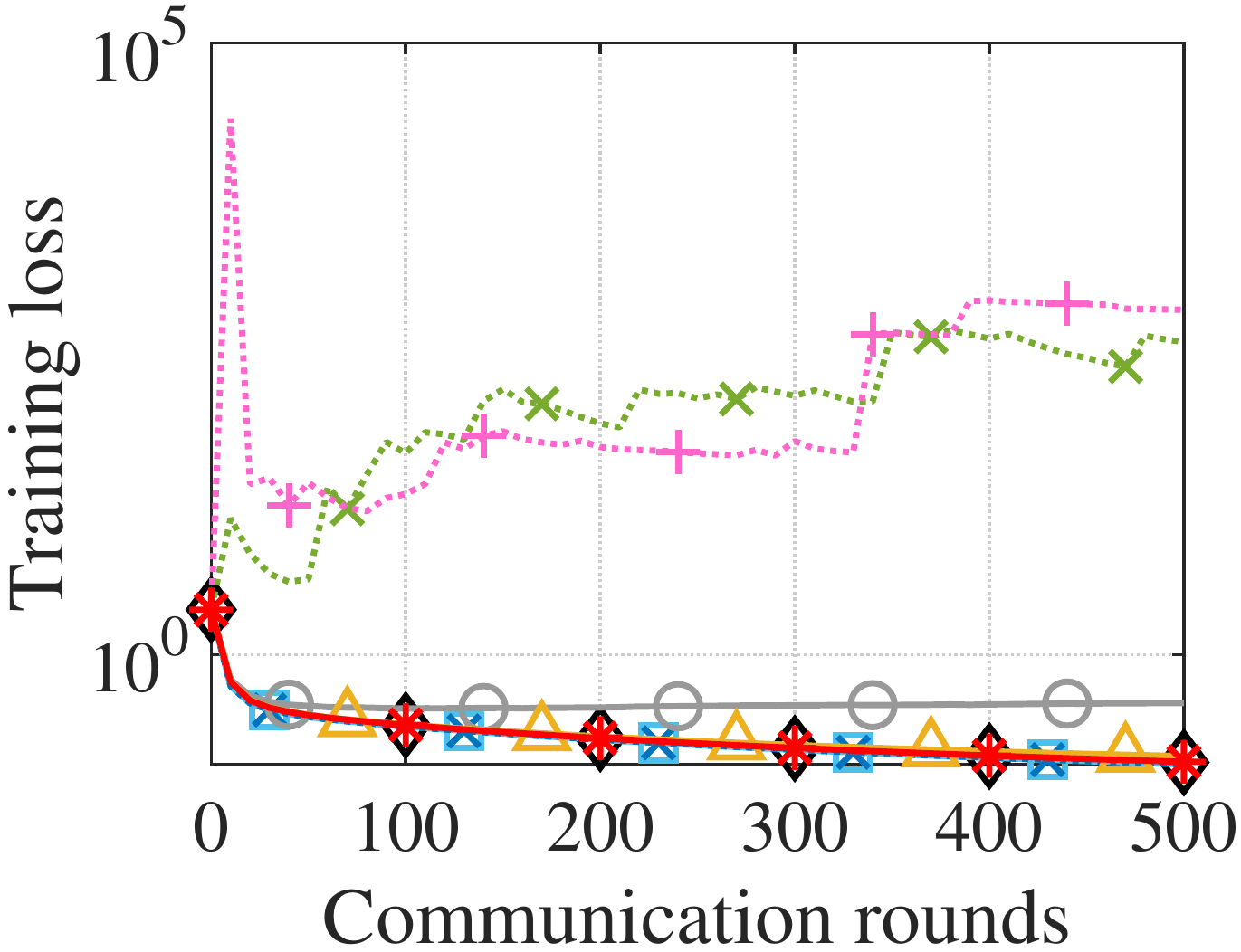}}
\subfigure[\scriptsize{$K=100$.}]{
\includegraphics[width= 2.2 in ]{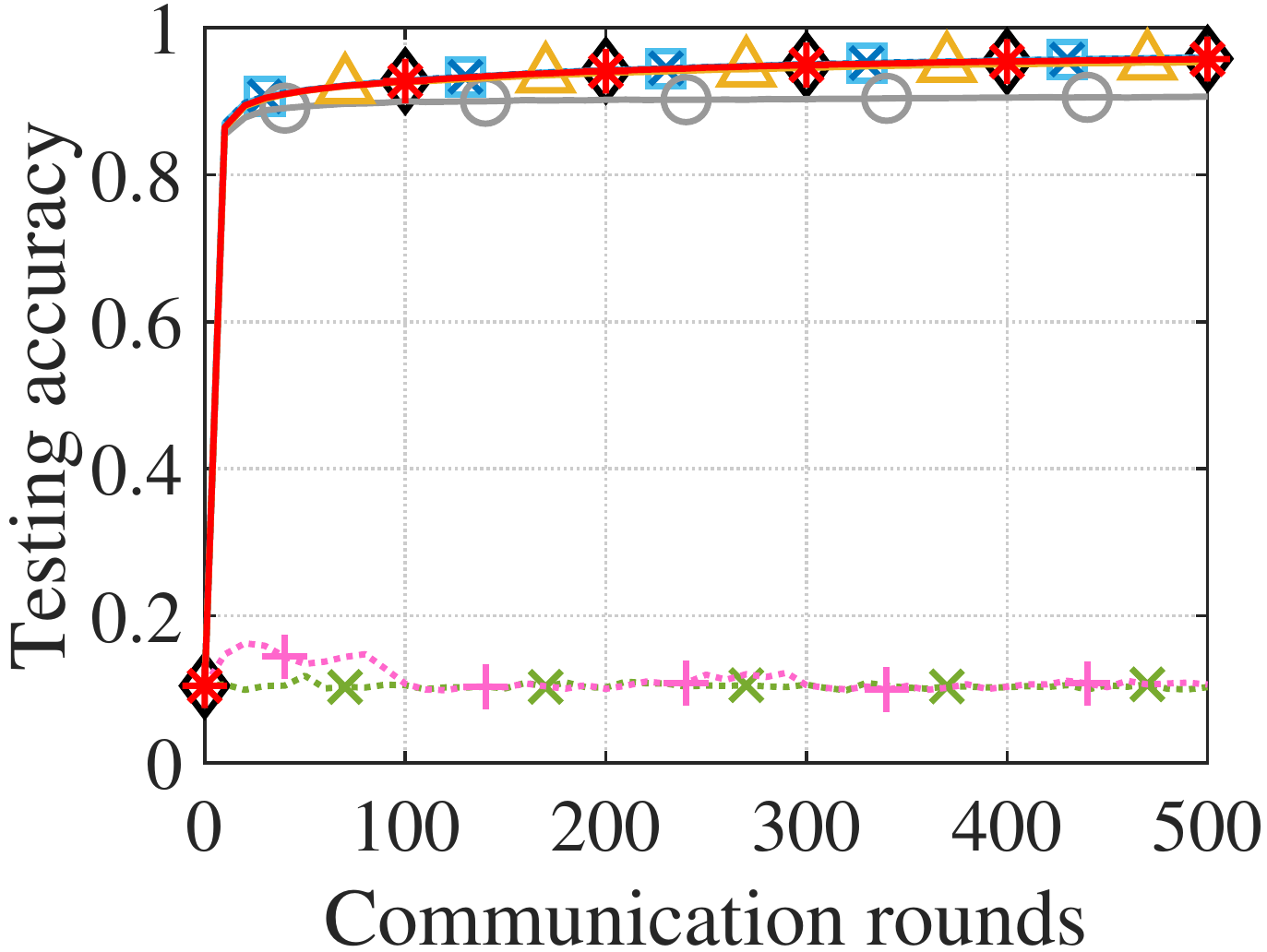}}
\caption{Comparison between baselines and \texttt{FedTOE} with $\tau_{\max}=50$ms for offline scheduling under the i.i.d. MNIST data.}
\label{fig:perforemance_uniform_iid}
\end{minipage}
\end{figure}

\begin{figure}[t]
\begin{minipage}[t]{1\linewidth}
\centering
\includegraphics[width= 3.5 in ]{4_legend.pdf}
\end{minipage}
\begin{minipage}{1\linewidth}
\centering
\subfigure[\scriptsize{$K=10$.}]{
\includegraphics[width= 2.2 in ]{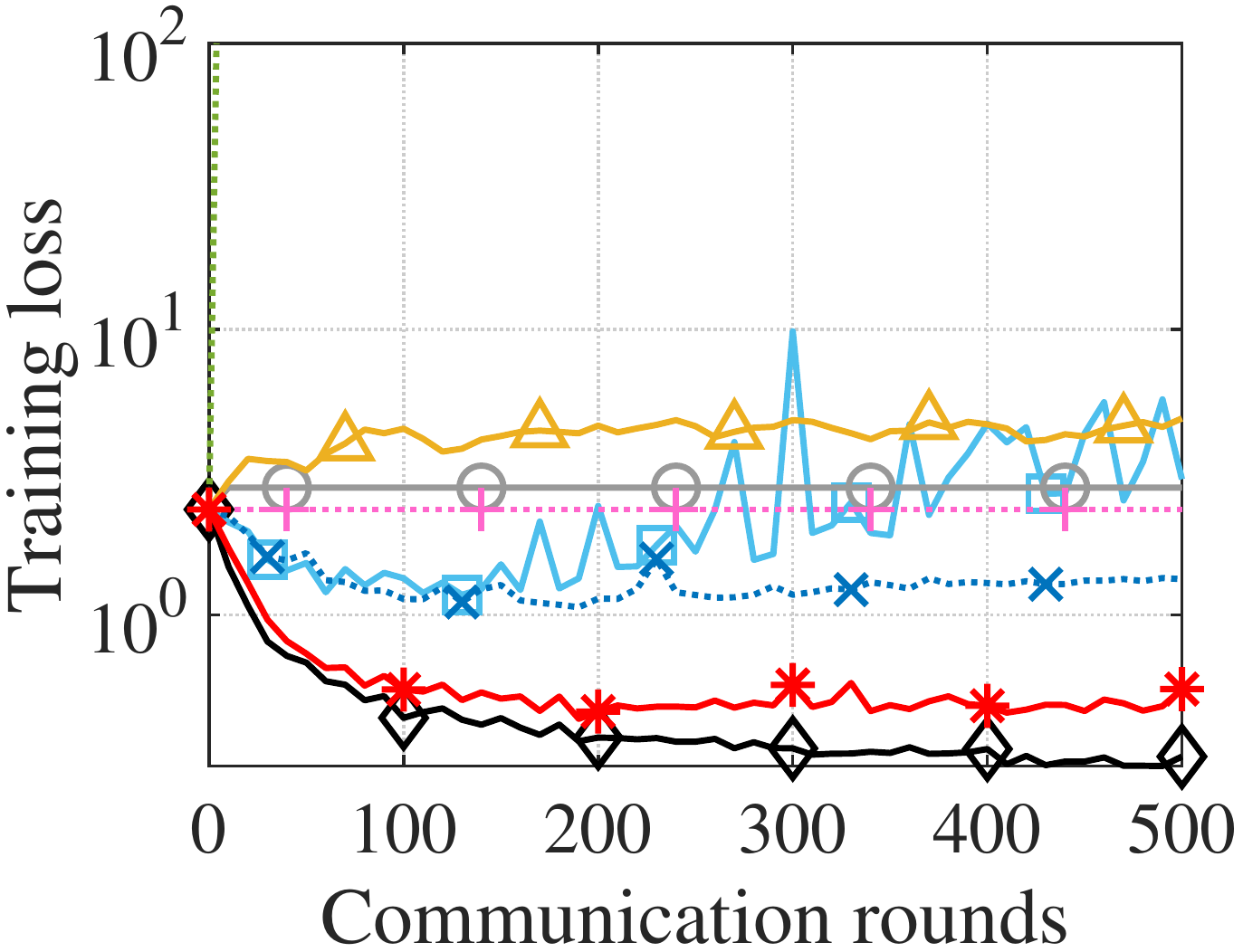}}
\subfigure[\scriptsize{$K=10$.}]{
\includegraphics[width= 2.2 in ]{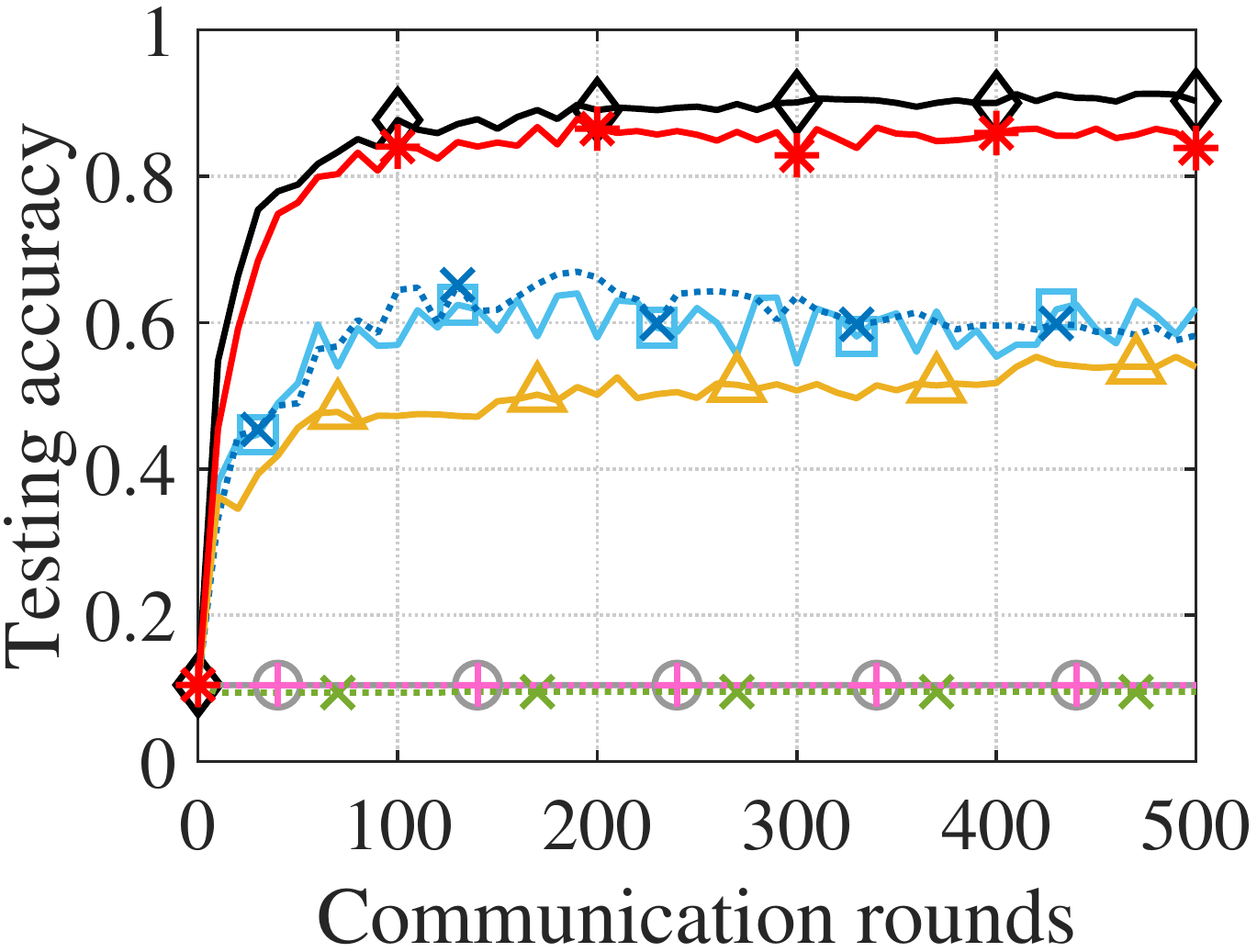}}
\subfigure[\scriptsize{$K=100$.}]{
\includegraphics[width= 2.2 in ]{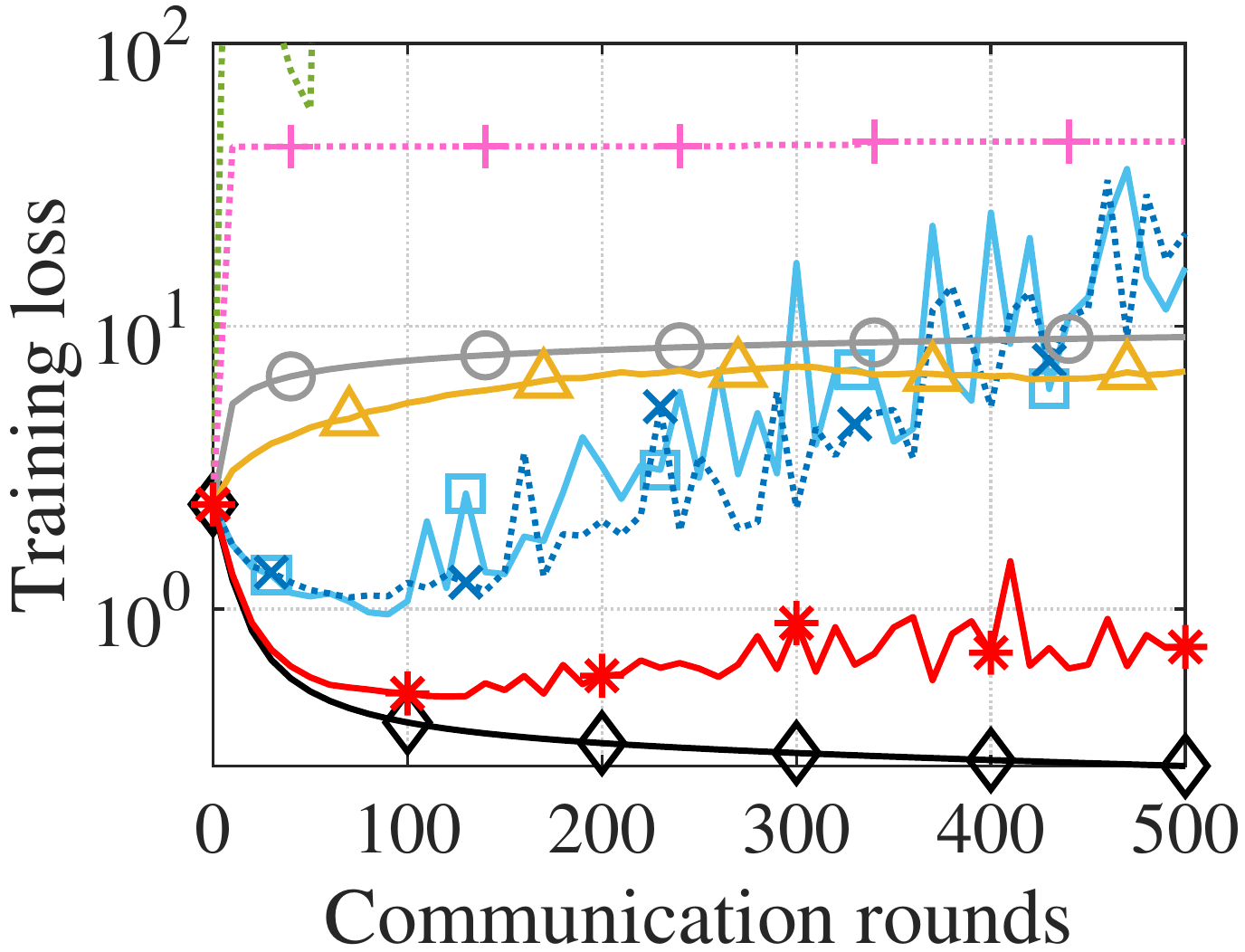}}
\subfigure[\scriptsize{$K=100$.}]{
\includegraphics[width= 2.2 in ]{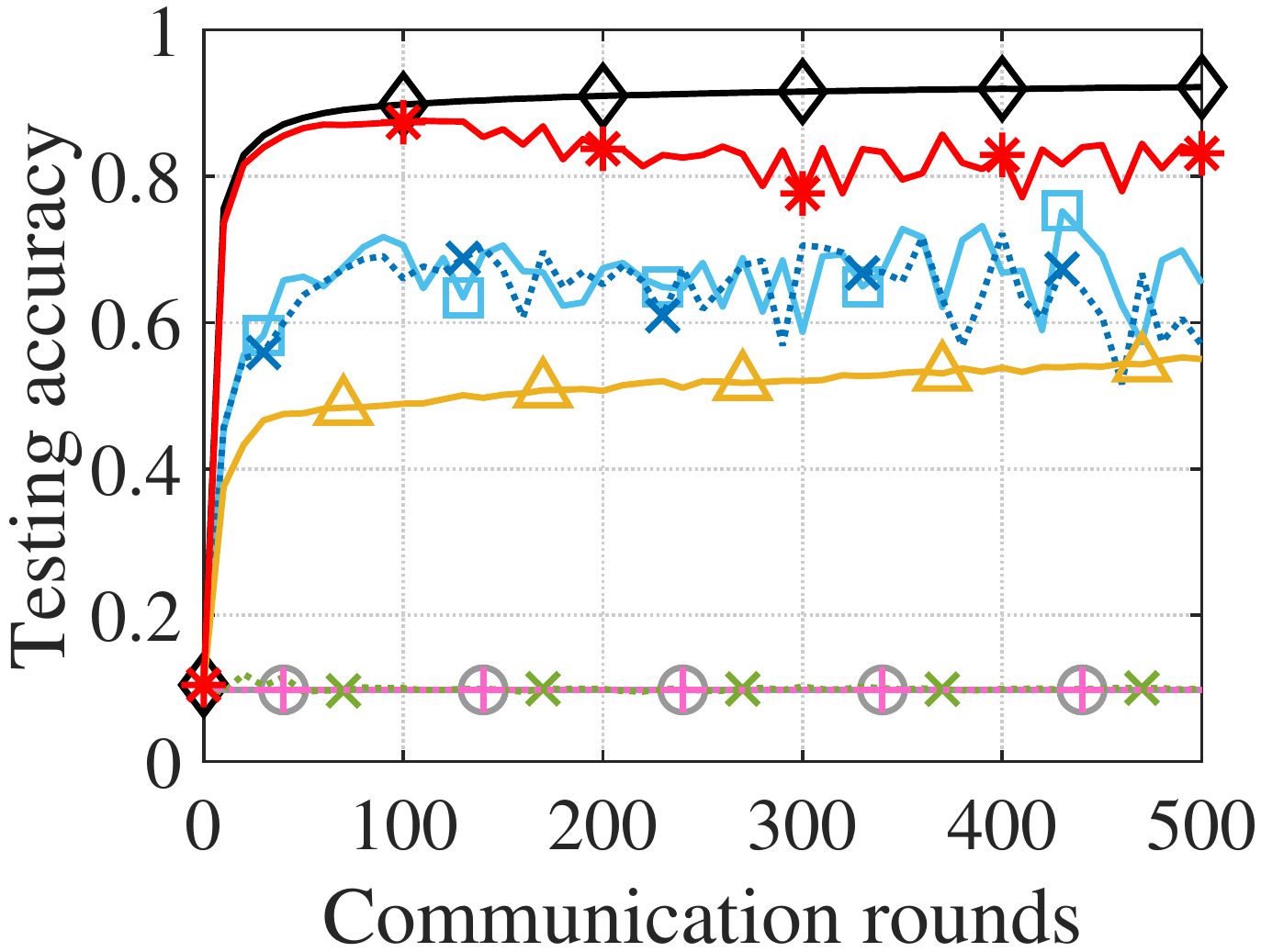}}
\caption{Comparison between baselines and \texttt{FedTOE} with $\tau_{\max}=50$ms for offline scheduling under the non-i.i.d. MNIST data.}
\label{fig:perforemance_uniform_noniid}
\end{minipage}
\end{figure}

\begin{figure}[t]
\begin{minipage}[t]{1\linewidth}
\centering
\subfigure[\scriptsize{Training loss.}]{
\includegraphics[width= 2.2 in ]{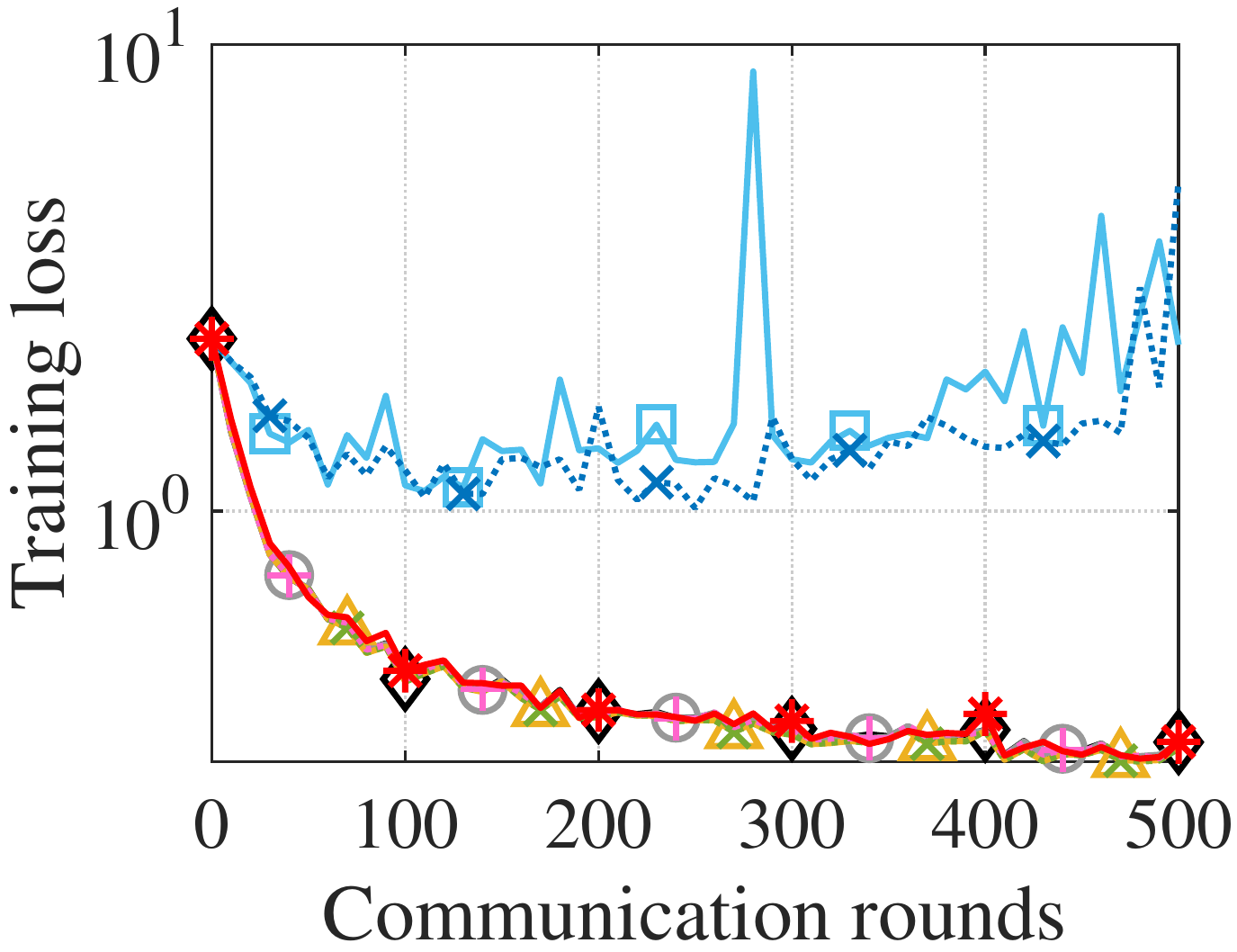}}
\subfigure[\scriptsize{Testing accuracy.}]{
\includegraphics[width= 2.2 in ]{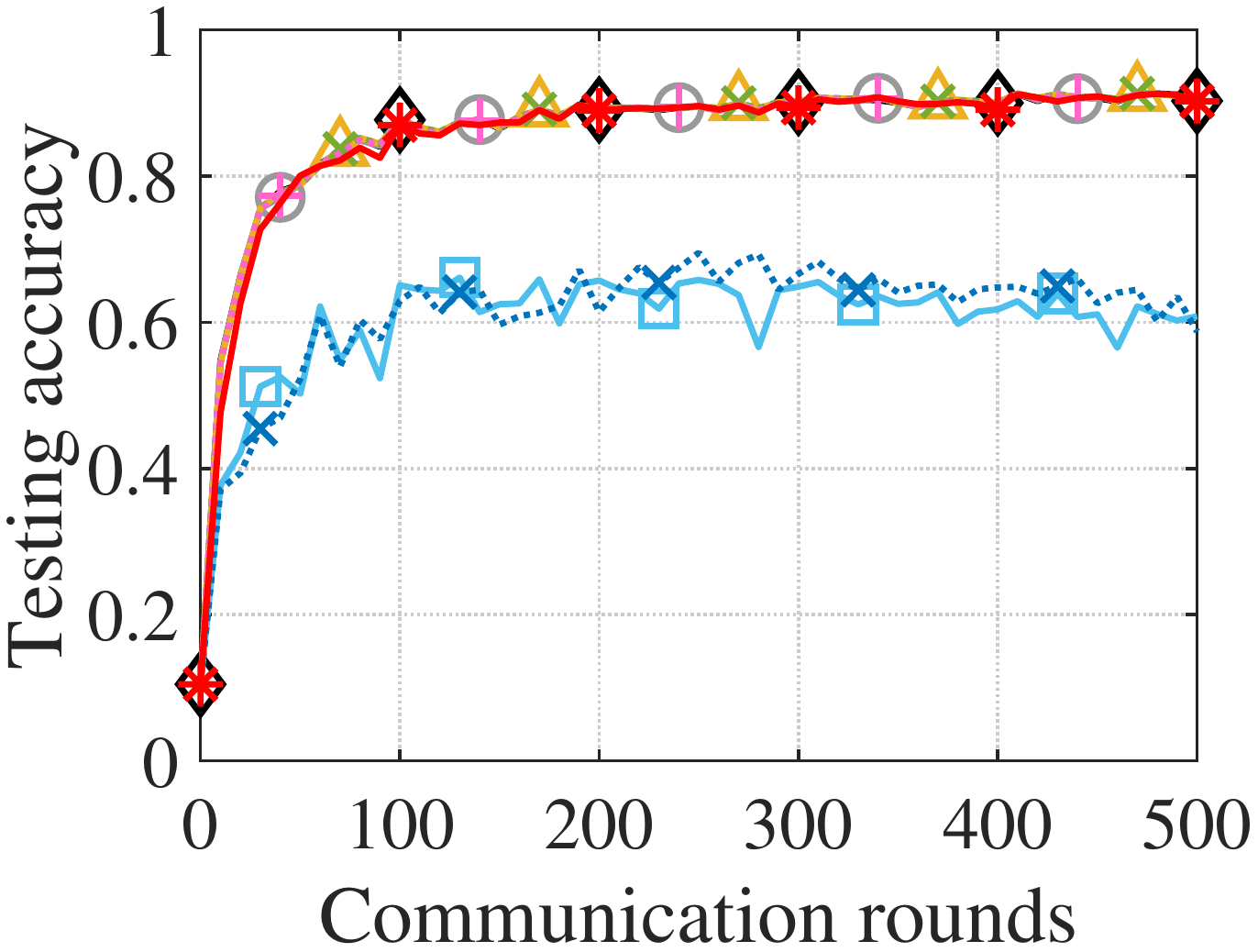}}
\caption{Comparison between baselines and \texttt{FedTOE} with $K = 10$ and $\tau_{\max}=200$ms for offline scheduling under the non-i.i.d. MNIST data (with the same legend as Fig. \ref{fig:perforemance_uniform_noniid}).}
\label{fig:perforemance_uniform_noniid_200}
\end{minipage}
\end{figure}

Next, we evaluate the performance of \texttt{FedTOE} with respect to the communication round.
From Fig. \ref{fig:perforemance_uniform_iid} to Fig. \ref{fig:perforemance_uniform_noniid_200}, the training loss and testing accuracy of the proposed \texttt{FedTOE}, Baseline 1 and Baseline 2 on MNIST dataset are compared.
The performance of the ideal scheme is also shown in the figures.
In the simulations, $K = 10$ refers to the partial participation with replacement and $K = N = 100$ corresponds to the full participation of all clients.
It should be pointed out that the retransmission rounds caused when all clients experience TO are also counted.

\textbf{The i.i.d.  data case.}
One can see from Fig. \ref{fig:perforemance_uniform_iid}(a) and Fig. \ref{fig:perforemance_uniform_iid}(b) that under the i.i.d. case, both \texttt{FedTOE} and Baseline 1 with smaller $B_i = 2,5$ perform closely to the ideal scheme.
Specifically, under the i.i.d. case with data variance $D_i^2 \approx 0$, the objective inconsistency in Theorem 1 will vanish and the learned model by Baseline 1 can converge in the right direction even with TO.
However, the TO probabilities will affect the average effective number of active clients ${\bar K}$, thus Baseline 1 with $B_i =10$ in Fig. \ref{fig:perforemance_uniform_iid}(a) and Fig. \ref{fig:perforemance_uniform_iid}(b) has a deteriorated performance due to the higher TO probabilities and large number of retransmission rounds.
Interestingly, as shown in Fig. \ref{fig:perforemance_uniform_iid}(c)-(d), with the number of selected clients $K$ increasing to $100$, the effect of outage probabilities in Baseline 1 will be alleviated since more clients can transmit their local model update successfully.

It can also be observed from Fig. \ref{fig:perforemance_uniform_iid} that Baseline 2 \cite{salehi2020federated} with $B_i=5$ and $10$ fails to learn the model.
This is because, for the partial participation with $K=10$, higher selection probabilities in Baseline 2 are allocated to the clients with larger TO probabilities, thus reducing the effective number of active clients ${\bar K}$ and consequently slowing down the convergence speed of FL.
Meanwhile, for the full participation with $K=100$, Baseline 2 with larger $B_i=5$ and $10$ still cannot correctly update the global model since the averaging scheme (\ref{baseline_2}) in Baseline 2 will be unstable if the outage probability $q_i$ is large.

\textbf{The non-i.i.d. data case.}
Comparing Fig. \ref{fig:perforemance_uniform_iid} with Fig. \ref{fig:perforemance_uniform_noniid}, we can find that non-i.i.d. degrades all curves, but the proposed \texttt{FedTOE} still performs closely to the ideal scheme and outperforms both Baseline 1 and 2.
Specifically, one can observe from Fig. \ref{fig:perforemance_uniform_noniid} that Baseline 1 and 2 with $B_i=2$ have a deteriorated performance, since the non-i.i.d. data amplifies the effect of QE and $B_i = 2$ is not enough to accurately represent the model update.
Different from the previous i.i.d. case, the reason why Baseline 1 with $B_i = 5$ and $10$ fails to learn the model with non-i.i.d. data is that not only the high TO probabilities decrease $\bar K$ but also the non-uniform TO probabilities among clients cause the objective inconsistency as discussed in Theorem 1.
Meanwhile, as shown in Fig. \ref{fig:perforemance_uniform_noniid}(c) and Fig. \ref{fig:perforemance_uniform_noniid}(d), the influence of non-uniform TO on Baseline 1 under the non-i.i.d. case cannot be alleviated with the number of selected clients $K$ increasing to $100$.
Besides, different from Baseline 1 and 2, \texttt{FedTOE} can adaptively determine the quantization levels via \eqref{obj_func_A1} to achieve superior performance.

Finally, it can be observed from Fig. \ref{fig:perforemance_uniform_noniid_200} that under a looser per-round delay constraint ($\tau_{\max}=200$ms), Baseline 1 and 2 with $B_i=5$ and $10$ can also perform well since the TO probabilities under $\tau_{\max}=200$ms are no longer high and become similar among clients as shown in Fig. \ref{fig:PER_uniform_distribution}(b).
In this situation, QE becomes a dominant factor in the performance for FL, thus Baseline 1 and 2 with $B_i=2$ still perform worse owing to large QE.

As a brief summary, the proposed \texttt{FedTOE} can automatically find the optimal bandwidth allocation $W_i$, quantization level $B_i$, and transmission rate $R_i$ for each client under different transmission delay constraints, and performs a robust FL performance for both the i.i.d. and non-i.i.d. cases.

\subsubsection{Necessity of optimization on bandwidth allocation}

\begin{figure}[t]
\begin{minipage}[t]{1\linewidth}
\centering
\includegraphics[width= 3.4 in ]{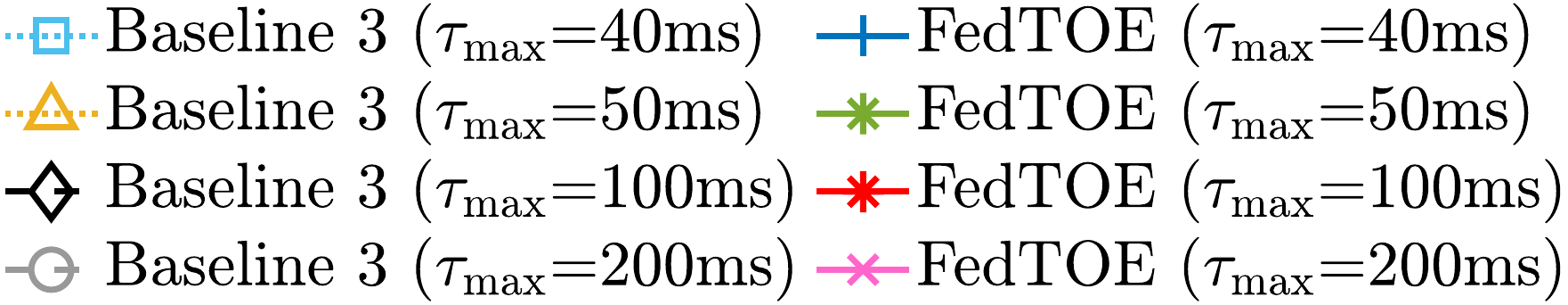}
\end{minipage}
\begin{minipage}[t]{1\linewidth}
\centering
\subfigure[\scriptsize{Training loss.}]{
\includegraphics[width= 2.2 in ]{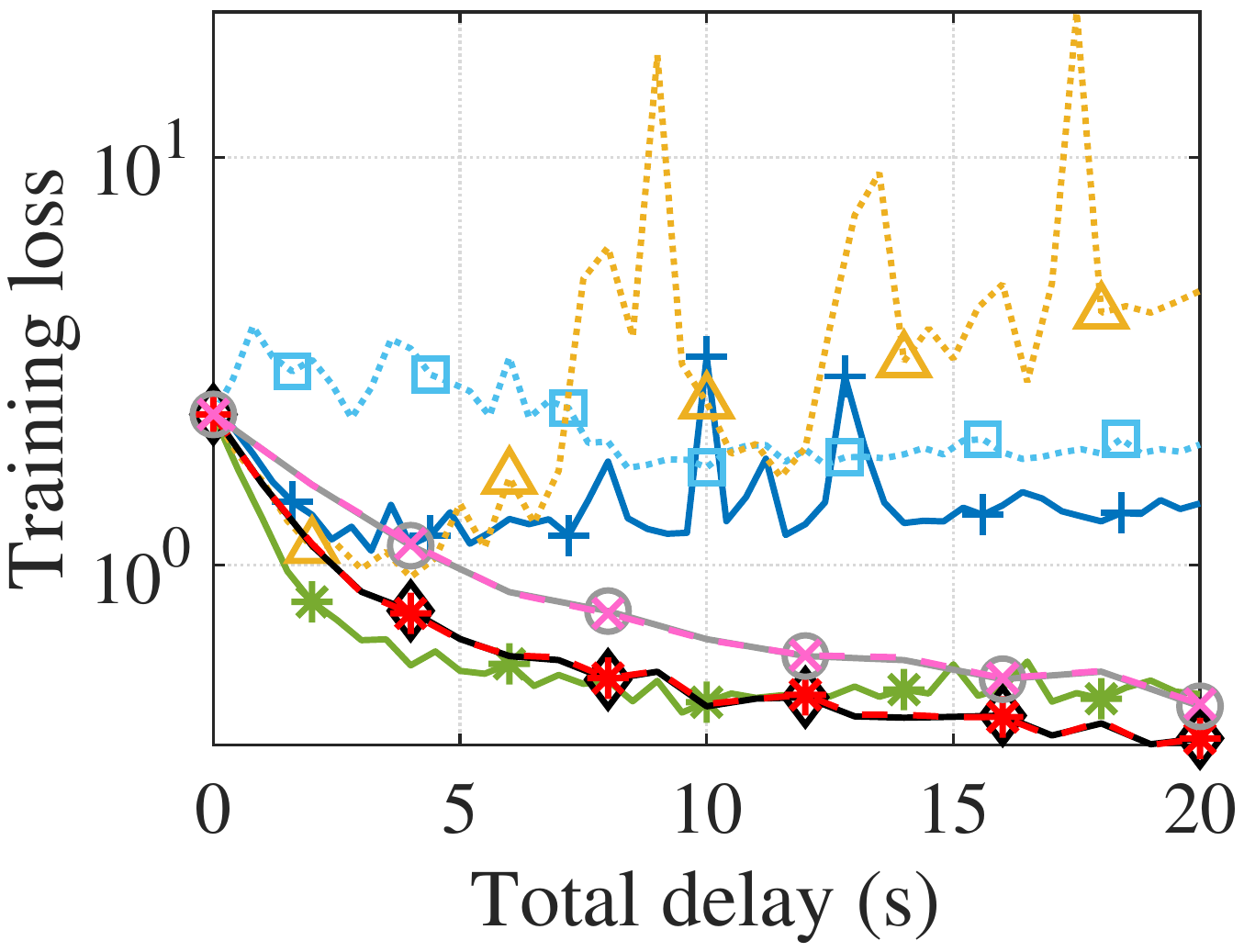}}
\subfigure[\scriptsize{Testing accuracy.}]{
\includegraphics[width= 2.2 in ]{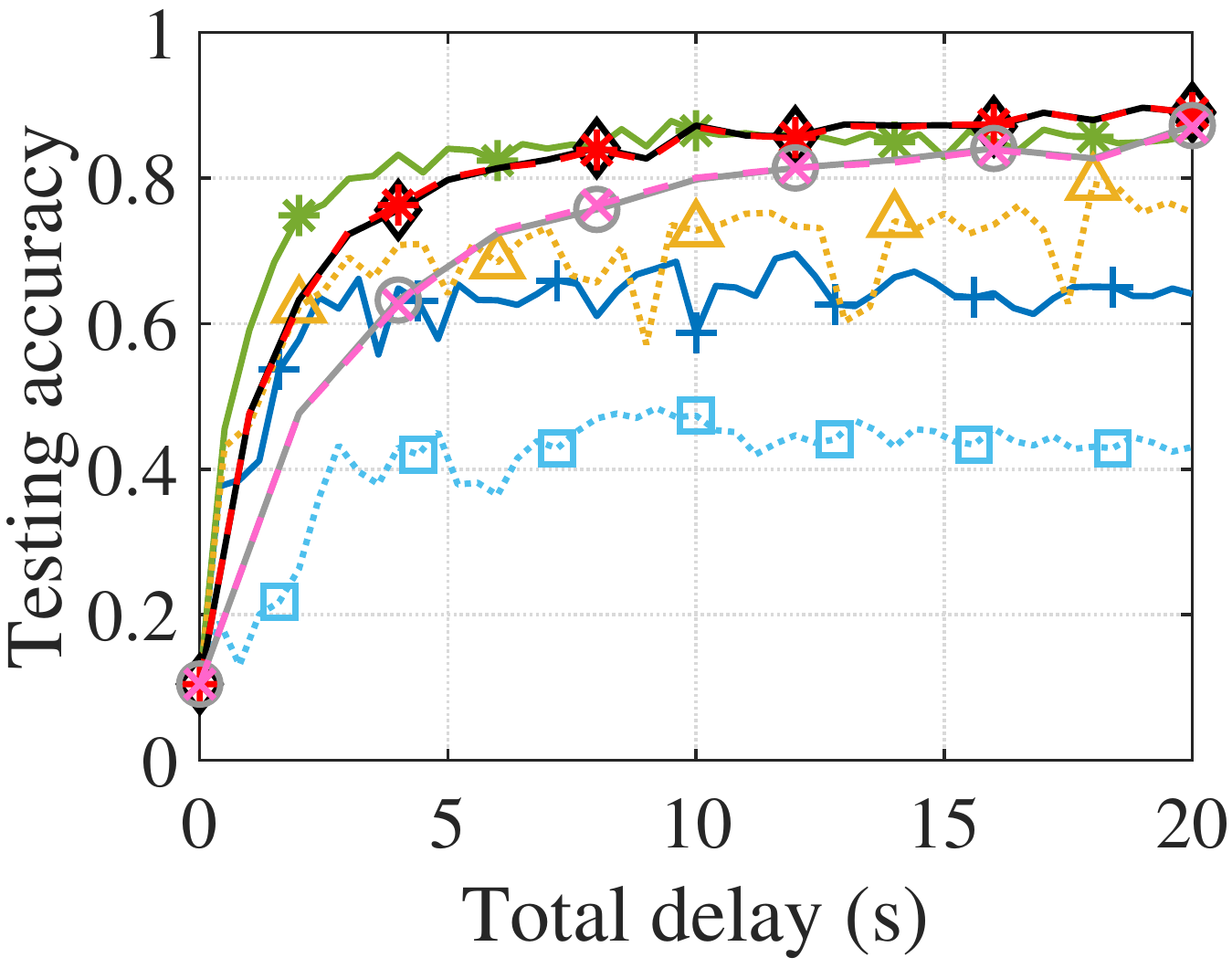}}
\caption{Comparison between Baseline 3 and \texttt{FedTOE} with $K = 10$ and different $\tau_{\max}$ for offline scheduling under the non-i.i.d. MNIST data.}
\label{fig:communication_time_uniform_noniid}
\end{minipage}
\end{figure}

In this part, we demonstrate the necessity of optimizing the bandwidth allocation for FL.
First of all, Fig. \ref{fig:communication_time_uniform_noniid} compares the training loss and testing accuracy of \texttt{FedTOE} and Baseline 3 with respect to the total uplink transmission time $\tau_{\rm total} = M\tau_{\max}$, under various  per-round delay constraints $\tau_{\max}$.
One can observe that for $\tau_{\max}=50$ms,  \texttt{FedTOE} performs significantly better than Baseline 3,  and for $\tau_{\max}\geq 100$ms, the two schemes perform comparably.  However,  both schemes don't converge well for $\tau_{\max}=40$ms due to the insufficient number of quantization bits under the stringent delay constraint.

\begin{figure}[t]
\begin{minipage}[t]{1\linewidth}
\centering
\subfigure[\scriptsize{Allocated bandwidth $W_i$.}]{
\includegraphics[width= 2.2 in ]{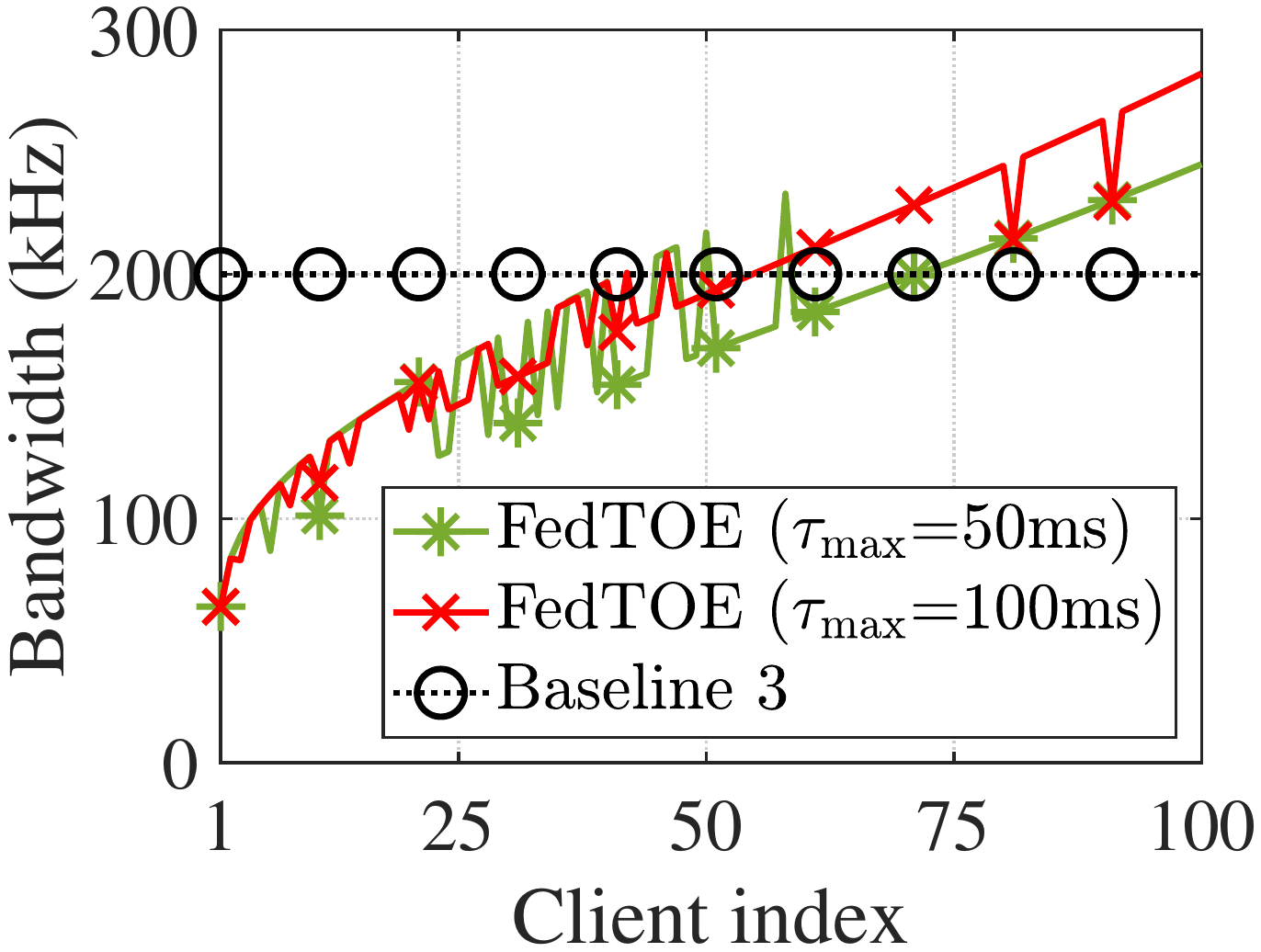}}
\subfigure[\scriptsize{Quantization level $B_i$.}]{
\includegraphics[width= 2.2 in ]{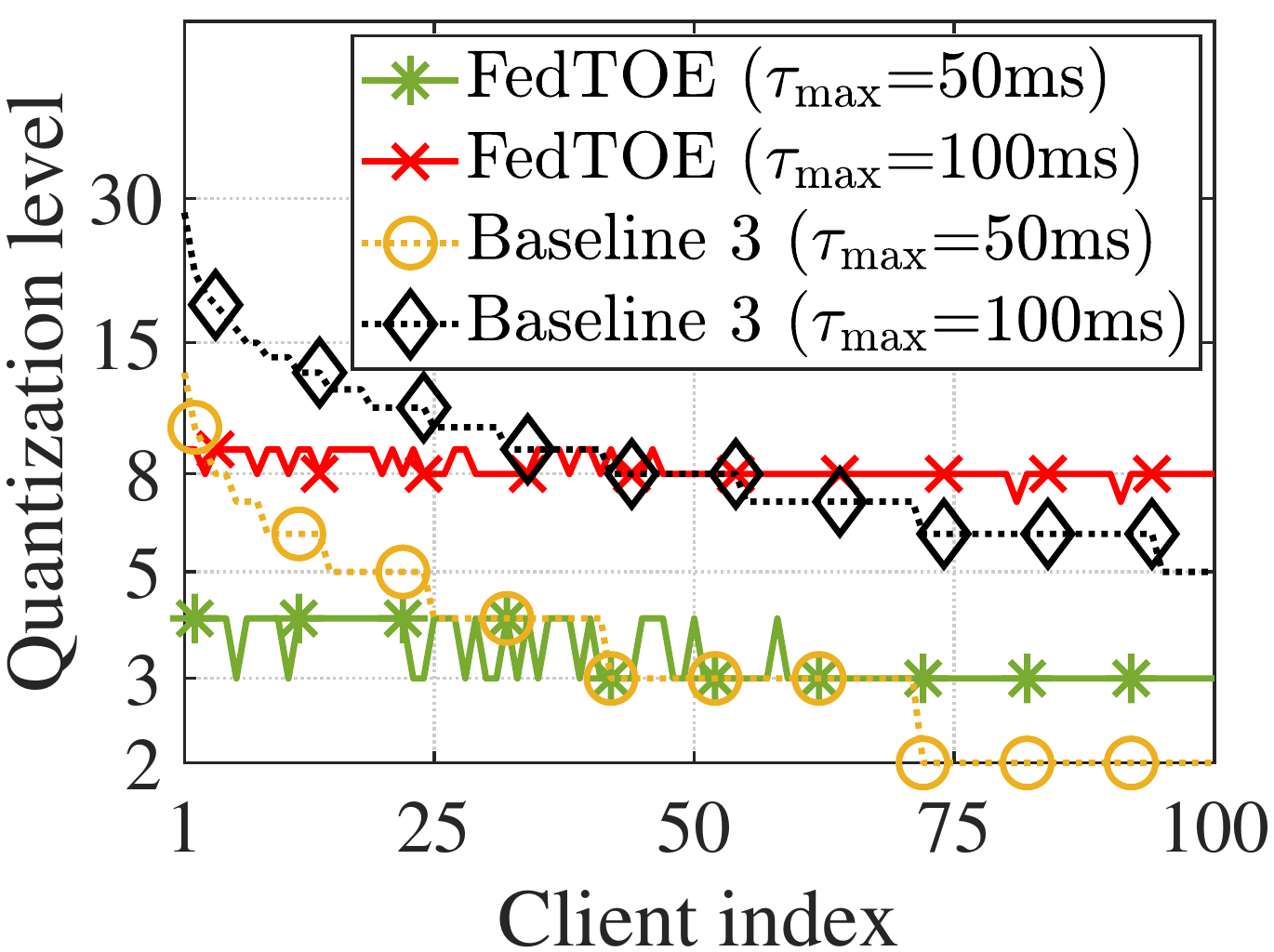}}
\caption{Allocated bandwidth and quantization level of each client for offline scheduling (Client with larger index is farther from the server).}
\label{fig:quantization_level_bandwdith_uniform}
\end{minipage}
\end{figure}

To analyze the cause why \texttt{FedTOE} outperforms Baseline 3,   we plot in Fig. \ref{fig:quantization_level_bandwdith_uniform} the uplink bandwidth and quantization level allocated to clients by the two schemes,
where the client with a larger index is farther from the server.
In the optimal wireless resource allocation scheme of both \texttt{FedTOE} and Baseline 3, the outage probabilities for all clients achieve $q_{\max} = 0.1$.
With this condition, it can be seen from Fig. \ref{fig:quantization_level_bandwdith_uniform}(a) that \texttt{FedTOE} prefers to allocate more bandwidth to the clients farther away from the server while less bandwidth to the clients close to the server, thus allowing a more uniform allocation of quantization bits as shown in Fig. \ref{fig:quantization_level_bandwdith_uniform}(b).
On the contrary, Baseline 3 (which has a uniform bandwidth allocation) allocates larger $B_i$ to the clients close to the server since they have larger channel capacity whereas Baseline 3 has to allocate smaller $B_i$ to the distant clients due to the delay constraint and it causes significant QE.
Therefore, when $\tau_{\max}$ is large, \texttt{FedTOE} and Baseline 3 perform equally well. However, when $\tau_{\max}$ is small,  \texttt{FedTOE} can greatly outperform Baseline 3 as seen in Fig. \ref{fig:communication_time_uniform_noniid}.

Lastly, one can see from Fig. \ref{fig:communication_time_uniform_noniid} that a tighter per-round delay $\tau_{\max}$ can speed up the learning process if the total uplink transmission time $\tau_{\rm total}$ is constrained.
For example, \texttt{FedTOE} under $\tau_{\max}=50$ms has a faster learning speed than those under $\tau_{\max}\geq 100$ms.
This is because a smaller $\tau_{\max}$ allows a larger number of communication rounds $M$ under a fixed $\tau_{\rm total}$.
Similarly, one can see that Baseline 3 under a smaller $\tau_{\max}$ converges faster than that under $\tau_{\max}\geq 100$ms.

\subsubsection{Influence of shadowing on learning performance}

\begin{figure}[t]
\begin{minipage}[t]{1\linewidth}
\centering
\subfigure[\scriptsize{Training loss.}]{
\includegraphics[width= 2.2 in ]{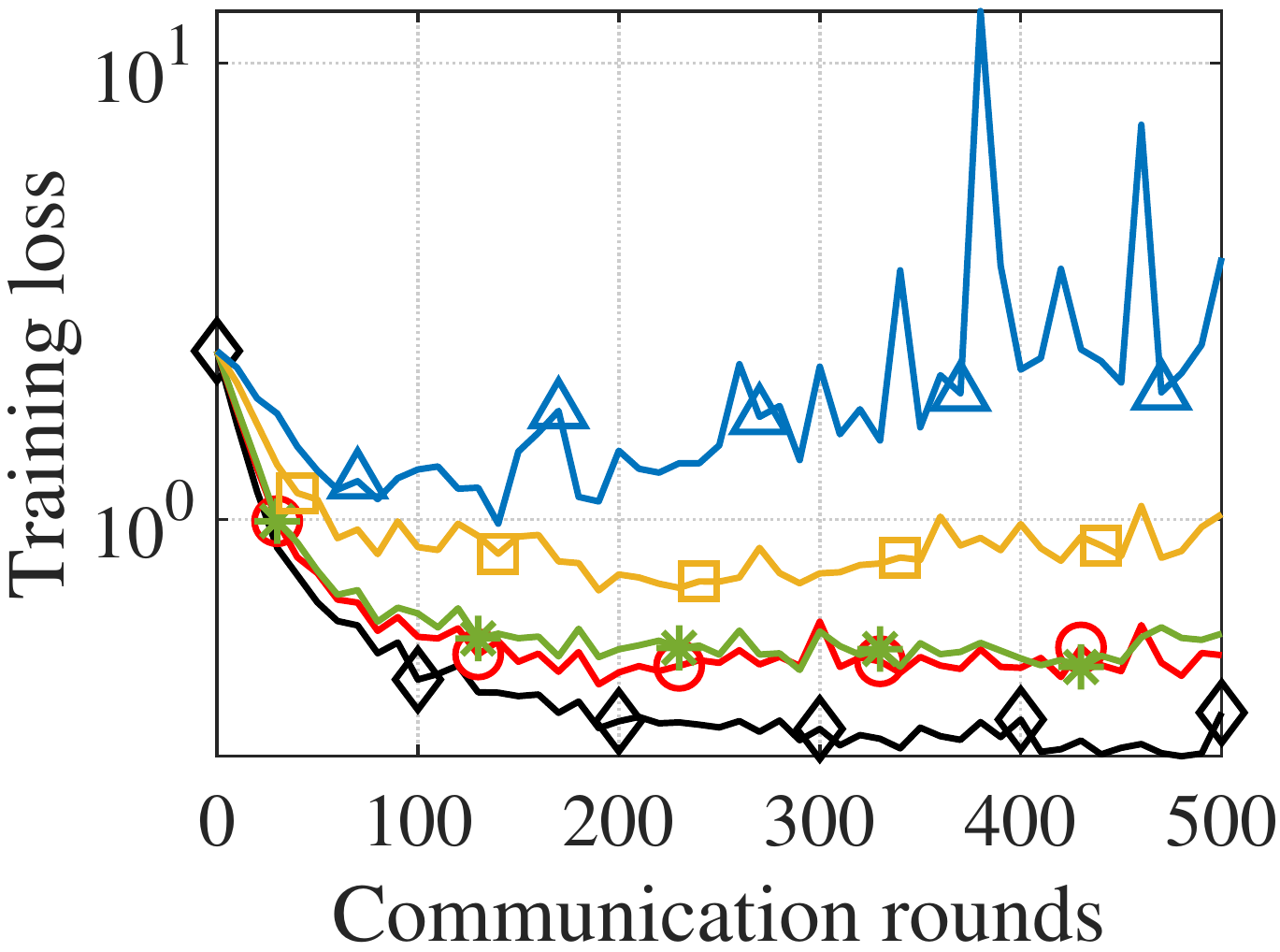}}
\subfigure[\scriptsize{Testing accuracy.}]{
\includegraphics[width= 2.2 in ]{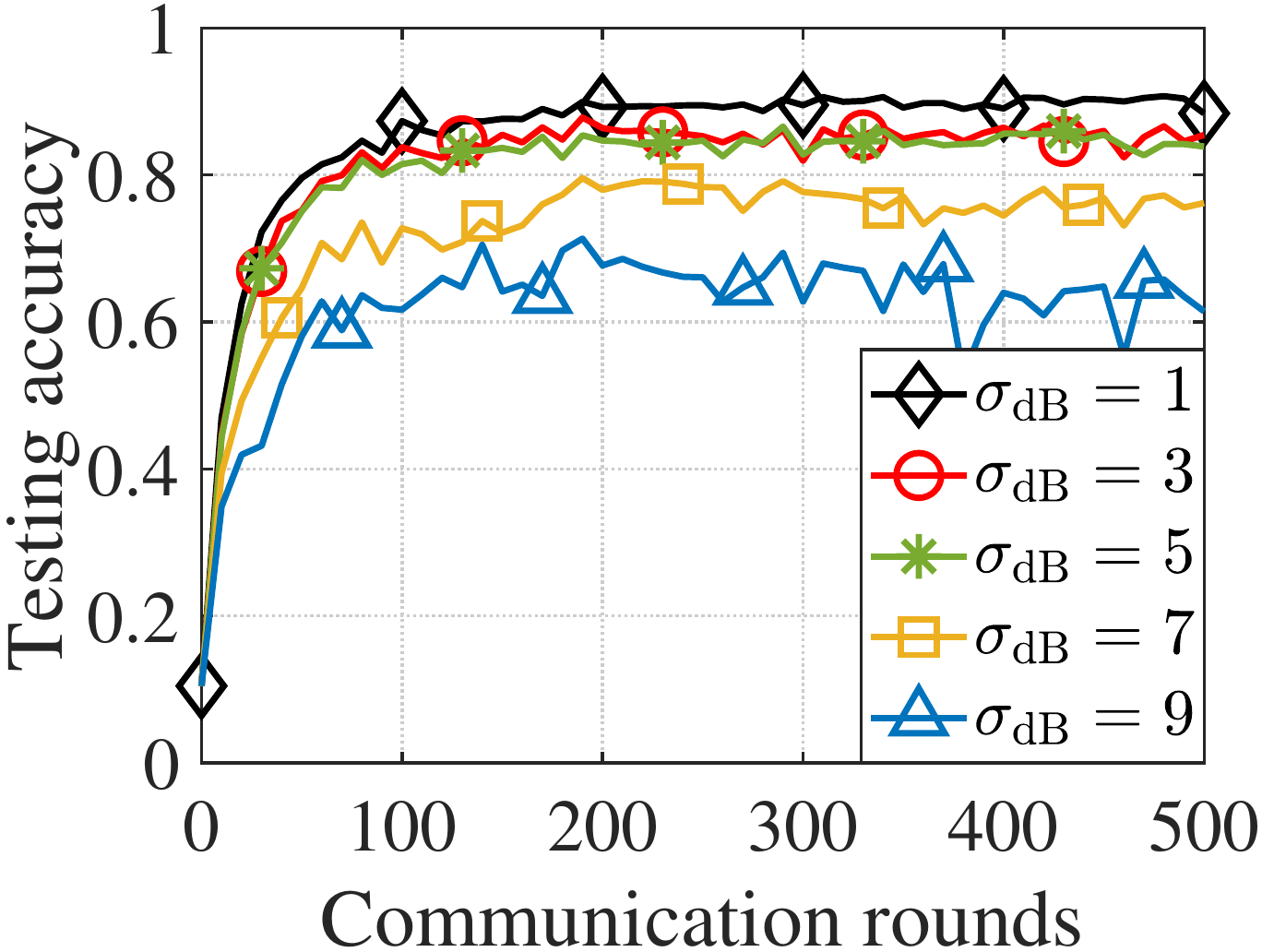}}
\caption{Performance of \texttt{FedTOE} under different values of  $\sigma_{\rm dB}$, for $K = 10$, $\tau_{\max}=50$ms, $q_{\max}=0.1$ and non-i.i.d. MNIST data.}
\label{fig:FedTOE_Performance_different_shadowing}
\end{minipage}
\end{figure}

We now discuss the influence of shadowing on \texttt{FedTOE}.
Fig. \ref{fig:FedTOE_Performance_different_shadowing} compares the learning performance of \texttt{FedTOE} under different values of $\sigma_{\rm dB}$, where the TO probability for \texttt{FedTOE} is $q_{\max} = 0.1$.
As can be seen from this figure, the performance of \texttt{FedTOE} decreases with an increasing shadowing power.
The reason is that, under the constraint of TO probability $q_{\max}$, a larger $\sigma_{\rm dB}$ reduces the achievable transmission rate,  and thus the clients should select a smaller quantization level $B_i$ in order to meet the transmission delay constraint.
As a result,  the increased QE damages the learning performance of \texttt{FedTOE}.
This result is consistent with Proposition \ref{optimal_condition}.

\subsection{Performance Comparison with Online Scheduling}

In this subsection, the performance of the proposed \texttt{FedTOE} with online scheduling is evaluated under both the MNIST and CIFAR-10 datasets.

\subsubsection{Performance on MNIST dataset}

\begin{figure}[t]
\begin{minipage}[t]{1\linewidth}
\centering
\includegraphics[width= 3.5 in ]{4_legend.pdf}
\end{minipage}
\begin{minipage}[t]{1\linewidth}
\centering
\subfigure[\scriptsize{Training loss.}]{
\includegraphics[width= 2.2 in ]{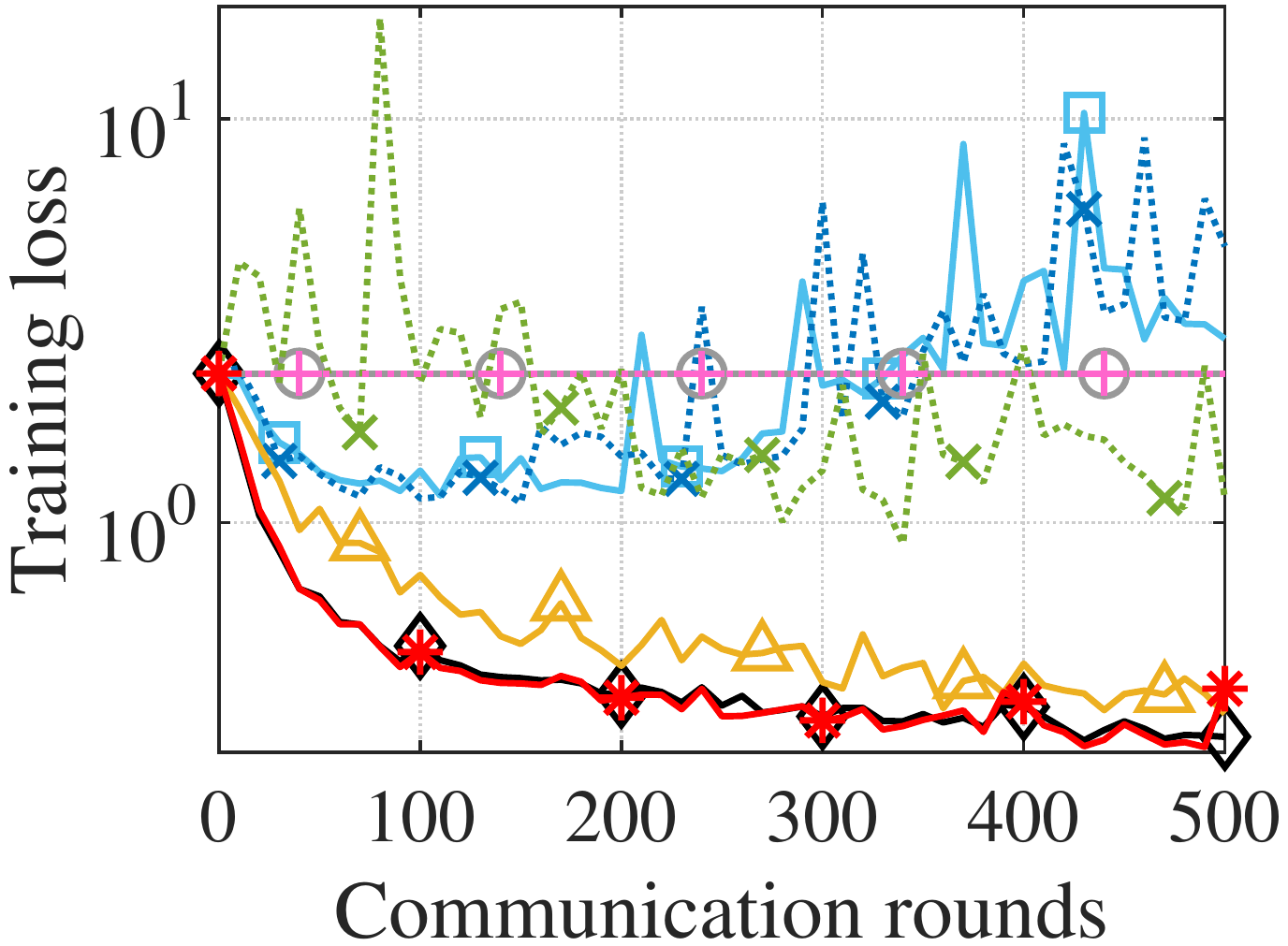}}
\subfigure[\scriptsize{Testing accuracy.}]{
\includegraphics[width= 2.2 in ]{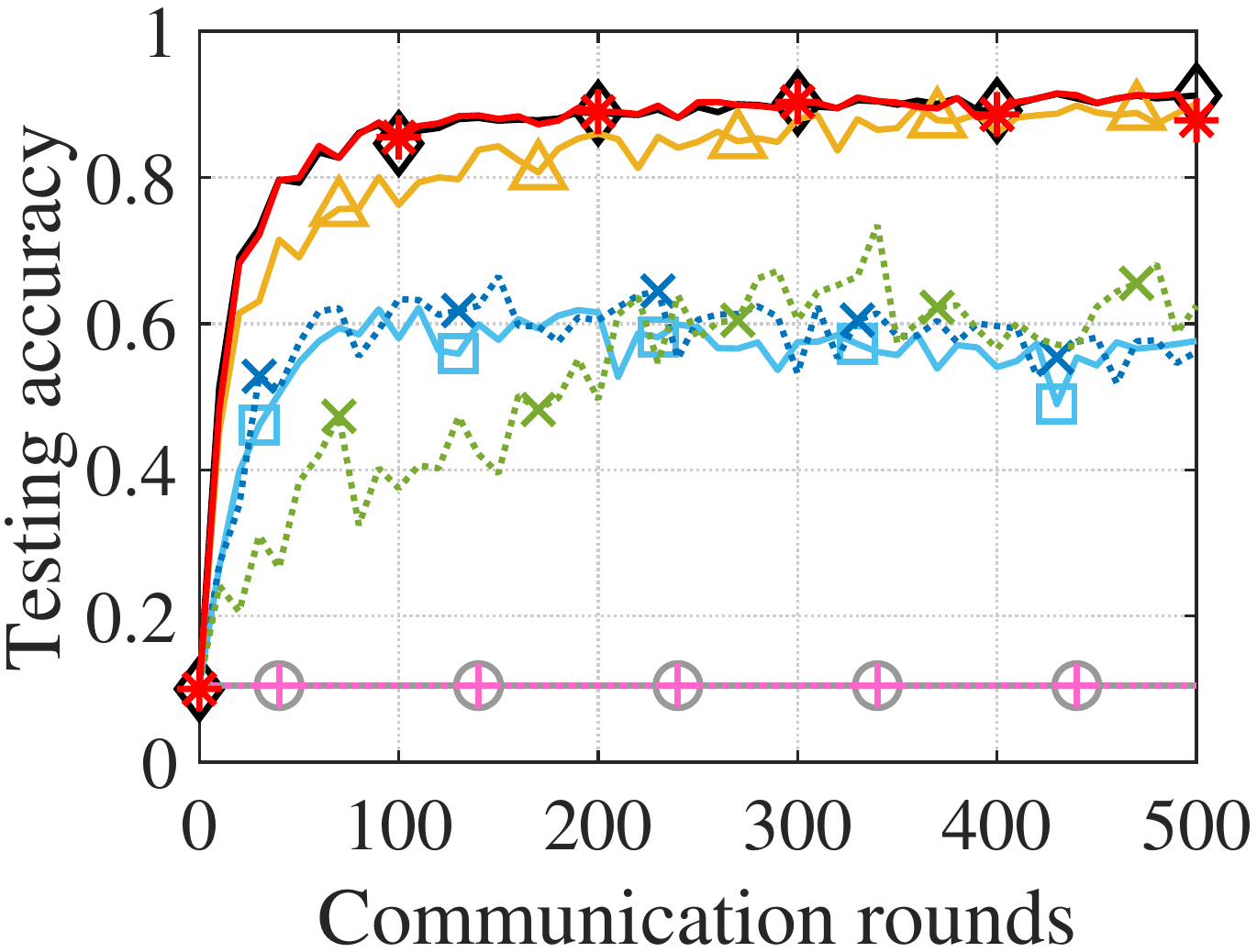}}
\caption{Comparison between baselines and \texttt{FedTOE} with $K = 10$ and $\tau_{\max}=9$ms for online scheduling under the non-i.i.d. MNIST data.}
\label{fig:perforemance_uniform_noniid_online}
\end{minipage}
\end{figure}

In online scheduling, the total 20MHz bandwidth is allocated to only the $K=10$ selected clients per round instead of to all the 100 clients in the offline scheme.
So a larger allocated bandwidth of clients can improve their transmission rates and then reduce the uplink transmission delay.
Thus, compared with the adopted per-round uplink delay constraint $\tau_{\max}$ for offline scheduling in Fig. \ref{fig:perforemance_uniform_noniid}, we choose a much tighter $\tau_{\max} = 9$ms to compare the training loss and testing accuracy of \texttt{FedTOE}, Baseline 1, and Baseline 2 in online scheduling.
It can be seen from Fig. \ref{fig:perforemance_uniform_noniid_online} that \texttt{FedTOE} still has superior performance than Baseline 1 and 2 in the online scheduling.
Specifically, Baseline 1 and 2 with $B_i = 2$ have poorer performance because of higher QE, while $B_i = 10$ fails to update the global model due to high TO probabilities.
Meanwhile, Baseline 2 with $B_i = 5$ converges slower and fluctuates a lot because of the unstable average scheme \eqref{baseline_2} under high TO probabilities.
While Baseline 1 with $B_i = 5$ gradually approaches to \texttt{FedTOE}, \texttt{FedTOE} has a faster convergence rate and can dynamically adjust the quantization levels by (\ref{obj_func_B1}) at each communication round.

\begin{figure}[t]
\begin{minipage}[t]{1\linewidth}
\centering
\includegraphics[width= 3.25 in ]{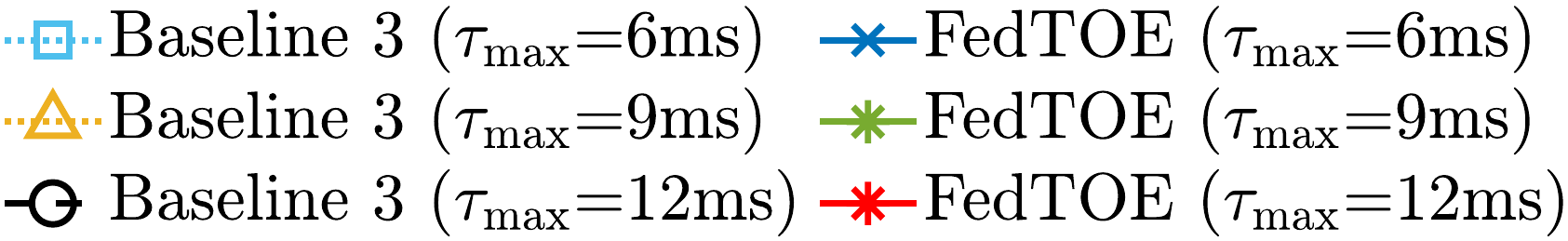}
\end{minipage}
\begin{minipage}[t]{1\linewidth}
\centering
\subfigure[\scriptsize{Training loss.}]{
\includegraphics[width= 2.2 in ]{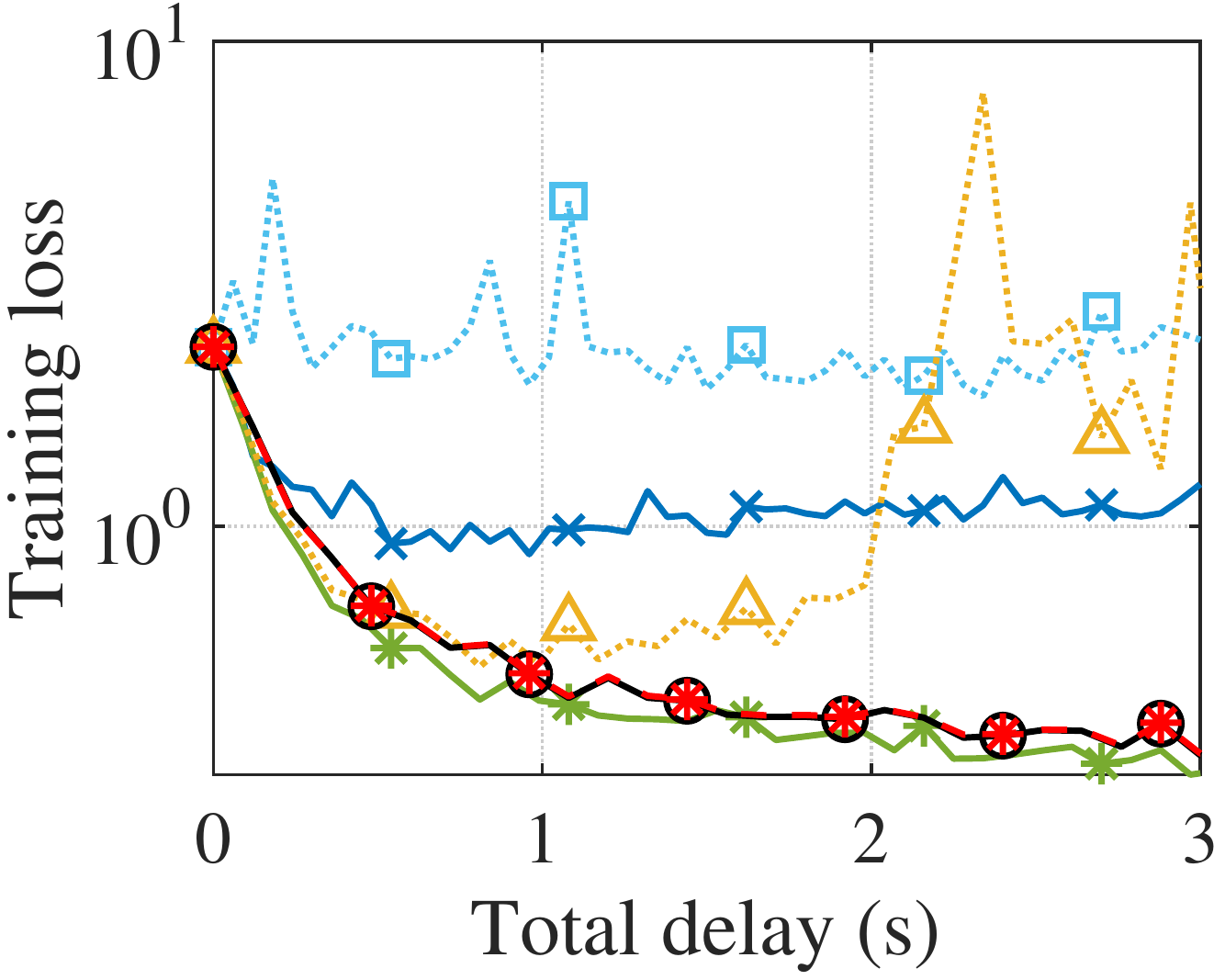}}
\subfigure[\scriptsize{Testing accuracy.}]{
\includegraphics[width= 2.2 in ]{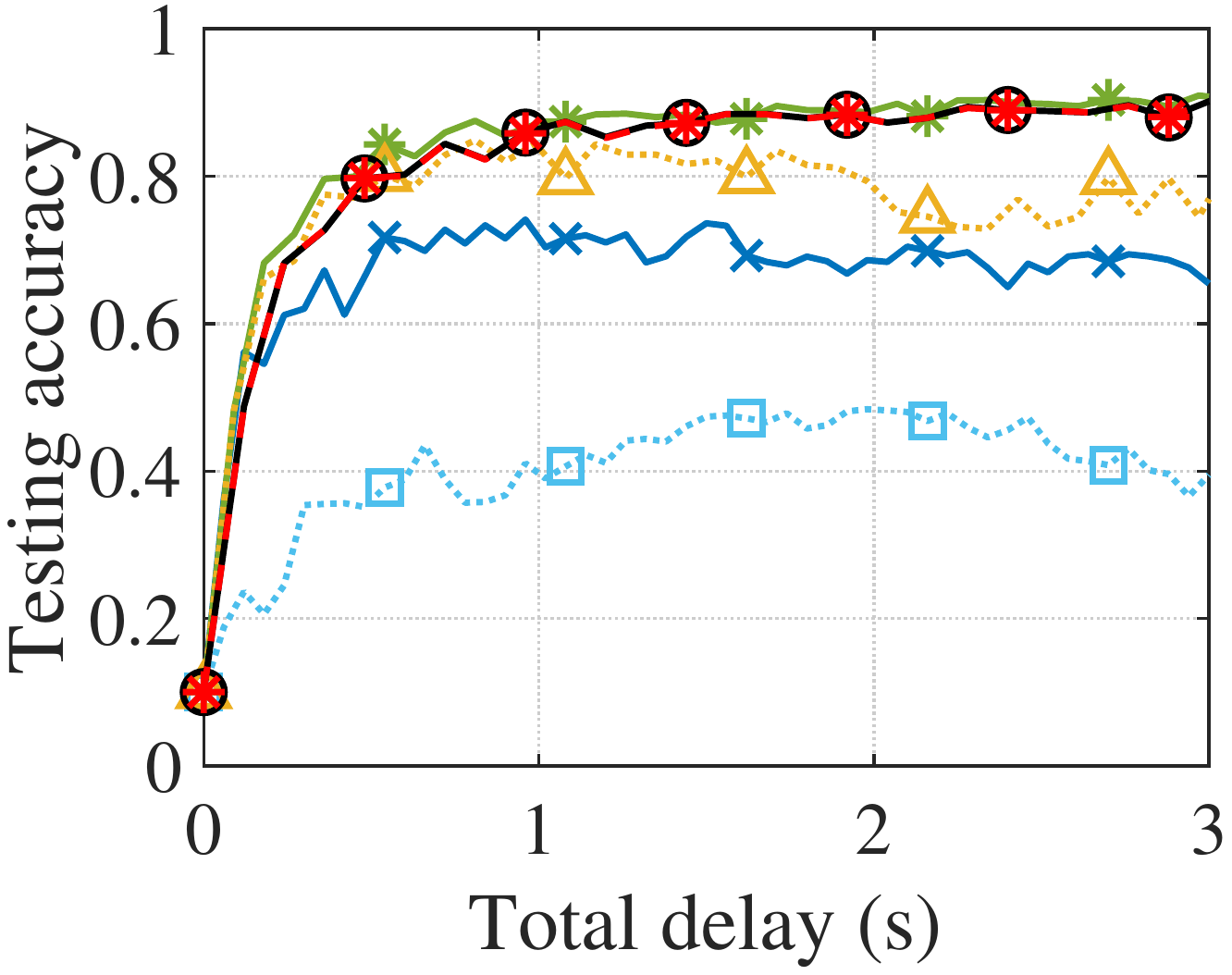}}
\caption{Comparison between Baseline 3 and \texttt{FedTOE} with $K = 10$ and different $\tau_{\max}$ for online scheduling under the non-i.i.d. MNIST data.}
\label{fig:communication_time_uniform_noniid_online}
\end{minipage}
\end{figure}

Next, Fig. \ref{fig:communication_time_uniform_noniid_online} compares the performance of \texttt{FedTOE} and Baseline 3 under online scheduling with different uplink delay constraints.
It can also be observed from Fig. \ref{fig:communication_time_uniform_noniid_online} that for a smaller uplink delay $\tau_{\max}=6$ms or $9$ms, \texttt{FedTOE} has a significant advantage over Baseline 3.

\subsubsection{Performance on CIFAR-10 dataset}

\begin{figure}[t]
\begin{minipage}[t]{1\linewidth}
\centering
\includegraphics[width= 3.5 in ]{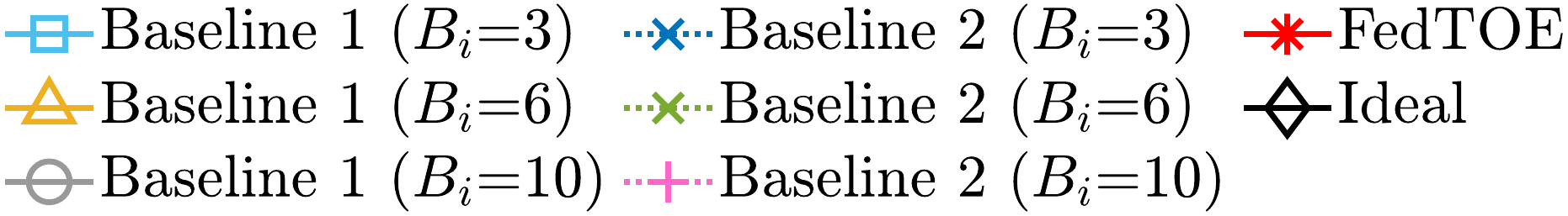}
\end{minipage}
\begin{minipage}[t]{1\linewidth}
\centering
\subfigure[\scriptsize{Training loss.}]{
\includegraphics[width= 2.2 in ]{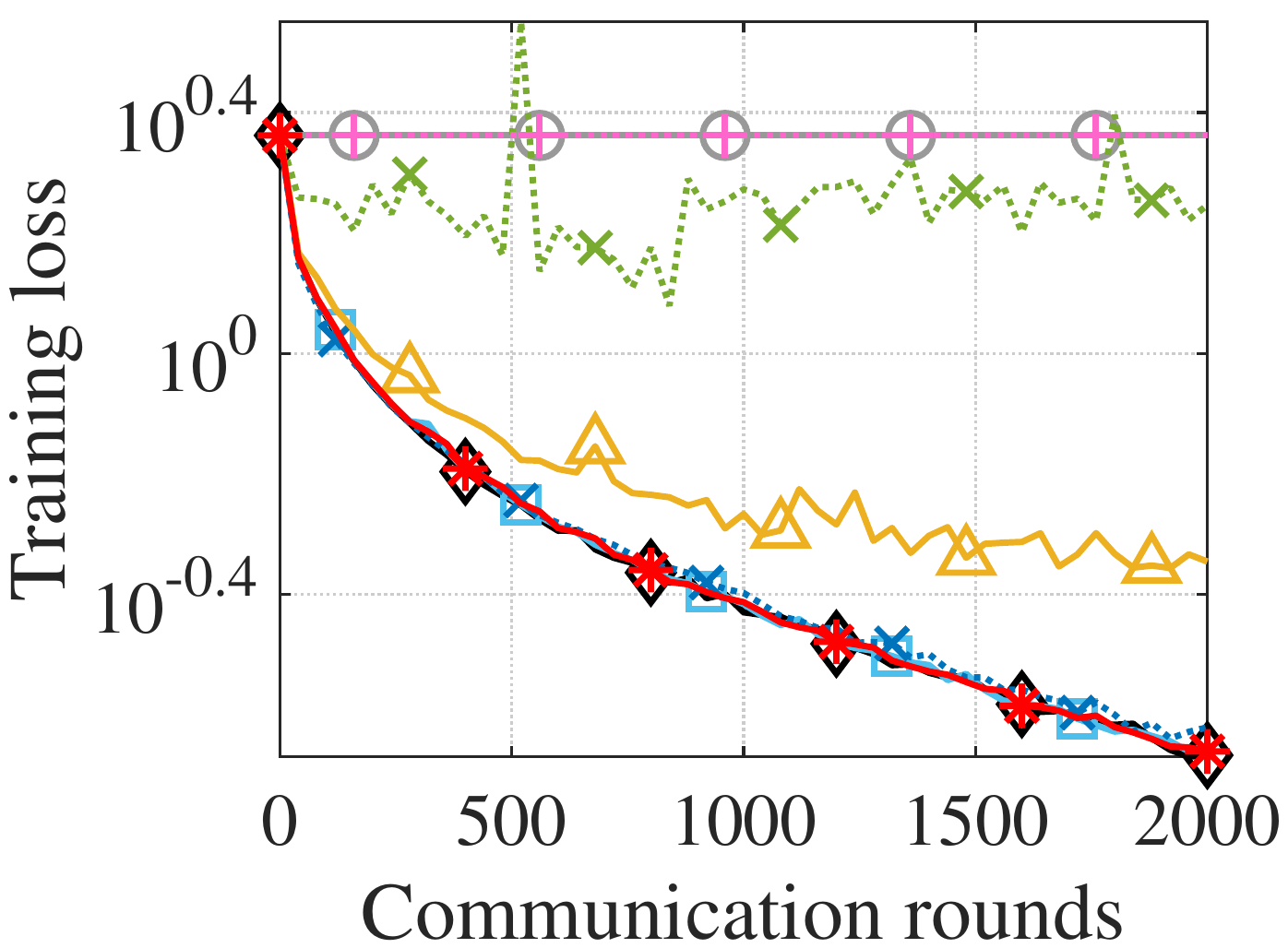}}
\subfigure[\scriptsize{Testing accuracy.}]{
\includegraphics[width= 2.2 in ]{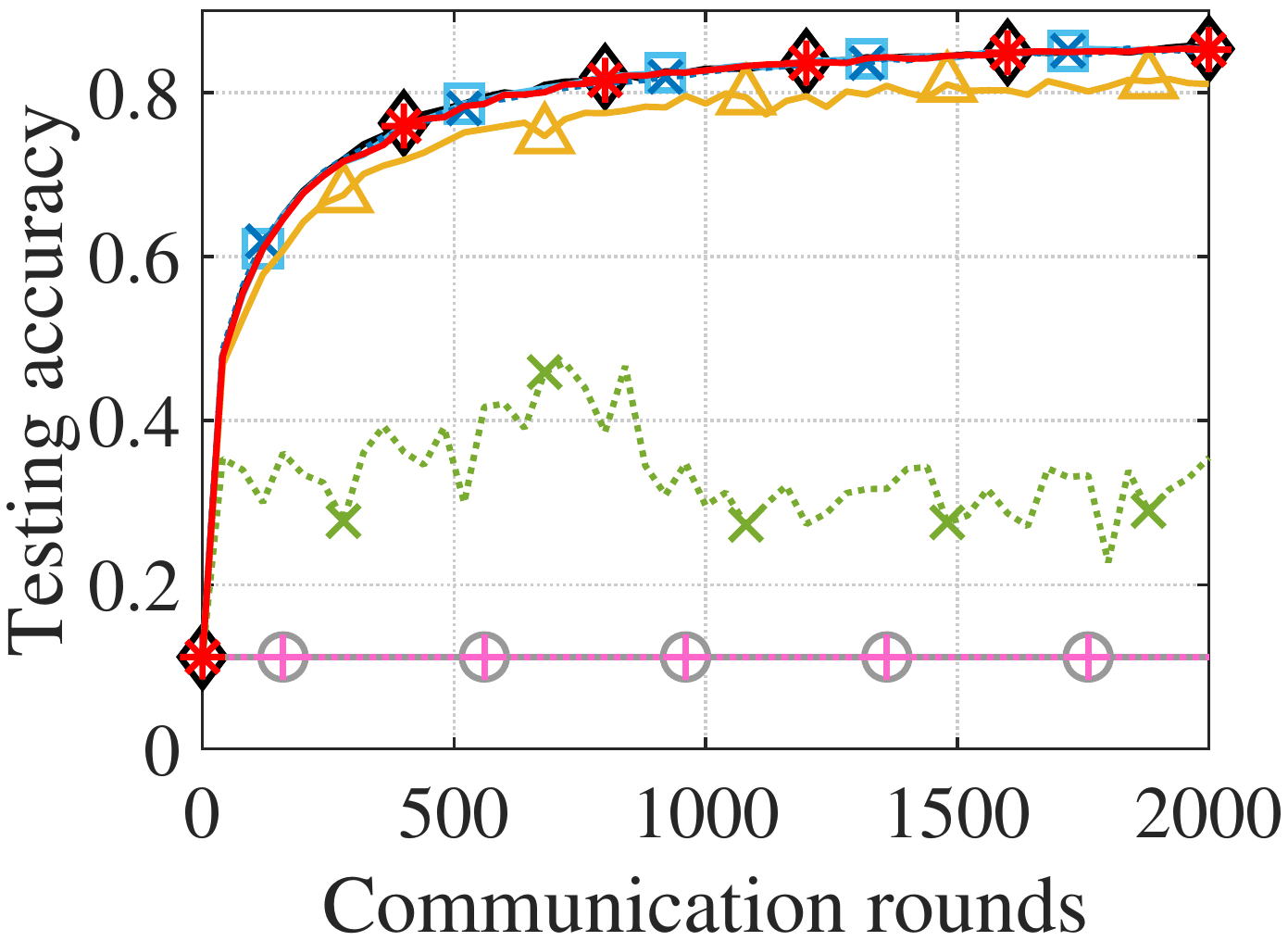}}
\caption{Comparison between baselines and \texttt{FedTOE} with $K = 10$ and $\tau_{\max}=90$ms for online scheduling under the i.i.d. CIFAR-10 data.}
\label{fig:perforemance_uniform_iid_online_CIFAR10}
\end{minipage}
\end{figure}

\begin{figure}[t]
\begin{minipage}[t]{1\linewidth}
\centering
\subfigure[\scriptsize{Training loss.}]{
\includegraphics[width= 2.2 in ]{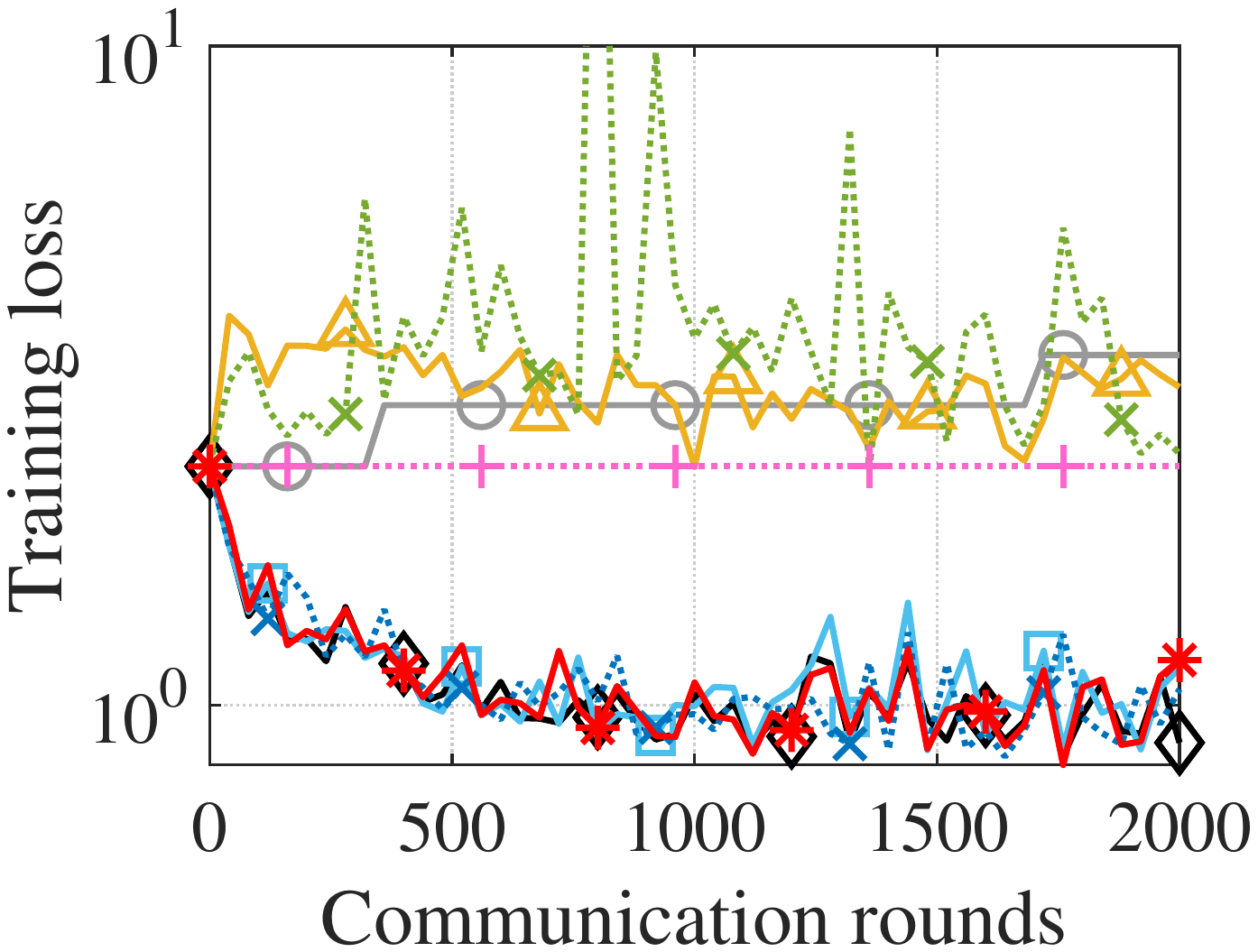}}
\subfigure[\scriptsize{Testing accuracy.}]{
\includegraphics[width= 2.2 in ]{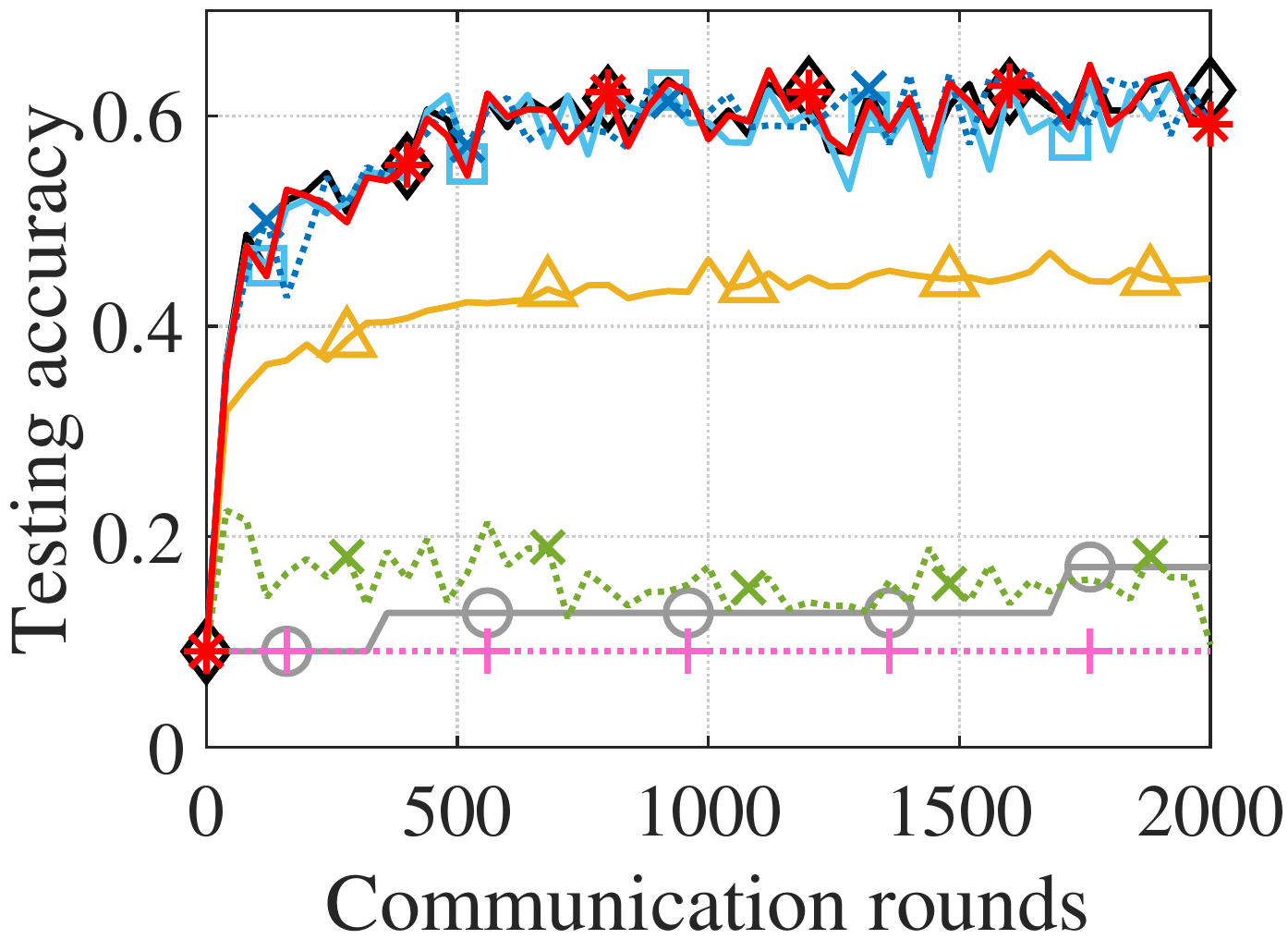}}
\caption{Comparison between baselines and \texttt{FedTOE} with $K = 10$ and $\tau_{\max}=90$ms for online scheduling under the non-i.i.d. CIFAR-10 data (with the same legend as Fig. \ref{fig:perforemance_uniform_iid_online_CIFAR10}).}
\label{fig:perforemance_uniform_noniid_online_CIFAR10}
\end{minipage}
\end{figure}

In this part, we examine the performance of \texttt{FedTOE} on the CIFAR-10 dataset, and compare it
with the baselines under the i.i.d. case and non-i.i.d. case.
Since the model size of ResNet-20 ($m = 271098$) is much larger,  we choose a larger delay constraint $\tau_{\max}=90$ms.
As shown in Fig. \ref{fig:perforemance_uniform_iid_online_CIFAR10} and Fig. \ref{fig:perforemance_uniform_noniid_online_CIFAR10}, the proposed \texttt{FedTOE} can still perform closely to the ideal case under both the i.i.d. and non-i.i.d. data distributions.
Meanwhile,  from Fig. \ref{fig:perforemance_uniform_iid_online_CIFAR10},  Baseline 1 with $B_i=6$ and $B_i=10$ has an impaired performance in the i.i.d. case due to high TO probabilities,  which becomes even worse in the non-i.i.d. case as shown in Fig. \ref{fig:perforemance_uniform_noniid_online_CIFAR10}.
In addition,  one can observe that Baseline 2 with $B_i=6$ cannot converge well and performs poorer than Baseline 1.
These results are consistent with those in Fig. \ref{fig:perforemance_uniform_iid} and Fig. \ref{fig:perforemance_uniform_noniid}.

\begin{figure}[t]
\begin{minipage}[t]{1\linewidth}
\centering
\includegraphics[width= 3.25 in ]{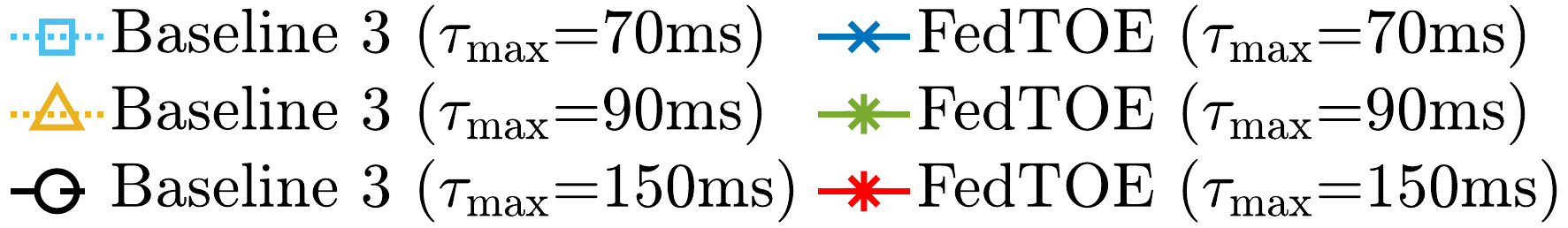}
\end{minipage}
\begin{minipage}[t]{1\linewidth}
\centering
\subfigure[\scriptsize{Training loss.}]{
\includegraphics[width= 2.2 in ]{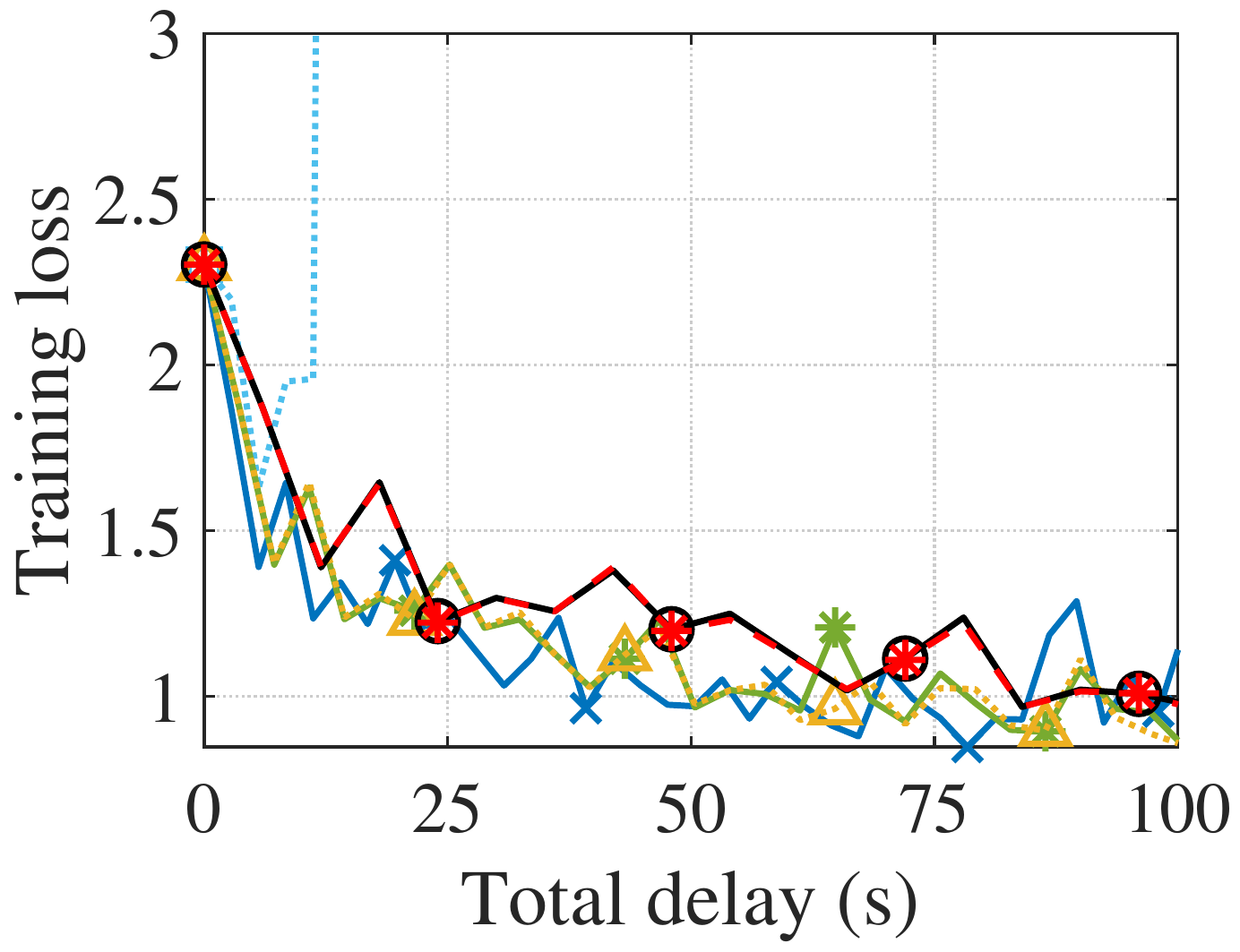}}
\subfigure[\scriptsize{Testing accuracy.}]{
\includegraphics[width= 2.2 in ]{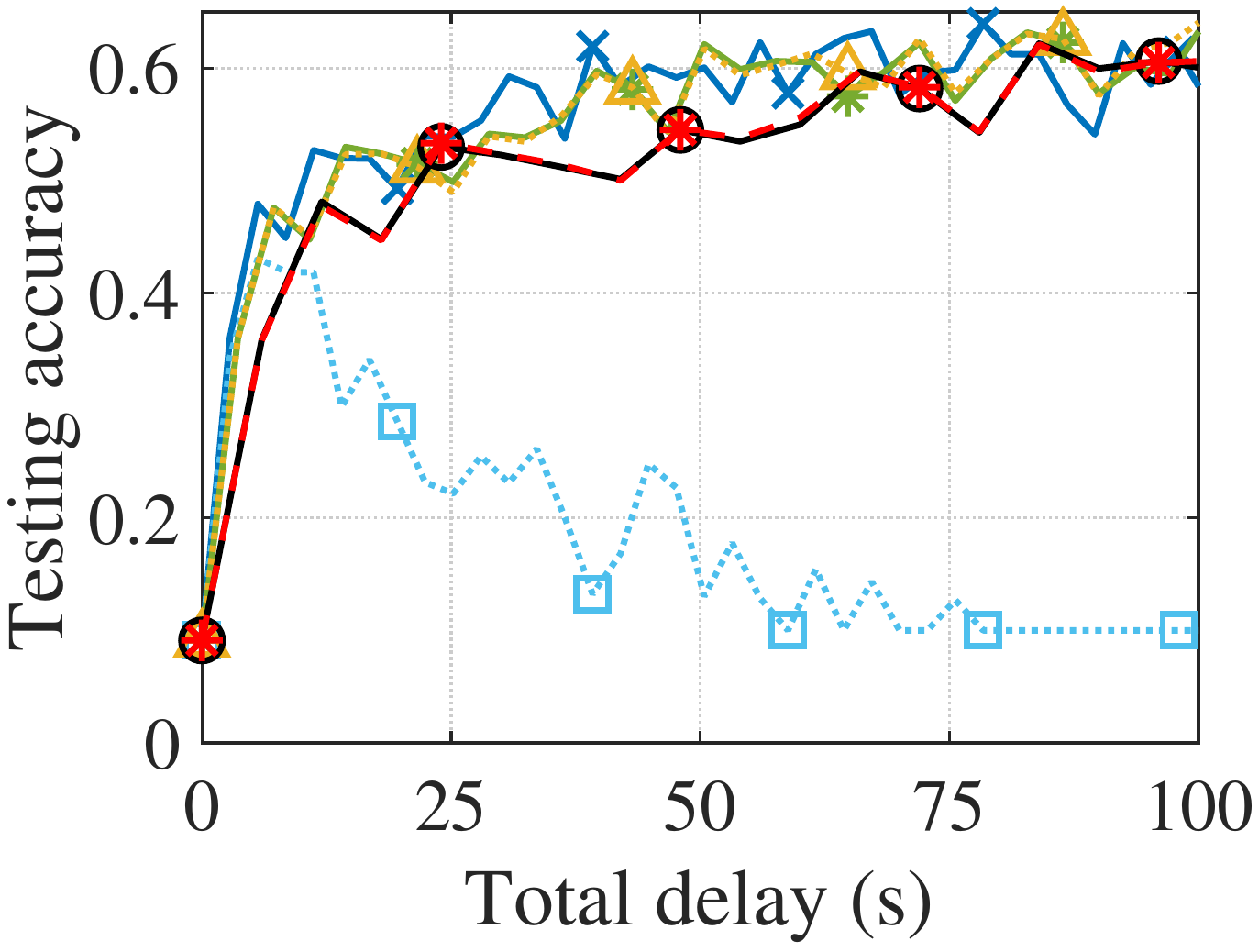}}
\caption{Comparison between Baseline 3 and \texttt{FedTOE} with $K = 10$ and different $\tau_{\max}$ for online scheduling under the non-i.i.d. CIFAR-10 data.}
\label{fig:communication_time_uniform_noniid_online_CIFAR}
\end{minipage}
\end{figure}

Finally, the performance of \texttt{FedTOE} and Baseline 3 in online scheduling is compared in Fig. \ref{fig:communication_time_uniform_noniid_online_CIFAR} under different transmission delay constraints.
It can be observed in Fig. \ref{fig:communication_time_uniform_noniid_online_CIFAR} that under a tighter $\tau_{\max} = 70$ms,  Baseline 3 cannot perform well whereas \texttt{FedTOE}  still achieves a good learning performance,
which again demonstrate the necessity of jointly optimization of bandwidth allocation and quantization level for achieving promising performance under tight delay constraints.

\section{Conclusion}\label{sec: Conclusion}

In this paper, we have investigated FL in non-ideal wireless channels in the presence of both TO and QE.
We have carried out a novel convergence analysis that shows TO and QE, together with non-i.i.d. data distribution,  can significantly impede the FL process.
In particular, we have shown that when the clients have heterogeneous TO probabilities,  not only the negative effects of QE and non-i.i.d data distribution can be enlarged but also the algorithm can converge to a biased solution. On the contrary,  when the clients have a uniform TO probability,  these issues can be alleviated and the algorithm achieves a linear speedup with the number of (effective) clients.
Inspired by this result, we have proposed  \texttt{FedTOE} which performs joint allocation of bandwidth and quantization bits to minimize the QE while satisfying the transmission delay constraint and uniform TO probabilities.
Based on the MNIST and CIFAR-10 datasets, the presented experiment results have demonstrated that \texttt{FedTOE} exhibits superior robustness against TO and QE when compared to the existing schemes.
Moreover, experiment results have also shown that a tighter transmission delay constraint per communication round may speed up the FL process.

There exist several interesting directions for future research.
While our current work has modeled the transmission outage by assuming no CSIT, it is equally interesting to consider the cases with CSIT error \cite{wang2014outage} or with finite blocklength transmission \cite{xu2020transmission}.
It will lead to different resource allocation problems and require new algorithm designs.
Another direction is to incorporate the outage probability $q_{\max}$ into the joint optimization of bandwidth and quantization levels in\eqref{obj_func_A1} and \eqref{obj_func_B1}.
In that case,  since the effective number of clients $\bar K$ depends on $q_{\max}$ (see Lemma 2), one may need to minimize not only the term about the QE but also the whole upper bound in the RHS of  \eqref{corollary_1}.
Lastly,  non-orthogonal multiple access techniques such as multiuser beamforming via massive MIMO and the over-the-air computation technique \cite{salehi2020federated,zhu2020one}  can further improve the link quality and uplink capacity,  which will further translate into higher communication efficiency for FL.  It is also worth study in the future.

\begin{appendices}

\section*{\huge Appendices}

\section{Proof of Lemma \ref{beta_alpha_K}} \label{Proof_beta_alpha_K_fixed}

\subsection{Proof of \eqref{average_beta_1} and \eqref{average_alpha_1}} \label{average_beta_alpha_fixed}

At each communication round, $K$ clients are selected independently and with replacement based on the probability distribution $\{ p_i \}_{i=1}^N$.
As a result, there are $N^K$ different possibilities for the set ${\mathcal{S}}_{r}$ (denoted by $\mathcal{S}_r^g,  g=1,\ldots,N^K$) and the appearance probability of each set ${\mathcal{S}}_r^g$ is $\Pr({\mathcal{S}}_{r} = {\mathcal{S}}_r^g) = \prod_{{i} \in {\mathcal{S}}_r^g} p_{i}$.
Meanwhile,  since TOs occur independently across the clients, we have
$\Pr\left[ {\sum}_{{i} \in {\mathcal{S}}_{r}} \mathds{1}^{r}_{i} \neq 0\right] = 1 - {\prod}_{{i} \in {\mathcal{S}}_{r}} q_{i}$.
Then, we can obtain \eqref{average_beta_1} for some non-negative $\bar \beta_i$, $i=1,\ldots,N$, according to the derivations in  \eqref{beta_proof},
{\small
\begin{subequations}\label{beta_proof}
\begin{align}
& \mathbb{E}
\left[  \left. \frac{ {\sum}_{{i} \in {\mathcal{S}}_{r}} \mathds{1}^{r}_{i}
\Delta \mathbf{w}^{r}_i }{ {\sum}_{{i} \in {\mathcal{S}}_{r}} \mathds{1}^{r}_{i}} \right|
{\sum}_{{i} \in {\mathcal{S}}_{r}} \mathds{1}^{r}_{i} \neq 0  \right] \label{beta_proof_a} \\
= & {\mathbb{E}}_{\mathcal{S}_{r}}
\left[ {\mathbb{E}}_{\rm TO}
\left[ \left. \frac{ {\sum}_{{i} \in {\mathcal{S}}_{r}} \mathds{1}^{r}_{i} \Delta \mathbf{w}^{r}_i }{ {\sum}_{{i} \in {\mathcal{S}}_{r}} \mathds{1}^{r}_{i}} \right| {\sum}_{{i} \in {\mathcal{S}}_{r}} \mathds{1}^{r}_{i} \neq 0  \right] \right] \\
= &  {\mathbb{E}}_{\mathcal{S}_{r}}
\Bigg[ \sum_{v=1}^K
\sum_{\mathcal{B}_r \bigcup {\bar{\mathcal{B}}}_r = \mathcal{S}_{r} \atop |\mathcal{B}_r|=v, |{\bar{\mathcal{B}}}_r|=K-v}
\Pr  \bigg(  \mathds{1}^{r}_{k_1}  =  1 \, \forall k_1  \in  \mathcal{B}_r, \mathds{1}^{r}_{k_2}  =  0 \, \forall k_2  \in  {\bar{\mathcal{B}}}_r \bigg|  {\sum}_{{i} \in {\mathcal{S}}_{r}} \mathds{1}^{r}_{i}  \neq  0  \bigg)
\cdot  \frac{ {\sum}_{k_1 \in \mathcal{B}_r}  \Delta \mathbf{w}^{r}_{k_1}}{v}
 \Bigg] \label{beta_proof_b} \\
= &  {\sum}_{g=1}^{N^K}
\Bigg(  {\prod}_{{i} \in {\mathcal{S}}_r^g}  p_{i}  \Bigg)  \cdot
\Bigg(  {\sum}_{v=1}^K  {\sum}_{\mathcal{B}_r^g \bigcup {\bar{\mathcal{B}}}_r^g = {\mathcal{S}}_r^g \atop |\mathcal{B}_r^g|=v, |{\bar{\mathcal{B}}}_r^g|=K-v}
\frac{ {\prod}_{k_1 \in \mathcal{B}_r^g}  (1  -  q_{{k_1}})
 {\prod}_{k_2 \in {\bar{\mathcal{B}}}_r^g}  q_{{k_2}}}{1- {\prod}_{{i} \in {\mathcal{S}}_r^{g}} q_{i}}  \cdot  \frac{{\sum}_{k_1 \in \mathcal{B}_r^g}  \Delta \mathbf{w}^{r}_{k_1}}{v}  \Bigg)  \label{beta_proof_c} \\
\triangleq & {\sum}_{i=1}^N {\bar \beta}_i \Delta \mathbf{w}^{r}_i \label{beta_proof_d}
\end{align}
\end{subequations}}

\vspace{-0.3cm}\noindent
where in \eqref{beta_proof_b}, $\mathcal{B}_r$ is the set of selected clients without TO while ${\bar{\mathcal{B}}}_r$ is the one of clients with TO, and in \eqref{beta_proof_c},
$\prod_{k_1 \in \mathcal{B}_r^g} (1-q_{{k_1}}) \prod_{k_2 \in {\bar{\mathcal{B}}}_r^g} q_{{k_2}} $ is the probability of the event that solely the clients in $\mathcal{B}_r^g$ have successful transmissions.
By letting $\Delta \mathbf{w}^{r}_i = 1$ in (\ref{beta_proof_a}), we then have $\sum_{i=1}^N {\bar \beta}_i = 1$.
In the same fashion as (\ref{beta_proof}), we can obtain  %\eqref{average_alpha_1} for some ${\bar \alpha}_i\geq 0$ $\forall i =1,\cdots,N$.
{\small
\begin{subequations}\label{alpha_proof}
\begin{align}
& \mathbb{E}
\left[ \left. \frac{ {\sum}_{{i} \in {\mathcal{S}}_{r}} \mathds{1}^{r}_{i} \Delta \mathbf{w}^{r}_i }{ ({\sum}_{{i} \in {\mathcal{S}}_{r}} \mathds{1}^{r}_{i} )^2 } \right|
{\sum}_{{i} \in {\mathcal{S}}_{r}} \mathds{1}^{r}_{i} \neq 0 \right] \label{alpha_proof_a} \\
= &  {\sum}_{g=1}^{N^K}
\Bigg(  {\prod}_{{i} \in {\mathcal{S}}_r^g}  p_{i}  \Bigg)  \cdot
\Bigg(  {\sum}_{v=1}^K  {\sum}_{\mathcal{B}_r^g \bigcup {\bar{\mathcal{B}}}_r^g = {\mathcal{S}}_r^g \atop |\mathcal{B}_r^g|=v, |{\bar{\mathcal{B}}}_r^g|=K-v}
 \frac{ {\prod}_{k_1 \in \mathcal{B}_r^g}  (1  -  q_{{k_1}})
 {\prod}_{k_2 \in {\bar{\mathcal{B}}}_r^g}  q_{{k_2}}}{1- \prod_{{i} \in {\mathcal{S}}_r^{g}} q_{i}}  \cdot  \frac{{\sum}_{k_1 \in \mathcal{B}_r^g}  \Delta \mathbf{w}^{r}_{k_1}}{v^2}  \Bigg)  \notag \\
\triangleq & {\sum}_{i=1}^N {\bar \alpha}_i
\Delta \mathbf{w}^{r}_i \label{alpha_proof_b}
\end{align}
\end{subequations}}

\vspace{-0.3cm}\noindent
for some ${\bar \alpha}_i\geq 0$ $\forall i =1,\cdots,N$, which is \eqref{average_alpha_1}.
\hfill $\blacksquare$

\subsection{Computing the values of ${\bar \beta}_i$, ${\bar \alpha}_i$ and ${\bar K}$ under uniform-TO}\label{paramter value_uniform_TO}
With the same TO probability $q$ for all clients, (\ref{beta_proof}) becomes
{\small
\begin{align}\label{same_outage probability_beta_proof_1}
\eqref{beta_proof_a}
= & {\mathbb{E}}_{\mathcal{S}_{r}}
\Bigg[ {\sum}_{v=1}^K
{\sum}_{\mathcal{B}_r \bigcup {\bar{\mathcal{B}}}_r = {\mathcal{S}}_r \atop |\mathcal{B}_r|=v, |{\bar{\mathcal{B}}}_r|=K-v}
\frac{ \left(1 - q\right)^{v} (q)^{K-v}}{1- (q)^K} \cdot \frac{ {\sum}_{k_1 \in \mathcal{B}_r} \Delta \mathbf{w}^{r}_{k_1} }{v}  \Bigg] \notag\\
= & {\mathbb{E}}_{\mathcal{S}_{r}}
\Bigg[ {\sum}_{v=1}^K
\frac{ \left(1  -  q\right)^{v} (q)^{K-v}}{1- (q)^K} \cdot \frac{1}{v}
 {\sum}_{\mathcal{B}_r \bigcup {\bar{\mathcal{B}}}_r = {\mathcal{S}}_r \atop |\mathcal{B}_r|=v, |{\bar{\mathcal{B}}}_r|=K-v}
{\sum}_{k_1 \in \mathcal{B}_r}  \Delta \mathbf{w}^{r}_{k_1} \Bigg] \notag\\
= & {\mathbb{E}}_{\mathcal{S}_{r}}
\Bigg[ {\sum}_{v=1}^K
\frac{ \left(1  -  q\right)^{v} (q)^{K-v}}{1- (q)^K} \cdot \frac{1}{v}
{\sum}_{i \in \mathcal{S}_r} \mathbb{C}_{K-1}^{v-1} \Delta \mathbf{w}^{r}_{i} \Bigg] \notag\\
\overset{(a)}{=} & {\mathbb{E}}_{\mathcal{S}_{r}}
\Bigg[ {\sum}_{v=1}^K
\frac{ \mathbb{C}_{K}^{v} \left(1  -  q\right)^{v} (q)^{K-v}}{1- (q)^K}
\cdot \frac{1}{K} {\sum}_{i \in \mathcal{S}_r} \Delta \mathbf{w}^{r}_{i} \Bigg]
\overset{(b)}{=} {\mathbb{E}}_{\mathcal{S}_{r}}
\left[ \frac{1}{K} {\sum}_{i \in \mathcal{S}_r} \Delta \mathbf{w}^{r}_{i} \right]\overset{(c)}{=} {\sum}^N_{i=1} p_i \Delta \mathbf{w}^{r}_i
\text{,}
\end{align}}

\vspace{-0.3cm}\noindent
where equality (a) follows from
$\frac{1}{v}\mathbb{C}_{K-1}^{v-1} = \frac{1}{v} \cdot \frac{(K-1)!}{(v-1)!(K-v)!} = \frac{1}{K} \cdot \frac{K!}{v!(K-v)!} = \frac{1}{K}\mathbb{C}_{K}^{v}$, equality (b) is by $\sum _{v=1}^K \frac{\mathbb{C}_K^v (1-q)^{v} (q)^{K-v}}{1- (q)^K} = 1$ since $\sum _{v=0}^K \mathbb{C}_K^v (1-q)^{v} (q)^{K-v} = 1$, and equality (c) is by the fact that the clients are independently sampled with replacement following distribution $\{ p_i \}_{i=1}^N$ \cite{li2019convergence}.
After comparing (\ref{beta_proof_d}) with \eqref{same_outage probability_beta_proof_1}, we have ${\bar \beta}_i = p_i$ $\forall i$ under the uniform-TO case.

Similar to the proof in (\ref{same_outage probability_beta_proof_1}),  with the same TO probability $q$ for all clients, \eqref{alpha_proof} becomes
\begin{small}
\begin{align}\label{uniform_q_alpha_proof}
\eqref{alpha_proof_a}
= {\mathbb{E}}_{\mathcal{S}_{r}}
\Bigg[ \sum_{v=1}^K
\sum_{\mathcal{B}_r \bigcup {\bar{\mathcal{B}}}_r = {\mathcal{S}}_r \atop |\mathcal{B}_r|=v, |{\bar{\mathcal{B}}}_r|=K-v}
\!\!\!\!\!\!
\frac{ \left(1 - q\right)^{v} (q)^{K-v}}{1- (q)^K} \cdot \frac{ {\sum}_{k_1 \in \mathcal{B}_r} \Delta \mathbf{w}^{r}_{k_1} }{v^2}  \Bigg]
= \sum_{v=1}^K \frac{\frac{1}{v} \mathbb{C}_K^v \left(1-q\right)^{v} (q)^{K-v}}{1- (q)^K} \left[ \sum^N_{i=1} p_i \Delta \mathbf{w}^{r}_i \right]
\text{,}
\end{align}
\end{small}

\vspace{-0.3cm}\noindent
and  letting $\Delta \mathbf{w}^{r}_i = 1$ in \eqref{alpha_proof_a}  and \eqref{uniform_q_alpha_proof} gives rise to
{\small
\begin{align}
\frac{1}{\bar K}
\! = {\mathbb{E}}
\bigg[ \left. \frac{1}{ \sum_{{i} \in {\mathcal{S}}_{r}} \!\! \mathds{1}^{r}_{i}} \right| {\sum}_{{i} \in {\mathcal{S}}_{r}} \mathds{1}^{r}_{i} \neq 0 \bigg]
\! = {\sum}_{v=1}^K \frac{\frac{1}{v} \mathbb{C}_K^v \left(1 - q\right)^{v} (q)^{K-v}}{1- (q)^K} \text{.} \notag
\end{align}}

\vspace{-0.3cm}\noindent
Finally,  by comparing \eqref{alpha_proof_b} and \eqref{uniform_q_alpha_proof}, we have ${\bar \alpha}_i = p_i /{\bar K}$ under the uniform-TO case.
\hfill $\blacksquare$

\section{Proof of Theorem 1}

%\textcolor{blue}{
%The proof of Theorem 1 is mainly based on the frameworks in \cite{lian2017can,yu2019parallel} and considers the presence of both TO and QE.}
Our analysis considers only the ``successful'' communication rounds where at least one client in $\mathcal{S}_r$ communicates with the server successfully, and therefore the derivations are all based on the conditional events that $\sum_{{i} \in \mathcal{S}_{r}} \mathds{1}^{r}_{i} \neq 0$ $\forall r = 1, \cdots, M$.
In the following proof,  without further clarification,   we simply write $\mathbb{E}[ \cdot ]$ and $\Pr[\cdot]$ for the conditional $\mathbb{E}[ \, \cdot \,  | \sum_{{i} \in \mathcal{S}_{r}} \mathds{1}^{r}_{i} \neq 0]$ and $\Pr[ \,  \cdot \,  | \sum_{{i} \in \mathcal{S}_{r}} \mathds{1}^{r}_{i} \neq 0]$, respectively.

\subsection{Proof of convergence rate}
With Assumption 1, we have
{\small
\begin{align}\label{Proof_Thm1_1}
\mathbb{E}[F({\bar {\mathbf{w}}}_{r})]
\leq & \mathbb{E}[F({\bar {\mathbf{w}}}_{r-1})]
+ \mathbb{E}\left[\langle\nabla  F({\bar {\mathbf{w}}}_{r-1}), {\bar {\mathbf{w}}}_{r} - {\bar {\mathbf{w}}}_{r-1}\rangle\right]
+ \frac{L}{2} \mathbb{E}\left[ \| {\bar {\mathbf{w}}}_{r} - {\bar {\mathbf{w}}}_{r-1}  \|^2 \right] \text{.}
\end{align}}

\vspace{-0.3cm}
We need the following three key lemmas which are proved in subsequent subsections.

\begin{lemma}\label{Proof_Thm1_lemma_1}
Under Assumptions 1 and 3, it holds that
{\small
\begin{align}\label{Proof_Thm1_lemma_1_formulation}
& \mathbb{E}\left[\langle\nabla  F({\bar {\mathbf{w}}}_{r-1}), {\bar {\mathbf{w}}}_{r} - {\bar {\mathbf{w}}}_{r-1}\rangle\right] \notag \\
\leq &
-\frac{\gamma E}{2} \mathbb{E} \left[\| \nabla  F({\bar {\mathbf{w}}}_{r-1}) \|^2 \right]
+ \gamma E \chi^2_{\bm{\beta}\|\mathbf{p}} {\sum}_{i = 1}^N p_i D_i^2
+ \gamma L^2 {\sum}_{i = 1}^N  {\bar \beta}_i {\sum}_{\ell = 1}^{E} \mathbb{E} \left[ \| \mathbf{w}^{r,\ell-1}_{i} - \mathbf{\bar w}_{r-1}  \|^2 \right]
\text{,}
\end{align}}

\vspace{-0.3cm}\noindent
where $\chi^2_{\bm{\beta}\|\mathbf{p}} = \sum_{i = 1}^N {({\bar \beta}_i - p_i)^2}/{ p_i}$ is the chi-square divergence between $\mathbf{p} =[p_1,\cdots,p_N]$ and $\bm{\beta} =[{\bar \beta}_1,\cdots,{\bar \beta}_N]$ \cite{wang2020tackling}.
\end{lemma}

\begin{lemma}\label{Proof_Thm1_lemma_2}
With $q_{\max} = \max\{q_1,\ldots,q_N\}$ and $\bar q = \sum_{i=1}^N p_i q_i$ as the maximum and the average TO probabilities, we have
{\small
\begin{align}\label{Proof_Thm1_lemma_2_formulation}
\mathbb{E} \left[  \| {\bar {\mathbf{w}}}_{r} - {\bar {\mathbf{w}}}_{r-1}  \|^2 \right]
\leq &
4 \gamma^2 E^2 \mathbb{E} \left[ \| \nabla F ({\bar {\mathbf{w}}}_{r-1}) \|^2 \right]
+ \gamma^2 \frac{E}{\bar K} \frac{\sigma^2}{b}
+ \gamma^2 {\sum}_{i = 1}^{N} {\bar \alpha}_i J_{ir}^2
+ 4 \gamma^2  E^2  \sum_{i=1}^N  {\bar \alpha}_i D_i^2
\notag \\
&
+  4 \gamma^2  E^2  \sum_{v=2}^K \frac{(q_{\max})^{K  -  v}\mathbb{C}^v_K}{1  -  (q_{\max})^K}
\sum_{i = 1 }^N p_{i} \|q_i - {\bar q}\|^2 D_i^2
\! + 2 \gamma^2 E L^2 \sum_{i = 1}^{N} {\bar \beta}_i \sum_{\ell = 1}^{E} \mathbb{E} \left[ \| \mathbf{w}^{r,\ell-1}_{i} - {\bar {\mathbf{w}}}_{r-1} \|^2  \right]
\text{.}
\end{align}}
\end{lemma}

\begin{lemma}\label{Proof_Thm1_lemma_3}
The difference between the local model at each round $r$ and the global model at the previous last round is bounded by
{\small
\begin{align}\label{Proof_Thm1_lemma_3_formulation}
& {\sum}_{\ell = 1}^{E} \mathbb{E} \left[ \| \mathbf{w}^{r,\ell-1}_{i} - {\bar {\mathbf{w}}}_{r-1} \|^2  \right] \notag \\
& \leq \frac{ \gamma^2 E^3 \frac{\sigma^2}{b}
+ 4 \gamma^2 E^3 D_i^2 + 4 \gamma^2 E^3 \mathbb{E} \left[ \| \nabla F(\mathbf{\bar w}_{r-1}) \|^2 \right] }{ 1 - 2 \gamma^2 E^2 L^2 } \, \text{.}
\end{align}}
\end{lemma}

By substituting (\ref{Proof_Thm1_lemma_1_formulation}) into the second term in the RHS of \eqref{Proof_Thm1_1}, (\ref{Proof_Thm1_lemma_2_formulation}) into the third term, and by (\ref{Proof_Thm1_lemma_3_formulation}), we have
{\small
\begin{align*}
\mathbb{E}[F({\bar {\mathbf{w}}}_{r})]
\leq & \mathbb{E}[F({\bar {\mathbf{w}}}_{r-1})]
-  \left( \frac{\gamma E}{2}  -  2 \gamma^2 E^2L  -  \frac{4 \gamma^3 E^3 L^2  +  4 \gamma^4 E^4 L^3}{1 - 2 \gamma^2 E^2 L^2} \right) \mathbb{E}\left[ \| \nabla  F({\bar {\mathbf{w}}}_{r-1}) \|^2 \right]
\notag \\
&
+  \left( \frac{\gamma^2 EL}{2 {\bar K}}
+  \frac{\gamma^3 E^3 L^2  +  \gamma^4 E^4 L^3}{1 - 2 \gamma^2 E^2 L^2} \right)  \frac{\sigma^2}{b}
+ \frac{\gamma^2L}{2} {\sum}_{i = 1}^{N} {\bar \alpha}_i J_{ir}^2
+ 2 \gamma^2 E^2L {\sum}_{i=1}^N {\bar \alpha}_i D_i^2 \notag \\
& + \frac{4 \gamma^3 E^3 L^2 + 4 \gamma^4 E^4 L^3}{1 - 2 \gamma^2 E^2 L^2} {\sum}_{i = 1}^{N} {\bar \beta}_i D_i^2
+ \gamma E \chi^2_{\bm{\beta}\|\mathbf{p}} {\sum}_{i = 1}^N p_i D_i^2 \notag \\
& + 2 \gamma^2 E^2L {\sum}_{v=2}^K  \frac{(q_{\max})^{K-v}\mathbb{C}^v_K}{1-(q_{\max})^K}
{\sum}_{i = 1 }^N   p_{i} \|q_i - {\bar q} \|^2 D_i^2
\, \text{.}
\end{align*}}

\vspace{-0.3cm}
Next, summing above items from ${r} = 1$ to $M$ and dividing both sides by the total number of local mini-batch SGD steps $T = ME$ yields
{\small
\begin{align}\label{Proof_Thm1_lemma_4_step3}
& \left( \frac{\gamma }{2} - 2 \gamma^2  E L - \frac{4 \gamma^3 E^2 L^2 + 4 \gamma^4  E^3 L^3}{1 - 2 \gamma^2 E^2 L^2}  \right)
\! \frac{{\sum}_{r = 1}^{M}  \mathbb{E}  \left[ \| \nabla F({\bar {\mathbf{w}}}_{r - 1}) \|^2 \right]}{M} \notag \\
\leq & \frac{\mathbb{E}[F({\bar {\mathbf{w}}}_{0})]
-  \mathbb{E}[F({\bar {\mathbf{w}}}_{M})]}{T}
+  \left(  \frac{\gamma^2 L}{2{\bar K}}
+  \frac{\gamma^3 E^2 L^2  +  \gamma^4 E^3  L^3}{1 - 2 \gamma^2 E^2 L^2}  \right)  \frac{\sigma^2}{b}
+ \frac{\gamma^2L}{2T} {\sum}_{r = 1}^{M} {\sum}_{i = 1}^{N} {\bar \alpha}_i J_{ir}^2
+ 2 \gamma^2 E L {\sum}_{i=1}^N {\bar \alpha}_i D_i^2 \notag \\
& + \frac{4 \gamma^3 E^2 L^2 + 4 \gamma^4 E^3 L^3}{1 - 2 \gamma^2 E^2 L^2} {\sum}_{i = 1}^{N} {\bar \beta}_i D_i^2
\! + \gamma \chi^2_{\bm{\beta}\|\mathbf{p}} {\sum}_{i = 1}^N p_i D_i^2
\! + 2 \gamma^2 E L {\sum}_{v=2}^K  \frac{(q_{\max})^{K-v}\mathbb{C}^v_K}{1-(q_{\max})^K}
{\sum}_{i = 1 }^N   p_{i}  \left \|q_i - {\bar q}\right\|^2 D_i^2
\, \text{.}
\end{align}}

\vspace{-0.5cm}\noindent
Further, dividing both sides in (\ref{Proof_Thm1_lemma_4_step3}) by $\gamma$ leads to
{\small
\begin{align}\label{Proof_Thm1_lemma_4_formulation}
& \underbrace{ \bigg( \frac{1 }{2} -  2 \gamma E L  - \frac{4 \gamma^2 E^2 L^2  +  4 \gamma^3 E^3 L^3}{1 - 2 \gamma^2 E^2 L^2} \bigg)}_{\triangleq H_1}
 \frac{{\sum}_{r = 1}^{M}  \mathbb{E}  \left[\| \nabla  F({\bar {\mathbf{w}}}_{r  -  1}) \|^2 \right]}{M} \notag \\
\leq & \underbrace{ \frac{1}{\gamma T} }_{\triangleq H_2} \left( \mathbb{E}[F({\bar {\mathbf{w}}}_{0})] - \mathbb{E}[F({\bar {\mathbf{w}}}_{M})] \right)
+  \underbrace{ \left( \frac{\gamma L}{2{\bar K}}
+  \frac{\gamma^2 E^2 L^2  +  \gamma^3 E^3 L^3}{1 - 2 \gamma^2 E^2 L^2} \right) }_{\triangleq H_3} \frac{\sigma^2}{b}
+  \underbrace{ \frac{\gamma L}{2T} }_{\triangleq H_4} \sum_{r = 1}^{M} \sum_{i = 1}^{N} {\bar \alpha}_i J_{ir}^2
+ \underbrace{ 2 \gamma E L  }_{\triangleq H_6} \sum_{i=1}^N {\bar \alpha}_i D_i^2\notag
\notag \\
&
+ \underbrace{ \frac{4 \gamma^2 E^2 L^2 + 4 \gamma^3 E^3 L^3}{1 - 2 \gamma^2 E^2 L^2} }_{\triangleq H_5} {\sum}_{i = 1}^{N} {\bar \beta}_i D_i^2
+ \chi^2_{\bm{\beta}\|\mathbf{p}} \sum\limits_{i = 1}^N p_i D_i^2
+ \underbrace{ 2 \gamma E L }_{\triangleq H_6} \sum_{v=2}^K  \frac{(q_{\max})^{K-v}\mathbb{C}^v_K}{1-(q_{\max})^K}
\! \sum_{i = 1 }^N   p_{i} \|q_i - {\bar q} \|^2 D_i^2
\, \text{.}
\end{align}}

\vspace{-0.3cm}
Let the learning rate $\gamma = {\bar K}^{\frac{1}{2}} / (8L{T}^{\frac{1}{2}}) $ and the number of local updating steps $E \leq T^{\frac{1}{4}}/{\bar K}^{\frac{3}{4}}$, where $T \geq \max \{ {\bar K}^{3}, 1/{\bar K} \}$ in order to guarantee $E \geq 1$.
By this,
$ H_2 = {8 L }{ ( T {\bar K})^{-\frac{1}{2}} }$ and $H_4 = {{\bar K}^{\frac{1}{2}}}{ T^{-\frac{3}{2}}}/16$.
Since $\gamma E L \leq ( T {\bar K})^{-\frac{1}{4}}/8 $, we have $ H_6 \leq ( T {\bar K})^{-\frac{1}{4}}/4 $  and
{\small
\begin{align*}
H_5
\leq & \frac{ \frac{4}{ 8^2 }( T {\bar K})^{-\frac{1}{2}}
+ \frac{4}{ 8^3}  ( T {\bar K})^{-\frac{3}{4}} }{1 - \frac{2}{ 8^2  } ( T {\bar K})^{-\frac{1}{2}} }
\overset{(a)}{\leq} \frac{ \frac{4}{ 8^2 }( T {\bar K})^{-\frac{1}{2}}
+ \frac{4}{ 8^3}  ( T {\bar K})^{-\frac{3}{4}} }{1 - \frac{2}{ 8^2  } }
= \frac{2 }{ 31 ( T {\bar K})^{\frac{1}{2}} }
+ \frac{1}{ 124 ( T {\bar K})^{\frac{3}{4}} }
\, \text{,}
\end{align*}}

\vspace{-0.3cm}\noindent
where inequality (a) is due to $T \geq 1/{\bar K}$. Then,
{\small
\begin{align*}
H_1
= & \frac{1 }{2}  -  H_6  -  H_5
\geq \frac{1}{2}
 -  \frac{1}{ 4 ( T {\bar K})^{\frac{1}{4}} }
 -  \frac{2}{ 31 ( T {\bar K})^{\frac{1}{2}} }
 -  \frac{1}{ 124 ( T {\bar K})^{\frac{3}{4}} }
\geq \frac{1}{2} - \frac{1}{4} - \frac{2}{31} - \frac{1}{124}
= \frac{11}{62}
\, \text{,}\\
H_3
= & \frac{\gamma L}{2 \bar K} + \frac{H_5}{4}
\leq \frac{L}{ 16 L ( T {\bar K})^{\frac{1}{2}} }
+ \frac{1}{ 62 ( T {\bar K})^{\frac{1}{2}} }
+ \frac{1}{ 496 ( T {\bar K})^{\frac{3}{4}} }
\leq \frac{ 39 }{ 496 ( T {\bar K})^{\frac{1}{2}} }
+ \frac{1}{ 496 ( T {\bar K})^{\frac{3}{4}} }
\, \text{.}
\end{align*}}

\vspace{-0.3cm}
Finally, by substituting above coefficients and $\mathbb{E}[F({\bar {\mathbf{w}}}_{M})] \geq \underline{F}$ in Assumption 1 into \eqref{Proof_Thm1_lemma_4_formulation}, Theorem 1 is proved.
\hfill $\blacksquare$

\subsection{Proof of Lemma \ref{Proof_Thm1_lemma_1}}
We have
{\small
\begin{align}\label{Proof_Thm1_lemma_1_step1}
& \mathbb{E}\left[\langle\nabla  F({\bar {\mathbf{w}}}_{r-1}), {\bar {\mathbf{w}}}_{r} - {\bar {\mathbf{w}}}_{r-1}\rangle\right] \notag \\
= & \mathbb{E}\left[\left\langle \nabla  F({\bar {\mathbf{w}}}_{r-1}), - \gamma \frac{ {\sum}_{{i} \in \mathcal{S}_{r}} {\mathds{1}}^{r}_{i} \mathcal{Q}(\Delta \mathbf{w}^{r}_{i}) }{ {\sum}_{{i} \in \mathcal{S}_{r}} {\mathds{1}}^{r}_{i}} \right\rangle\right] \notag \\
\overset{(a)}{=} & \mathbb{E}\left[\left\langle\nabla  F({\bar {\mathbf{w}}}_{r-1}),- \gamma \frac{{\sum}_{{i} \in \mathcal{S}_{r}} {\mathds{1}}^{r}_{i} {\sum}_{\ell = 1}^{E} \nabla F_{i} (\mathbf{w}^{r,\ell-1}_{i}, {\bm \xi}^{r,\ell}_{i}) }{{\sum}_{{i} \in \mathcal{S}_{r}} {\mathds{1}}^{r}_{i}} \right\rangle\right] \notag \\
\overset{(b)}{=} & \mathbb{E}\left[\left\langle\nabla  F({\bar {\mathbf{w}}}_{r-1}),- \gamma \frac{{\sum}_{{i} \in \mathcal{S}_{r}} {\mathds{1}}^{r}_{i} {\sum}_{\ell = 1}^{E} \nabla F_{i} (\mathbf{w}^{r,\ell-1}_{i}) }{{\sum}_{{i} \in \mathcal{S}_{r}} {\mathds{1}}^{r}_{i}} \right\rangle\right] \notag \\
\overset{(c)}{=} & - \gamma {\sum}_{\ell = 1}^{E} \mathbb{E}\left[\left\langle\nabla  F({\bar {\mathbf{w}}}_{r-1}), {\sum}_{i = 1}^N {\bar \beta}_i \nabla F_{i} (\mathbf{w}^{r,\ell-1}_{i}) \right \rangle\right] \notag \\
\overset{(d)}{=} & -  \frac{\gamma}{2}  {\sum}_{\ell = 1}^{E}   \mathbb{E}  \left[\| \nabla F({\bar {\mathbf{w}}}_{r-1}) \|^2 \right]
 -  \frac{\gamma}{2}  {\sum}_{\ell = 1}^{E}  \mathbb{E}  \Big[ \Big\|  {\sum}_{i = 1}^N {\bar \beta}_i \nabla  F_{i} (\mathbf{w}^{r,\ell-1}_{i}) \Big\|^2 \Big]  \notag \\
& + \frac{\gamma}{2} {\sum}_{\ell = 1}^{E} \mathbb{E} \Big[ \Big\| \nabla  F({\bar {\mathbf{w}}}_{r-1}) -  {\sum}_{i = 1}^N {\bar \beta}_i \nabla F_{i} (\mathbf{w}^{r,\ell-1}_{i}) \Big\|^2 \Big] \notag \\
\leq & -\frac{\gamma E}{2} \mathbb{E}\left[ \| \nabla  F({\bar {\mathbf{w}}}_{r-1}) \|^2 \right]
+ \frac{\gamma}{2} {\sum}_{\ell = 1}^{E} \mathbb{E}\left[\Big\| \nabla  F({\bar {\mathbf{w}}}_{r-1}) -  {\sum}_{i = 1}^N {\bar \beta}_i \nabla F_{i} (\mathbf{w}^{r,\ell-1}_{i}) \right\|^2 \Big] \notag \\
\overset{(e)}{\leq} & -\frac{\gamma E}{2} \mathbb{E}\left[ \| \nabla  F({\bar {\mathbf{w}}}_{r-1}) \|^2 \right]
+ \gamma {\sum}_{\ell = 1}^{E} \underbrace{\mathbb{E}\Big[\left\| \nabla  F({\bar {\mathbf{w}}}_{r-1}) -  {\sum}_{i = 1}^N {\bar \beta}_i \nabla F_i({\bar {\mathbf{w}}}_{r-1}) \right\|^2 \Big]}_{\triangleq A_1} \notag \\
& + \gamma {\sum}_{\ell = 1}^{E} \underbrace{\mathbb{E}\Big[\left\| {\sum}_{i = 1}^N {\bar \beta}_i (\nabla F_i({\bar {\mathbf{w}}}_{r-1}) -  \nabla F_i(\mathbf{w}^{r,\ell-1}_{i}) ) \right\|^2 \Big]}_{\triangleq A_2}
\text{,}
\end{align}}

\vspace{-0.3cm}
\noindent
where equality (a) is due to the unbiased quantization in (\ref{unbiased_estimation}) and the definition of $\Delta \mathbf{w}^{r}_{i}$ in (\ref{local_model_update}), equality (b) is due to $ \mathbb{E}[ \nabla F_{i}(\mathbf{w}^{r,\ell-1}_{i}, {\bm \xi}^{r,\ell}_{i}) ] = \nabla F_{i}(\mathbf{w}^{r,\ell-1}_{i}  ) $ in Assumption 2,
equality (c) is obtained by (\ref{average_beta_1}),
equality (d) follows from the basic identity $\langle {\mathbf{x}}_1,{\mathbf{x}}_2 \rangle = \frac{1}{2}( \| {\mathbf{x}}_1 \|^2 + \| {\mathbf{x}}_2 \|^2 - \| {\mathbf{x}}_1 - {\mathbf{x}}_2 \|^2 )$,
and inequality (e) is due to $\|x_1+x_2\|^2 \leq 2\|x_1\|^2 + 2\|x_2\|^2$.

In \eqref{Proof_Thm1_lemma_1_step1}, the term $A_1$ can be further bounded as
{\small
\begin{align}\label{Proof_Thm1_lemma_1_step2}
A_1
= & \mathbb{E}\left[\left\| {\sum}_{i = 1}^N p_i \nabla  F_i({\bar {\mathbf{w}}}_{r-1}) -  {\sum}_{i = 1}^N {\bar \beta}_i \nabla F_i({\bar {\mathbf{w}}}_{r-1}) \right\|^2 \right] \notag \\
\overset{(a)}{=} & \mathbb{E}\Bigg[\left\| \sum_{i = 1}^N ( p_i - {\bar \beta}_i ) \nabla F_i(  {\bar {\mathbf{w}}}_{r  -  1}) - \sum_{i = 1}^N ( p_i - {\bar \beta}_i ) \nabla F(  {\bar {\mathbf{w}}}_{r  -  1}) \right\|^2 \Bigg] \notag \\
= & \mathbb{E}\left[\left\| {\sum}_{i = 1}^N \frac{p_i - {\bar \beta}_i}{\sqrt{p_i}}\sqrt{p_i} \left( \nabla F_i({\bar {\mathbf{w}}}_{r-1}) -  \nabla  F ({\bar {\mathbf{w}}}_{r-1}) \right) \right\|^2 \right] \notag \\
\overset{(b)}{\leq} & \left(  \sum_{i = 1}^N  \frac{({\bar \beta}_i -  p_i)^2}{ p_i}  \right)  \sum_{i = 1}^N p_i \mathbb{E} \left[ \| \nabla F_i({\bar {\mathbf{w}}}_{r - 1})  -  \nabla  F({\bar {\mathbf{w}}}_{r - 1}) \|^2 \right] \notag \\
\overset{(c)}{\leq} & \chi^2_{\bm{\beta}\|\mathbf{p}} {\sum}_{i = 1}^N p_i D_i^2
\, \text{,}
\end{align}}

\vspace{-0.3cm}
\noindent
where equality (a) is because $ \sum_{i = 1}^N  (p_i - {\bar \beta}_i) = 0 $,
inequality (b) is due to the Cauchy-Schwarz Inequality,
and inequality (c) is due to Assumption 3 and the definition of $\chi^2_{\bm{\beta}\|\mathbf{p}}$ in Lemma \ref{Proof_Thm1_lemma_1}.
Besides, $A_2$ is bounded as
{\small
\begin{align}\label{Proof_Thm1_lemma_1_step3}
A_2
\overset{(a)}{\leq} {\sum}_{i = 1}^N  {\bar \beta}_i \mathbb{E}\left[ \| \nabla F_i({\bar {\mathbf{w}}}_{r-1}) -  \nabla F_i(\mathbf{w}^{r,\ell-1}_{i}) \|^2 \right]
\overset{(b)}{\leq} L^2 {\sum}_{i = 1}^N  {\bar \beta}_i \mathbb{E}\left[ \| \mathbf{w}^{r,\ell-1}_{i} - \mathbf{\bar w}_{r-1} \|^2 \right]
\text{,}
\end{align}}

\vspace{-0.3cm}
\noindent
where inequality (a) is by the Jensen's Inequality and inequality (b) is due to Assumption 1.

Finally, by substituting (\ref{Proof_Thm1_lemma_1_step2}) and (\ref{Proof_Thm1_lemma_1_step3}) into (\ref{Proof_Thm1_lemma_1_step1}), we can obtain Lemma \ref{Proof_Thm1_lemma_1} directly.
\hfill $\blacksquare$

\subsection{Proof of Lemma \ref{Proof_Thm1_lemma_2}}
We have
{\small
\begin{align}\label{Proof_Thm1_lemma_2_step1}
& \mathbb{E}[ \| {\bar {\mathbf{w}}}_{r} - {\bar {\mathbf{w}}}_{r-1}  \|^2 ] \notag \\
= & \mathbb{E}\left[ \left\| - \gamma \frac{ {\sum}_{{i} \in \mathcal{S}_{r}} {\mathds{1}}^{r}_{i} \mathcal{Q}(\Delta \mathbf{w}^{r}_{i}) }{{\sum}_{{i} \in \mathcal{S}_{r}} {\mathds{1}}^{r}_{i}} \right\|^2 \right] \notag \\
\overset{(a)}{=} & \gamma^2 \mathbb{E}\left[ \left\| \frac{ {\sum}_{{i} \in \mathcal{S}_{r}}  {\mathds{1}}^{r}_{i} \Delta \mathbf{w}^{r}_{i} }{ \sum _{{i} \in \mathcal{S}_{r}} {\mathds{1}}^{r}_{i}} \right\|^2
 +  \left\| \frac{ {\sum}_{{i} \in \mathcal{S}_{r}}  {\mathds{1}}^{r}_{i} ( \mathcal{Q}(\Delta \mathbf{w}^{r}_{i}) -  \Delta \mathbf{w}^{r}_{i} )  }{ {\sum}_{{i} \in \mathcal{S}_{r}} {\mathds{1}}^{r}_{i}} \right\|^2 \right] \notag \\
\overset{(b)}{=} & \gamma^2  \underbrace{ \mathbb{E}  \left[  \left\|  \frac{ {\sum}_{{i} \in \mathcal{S}_{r}}   {\mathds{1}}^{r}_{i}  {\sum}_{\ell = 1}^{E}  ( \nabla  F_{i} (\mathbf{w}^{r,\ell-1}_{i},  {\bm \xi}^{r,\ell}_{i}) -  \nabla  F_{i} (\mathbf{w}^{r,\ell-1}_{i} ) )}{ {\sum}_{{i} \in \mathcal{S}_{r}} {\mathds{1}}^{r}_{i} }  \right\|^2 \right] }_{\triangleq G_1 {\rm \ (caused \ by \ SGD)} }
+ \gamma^2 \underbrace{ \mathbb{E}\left[ \left\| \frac{ {\sum}_{{i} \in \mathcal{S}_{r}} {\mathds{1}}^{r}_{i} {\sum}_{\ell = 1}^{E} \nabla F_{i} (\mathbf{w}^{r,\ell-1}_{i}) }{{\sum}_{{i} \in \mathcal{S}_{r}} {\mathds{1}}^{r}_{i}} \right\|^2 \right] }_{\triangleq G_2} \notag \\
& + \gamma^2 \underbrace{ \mathbb{E}\left[ \left\| \frac{ {\sum}_{{i} \in \mathcal{S}_{r}} {\mathds{1}}^{r}_{i} ( \mathcal{Q}(\Delta \mathbf{w}^{r}_{i}) - \Delta \mathbf{w}^{r}_{i} ) }{ {\sum}_{{i} \in \mathcal{S}_{r}} {\mathds{1}}^{r}_{i}} \right\|^2 \right] }_{\triangleq G_3 {\rm \; (caused \ by \ quantization \ error)} }
\text{,}
\end{align}}

\vspace{-0.3cm}
\noindent
where equality (a) is by $\mathbb{E}[ \| \mathbf{x} \|^2] = \mathbb{E}[ \| \mathbf{x} - \mathbb{E}[\mathbf{x}]\|^2] + \| \mathbb{E}[\mathbf{x}]\|^2$ and \eqref{unbiased_estimation};
equality (b) is obtained similarly but using $\Delta \mathbf{w}^{r}_{i} = \sum_{\ell = 1}^{E}  \nabla  F_{i} (\mathbf{w}^{r,\ell-1}_{i} , {\bm \xi}^{r,\ell}_{i})$ in (\ref{local_model_update}) and $\mathbb{E}[\nabla  F_{i} (\mathbf{w}^{r,\ell-1}_{i},  {\bm \xi}^{r,\ell}_{i})] = \nabla  F_{i} (\mathbf{w}^{r,\ell-1}_{i} )$ in Assumption 2.

In \eqref{Proof_Thm1_lemma_2_step1}, the term $G_1$ can be shown as
{\small
\begin{align}\label{G_1}
G_1
\overset{(a)}{=}  & \mathbb{E}  \left[  \frac{{\sum}_{{i} \in \mathcal{S}_{r}}  \mathds{1}^{r}_{i}  {\sum}_{\ell = 1}^{E}  \| \nabla  F_{i} (\mathbf{w}_{i}^{r,\ell-1} , {\bm \xi}^{r,\ell}_{i})  -  \nabla  F_{i} (\mathbf{w}_{i}^{r,\ell-1}) \|^2 }{\left( \sum_{{i} \in \mathcal{S}_{r}} \mathds{1}^{r}_{i} \right)^2 }  \right] \notag \\
\overset{(b)}{=} & \mathbb{E}  \left[  \frac{ {\sum}_{{i} \in \mathcal{S}_{r}}  \mathds{1}^{r}_{i} {\sum}_{\ell = 1}^{E}  \frac{\sigma^2}{b} }{\left( {\sum}_{{i} \in \mathcal{S}_{r}} \mathds{1}^{r}_{i} \right)^2 }  \right]
 =  \frac{E \sigma^2}{b} \mathbb{E}  \left[  \frac{1}{ {\sum}_{{i} \in \mathcal{S}_{r}}  \mathds{1}^{r}_{i} }  \right]
\overset{(c)}{=} \frac{E \sigma^2}{{\bar K}b }
\, \text{,}
\end{align}}

\vspace{-0.1cm}
\noindent
where equality (a) is due to ${\mathbb{E}} [\nabla F_{i} (\mathbf{w}^{r,\ell-1}_{i}, {\bm \xi}^{r,\ell}_{i})] = \nabla F_{i} (\mathbf{w}^{r,\ell-1}_{i})$ in Assumption 2, equality (b) is due to the bounded variance of SGD in Assumption 2, and equality (c) is due to (\ref{bar_K_1}).
For $G_2$ in \eqref{Proof_Thm1_lemma_2_step1}, we have
{\small
\begin{align}\label{G_2}
G_2
\leq & 2 \underbrace{ \mathbb{E} \left[ \left \|
\frac{ {\sum}_{{i} \in \mathcal{S}_{r}} {\mathds{1}}^{r}_{i}  {\sum}_{\ell = 1}^{E} ( \nabla F_{i} (\mathbf{w}^{r,\ell-1}_{i}) - \nabla F_{i} ({\bar {\mathbf{w}}}_{r-1}) ) }{{\sum}_{{i} \in \mathcal{S}_{r}} {\mathds{1}}^{r}_{i}}
\right\|^2 \right]  }_{\triangleq G_{21}}
+ 2 \underbrace{ \mathbb{E} \left[ \left \| \frac{ {\sum}_{{i} \in \mathcal{S}_{r}} {\mathds{1}}^{r}_{i} {\sum}_{\ell = 1}^{E} \nabla F_{i} ({\bar {\mathbf{w}}}_{r-1}) }{\sum_{{i} \in \mathcal{S}_{r}} {\mathds{1}}^{r}_{i}} \right\|^2 \right] }_{\triangleq G_{22}}
\text{,}
\end{align}}

\vspace{-0.3cm}
\noindent
where
{\small
\begin{align}\label{G_21}
G_{21}
\leq & E \cdot \mathbb{E}  \left[
\frac{ {\sum}_{{i} \in \mathcal{S}_{r}} {\mathds{1}}^{r}_{i} {\sum}_{\ell = 1}^{E}  \| \nabla F_{i} (\mathbf{w}^{r,\ell-1}_{i})  - \nabla F_{i} ({\bar {\mathbf{w}}}_{r-1}) \|^2 }{{\sum}_{{i} \in \mathcal{S}_{r}} {\mathds{1}}^{r}_{i}}  \right]\notag \\
\overset{(a)}{=} & E {\sum}_{i = 1}^{N} {\bar \beta}_i {\sum}_{\ell = 1}^{E} \mathbb{E} \left[ \| \nabla F_{i} (\mathbf{w}^{r,\ell-1}_{i}) - \nabla F_{i} ({\bar {\mathbf{w}}}_{r-1}) \|^2  \right]
\overset{(b)}{\leq} E L^2 {\sum}_{i = 1}^{N} {\bar \beta}_i {\sum}_{\ell = 1}^{E} \mathbb{E} \left[ \| \mathbf{w}^{r,\ell-1}_{i} - {\bar {\mathbf{w}}}_{r-1} \|^2  \right]
\text{,}
\end{align}}

\vspace{-0.3cm}
\noindent
in which equality (a) is due to \eqref{average_beta_1} in Lemma \ref{beta_alpha_K}, and inequality (b) is due to Assumption 1.

\vspace{-0.4cm}
{\small
\begin{align}\label{G_22}
G_{22}
= & E^2 \mathbb{E} \left[ \left \| \frac{ {\sum}_{{i} \in \mathcal{S}_{r}} {\mathds{1}}^{r}_{i} \nabla F_{i} ({\bar {\mathbf{w}}}_{r-1}) }{\sum_{{i} \in \mathcal{S}_{r}} {\mathds{1}}^{r}_{i}} \right\|^2 \right] \notag \\
= & 2 E^2 \underbrace{ \mathbb{E} \left[ \left \| \frac{ {\sum}_{{i} \in \mathcal{S}_{r}} {\mathds{1}}^{r}_{i} \left( \nabla F_{i} ({\bar {\mathbf{w}}}_{r-1}) -  \nabla F ({\bar {\mathbf{w}}}_{r-1}) \right) }{ {\sum}_{{i} \in \mathcal{S}_{r}} {\mathds{1}}^{r}_{i}} \right\|^2 \right]}_{ {\rm \; (caused \ by \ partial \ participation)} }
+ 2 E^2 \mathbb{E} \left[ \left \| \nabla F ({\bar {\mathbf{w}}}_{r-1}) \right\|^2 \right]\notag \\
= & 2 E^2 \underbrace{ \mathbb{E} \left[ \frac{{\sum}_{{i} \in \mathcal{S}_{r}} \mathds{1}^{r}_{i} \left\|  \nabla F_{i} ({\bar {\mathbf{w}}}_{r-1}) - \nabla F ({\bar {\mathbf{w}}}_{r-1}) \right\|^2}{ \left( \sum_{{i} \in \mathcal{S}_{r}} \mathds{1}^{r}_{i}\right)^2 }
 \right]}_{\triangleq G_{23}} \notag \\
& +  2 E^2  \underbrace{\mathbb{E}  \left[  \frac{
{\sum}_{k' \in \mathcal{S}_{r}} {\sum}_{k \in \mathcal{S}_{r} \atop k \neq k'}  \mathds{1}^{r}_{k} {\mathds{1}}^{r}_{k'} \Big( ( \nabla F_{k} ({\bar {\mathbf{w}}}_{r-1}) -
\nabla  F ({\bar {\mathbf{w}}}_{r-1}) ) ( \nabla  F_{k'} ({\bar {\mathbf{w}}}_{r-1})  -  \nabla  F ({\bar {\mathbf{w}}}_{r-1}) ) \Big)
}{\left( \sum_{{i} \in \mathcal{S}_{r}} \mathds{1}^{r}_{i}\right)^2}
 \right]}_{\triangleq G_{24}} \notag \\
& + 2 E^2 \mathbb{E} \left[ \| \nabla F ({\bar {\mathbf{w}}}_{r-1}) \|^2 \right]
\text{.}
\end{align}}

\vspace{-0.3cm}
Next, we bound $G_{23}$ and $G_{24}$ in \eqref{G_22} as follows.
Firstly
{\small
\begin{align}\label{G_23}
G_{23}
\overset{(a)}{=} {\sum}_{i=1}^N {\bar \alpha}_i \mathbb{E}  \left[ \|  \nabla F_{i} ({\bar {\mathbf{w}}}_{r-1})  -  \nabla F ({\bar {\mathbf{w}}}_{r-1}) \|^2 \right]
\overset{(b)}{\leq}  {\sum}_{i=1}^N   {\bar \alpha}_i D_i^2
\text{,}
\end{align}}

\vspace{-0.5cm}
\noindent
where equality (a) is due to (\ref{average_alpha_1}) in Lemma \ref{beta_alpha_K}, and inequality (b) is due to Assumption 3.
Secondly,
{\small
\begin{align*}
G_{24}
= & \mathbb{E} \bigg [ {\sum}_{v=1}^K \Pr  \Big( {\sum}_{{i} \in \mathcal{S}_{r}}{\mathds{1}}^{r}_{i}  =  v \Big) \cdot \frac{1}{v^2}
{\sum}_{k \in \mathcal{S}_{r}}   {\sum}_{k' \in \mathcal{S}_{r} \atop k' \neq k }
 \mathbb{E} \Big[ \mathds{1}^{r}_{k} {\mathds{1}}^{r}_{{k'}} ( \nabla  F_{k} ({\bar {\mathbf{w}}}_{{r} - 1} ) \notag \\
& \quad -  \nabla  F ({\bar {\mathbf{w}}}_{{r} - 1})  )  ( \nabla  F_{{k'}} ({\bar {\mathbf{w}}}_{{r} - 1})  -  \nabla  F ({\bar {\mathbf{w}}}_{{r} - 1}) ) \Big| {\sum}_{{i} \in \mathcal{S}_{r}}  {\mathds{1}}^{r}_{i}  =  v \Big]   \bigg]  \notag \\
\overset{(a)}{=} & \mathbb{E}  \bigg[  {\sum}_{v=1}^K  \frac{1}{v^2}
 {\sum}_{k \in \mathcal{S}_{r}}  {\sum}_{k' \in \mathcal{S}_{r} \atop k' \neq k }
\bigg( \Pr  \Big({\mathds{1}}^{r}_{k} = 1, {\mathds{1}}^{r}_{{k'}} = 1, {\sum}_{{i} \in \mathcal{S}_{r}}  {\mathds{1}}^{r}_{i}  =  v \Big) \notag \\
& \quad \cdot  ( \nabla  F_{k} ({\bar {\mathbf{w}}}_{{r} - 1} )  -  \nabla  F ({\bar {\mathbf{w}}}_{{r} - 1})  )  ( \nabla  F_{{k'}} ({\bar {\mathbf{w}}}_{{r} - 1})  -  \nabla F ({\bar {\mathbf{w}}}_{{r} - 1}) ) \bigg) \bigg]
\text{,}
\end{align*}}

\vspace{-0.3cm}
\noindent
where equality (a) follows because if $\mathds{1}^{r}_{k} =0$ or ${\mathds{1}}^{r}_{{k'}}=0$, then ${\mathds{1}}^{r}_{k} {\mathds{1}}^{r}_{{k'}} ( \nabla F_{k} ({\bar {\mathbf{w}}}_{r-1}) - \nabla F ({\bar {\mathbf{w}}}_{r-1}) ) ( \nabla F_{{k'}} ({\bar {\mathbf{w}}}_{r-1}) - \nabla F ({\bar {\mathbf{w}}}_{r-1}) )$ $= 0$.
In addition, when $v=1$, there is only one selected client with successful transmission, and $\mathds{1}^{r}_{k}$ and ${\mathds{1}}^{r}_{{k'}}$ cannot equal to 1 at the same time, thus $\Pr ({\mathds{1}}^{r}_{k}=1, {\mathds{1}}^{r}_{{k'}}=1, \sum_{{i} \in \mathcal{S}_{r}} {\mathds{1}}^{r}_{i}  = 1 ) = 0$.
When $v \geq 2$,
{\small
\begin{align}\label{Pr_1_fixed}
& \Pr\left({\mathds{1}}^{r}_{k}=1, {\mathds{1}}^{r}_{{k'}}=1, {\sum}_{{i} \in \mathcal{S}_{r}}{\mathds{1}}^{r}_{i}=v \right) \notag \\
= & \frac{(1  -  q_{k})(1  -  q_{{k'}})  {\sum}_{\mathcal{B}_r \bigcup  {\bar{\mathcal{B}}}_r = \{ \mathcal{S}_{r} \setminus \{{k},{k'}\} \} \atop |\mathcal{B}_r|=v-2, |{\bar{\mathcal{B}}}_r|=K -v}
\bigg(  \prod \limits_{k_1 \in \mathcal{B}_r}  (1  -  q_{{k_1}})  \prod \limits_{k_2 \in {\bar{\mathcal{B}}}_r}  q_{{k_2}}  \bigg) }{1- \prod_{{i} \in \mathcal{S}_{r}}  q_{k} } \notag \\
\overset{(a)}{\leq} & \frac{(1-q_{k})(1-q_{{k'}}) {\sum}_{\mathcal{B}_r \bigcup {\bar{\mathcal{B}}}_r = \{ \mathcal{S}_{r} \setminus \{{k},{k'}\} \} \atop |\mathcal{B}_r|=v-2, |{\bar{\mathcal{B}}}_r|=K-v} (q_{\max})^{K-v} }{1-(q_{\max})^K}
\overset{(b)}{=} \frac{(1-q_{k})(1-q_{{k'}}) (q_{\max})^{K-v} \mathbb{C}_{K-2}^{v-2} }{1-(q_{\max})^K}
\, \text{,}
\end{align}}

\vspace{-0.3cm}
\noindent
where $\mathcal{B}_r$ is the set of selected clients (except $k$ and $k'$) in $\mathcal{S}_r$ transmitting their local model updates successfully while ${\bar{\mathcal{B}}}_r$ is the one that suffers from TO;
inequality (a) is due to $1-q_{k_1} \leq 1$ and $q_{k_2} \leq q_{\max} = \max\{q_1,\ldots,q_N\}$,
and in equality (b), $\mathbb{C}_{K-2}^{v-2} = \frac{(K-2)!}{(v-2)!(K-v)!}$.
Thus,
{\small
\begin{align}\label{G_24_step1}
G_{24}
\leq & \mathbb{E} \bigg[ \sum_{v=2}^K  \frac{(q_{\max})^{K-v}\mathbb{C}_{K-2}^{v-2}}{(1  -  (q_{\max})^K)v^2}
\sum_{k \in \mathcal{S}_{r}} \sum_{k' \in \mathcal{S}_{r} \atop k' \neq k }
(1 - q_{k})(1 - q_{{k'}})
( \nabla F_{k} ({\bar {\mathbf{w}}}_{ {r}  -  1} ) - \nabla F ({\bar {\mathbf{w}}}_{ {r}  -  1})  )  ( \nabla F_{{k'}} ({\bar {\mathbf{w}}}_{ {r}  -  1}) - \nabla F ({\bar {\mathbf{w}}}_{ {r}  -  1}) ) \bigg] \notag \\
= &
\mathbb{E} \bigg[ \sum_{v=2}^K \frac{(q_{\max})^{K-v}\mathbb{C}_{K-2}^{v-2}}{(1  - (q_{\max})^K)v^2} \!\! \sum_{k \in \mathcal{S}_{r}} (1 - q_{{k}})   ( \nabla  F_{{k}} ({\bar {\mathbf{w}}}_{ {r}  -  1}) - \nabla F ({\bar {\mathbf{w}}}_{ {r}  -  1}) ) \sum_{k' \in \mathcal{S}_{r} \atop k' \neq k} (1  -  q_{k'}) ( \nabla F_{k'} ({\bar {\mathbf{w}}}_{ {r}  -  1} )  -  \nabla  F ({\bar {\mathbf{w}}}_{ {r}  -  1})  )  \bigg] \notag \\
\overset{(a)}{=} &
\mathbb{E} \bigg[  \sum_{v=2}^K  \frac{(q_{\max})^{K-v} K (K  -  1) \mathbb{C}_{K-2}^{v-2}}{(1-(q_{\max})^K)v^2}  \sum_{j = 1 }^N  p_{j} (1  -  q_{j}) ( \nabla F_{j} ({\bar {\mathbf{w}}}_{{r} - 1} ) - \nabla F ({\bar {\mathbf{w}}}_{r  -  1})  ) \notag \\
&
\quad\;
\cdot \sum_{j' = 1 }^N  p_{j'}  (1  -  q_{j'})  ( \nabla F_{{j'}} ({\bar {\mathbf{w}}}_{r  -  1})  -  \nabla F ({\bar {\mathbf{w}}}_{r  -  1}) ) \bigg] \notag \\
\overset{(b)}{\leq} & \mathbb{E} \bigg[ \sum_{v=2}^K  \frac{(q_{\max})^{K-v}\mathbb{C}^v_K}{1-(q_{\max})^K} \sum_{j = 1 }^N  \sum_{j' = 1 }^N p_{j}p_{j'} (1 -\! q_{j})(1 - q_{j'}) ( \nabla F_{j} ({\bar {\mathbf{w}}}_{{r} - 1} )  - \nabla F ({\bar {\mathbf{w}}}_{r  -  1})  )  ( \nabla F_{{j'}} ({\bar {\mathbf{w}}}_{r  -  1})  -  \nabla F ({\bar {\mathbf{w}}}_{r  -  1}) ) \bigg]
\text{,}
\end{align}}

\vspace{-0.3cm}
\noindent
where
equality (a) can be obtained based on the same reason as obtaining (c) in \eqref{same_outage probability_beta_proof_1} since the clients $k,k' \in \mathcal{S}_{r}$ are selected independently and with replacement.
The above inequality (b) is obtained by $\frac{K(K-1)\mathbb{C}_{K-2}^{v-2}}{v^2} \leq \frac{K(K-1)}{v(v-1)} \mathbb{C}_{K-2}^{v-2} = \mathbb{C}^v_K$ for $v \geq 2$.
Then, with the average TO probability ${\bar q} = \sum_{i = 1 }^N p_i q_i$, we have
$ (1-q_{j})(1-q_{j'})
=  (1- {\bar q} + {\bar q} - q_{j})(1- {\bar q} + {\bar q} -q_{j'})
=  (1  -  {\bar q})^2  +  (1  -  {\bar q})({\bar q}  -  q_j)  +  (1 -  {\bar q})({\bar q}  -  q_{j'})  +  ({\bar q}  -  q_j)({\bar q}  - q_{j'})$.
Thus, with $\nabla F ({\bar {\mathbf{w}}}_{r - 1}) = {\sum}_{i = 1}^N p_i \nabla  F_{i} ({\bar {\mathbf{w}}}_{r-1})$, \eqref{G_24_step1} turns into
{\small
\begin{align}\label{G_24}
G_{24}
\leq & \mathbb{E} \bigg[ {\sum}_{v=2}^K  \frac{(q_{\max})^{K-v}\mathbb{C}^v_K}{1 - (q_{\max})^K} \notag \\
& \cdot \bigg\{
 (1 - {\bar q})^2 \underbrace{\sum_{j = 1 }^N p_{j} \left( \nabla  F_{j} ({\bar {\mathbf{w}}}_{r-1}) - \sum_{i = 1}^N p_i \nabla  F_{i} ({\bar {\mathbf{w}}}_{r-1}) \right)}_{=0}
\underbrace{\sum_{j' = 1 }^N p_{j'}  \left( \nabla F_{j'} ({\bar {\mathbf{w}}}_{r-1}) - \sum_{i = 1}^N p_i \nabla F_{i} ({\bar {\mathbf{w}}}_{r-1}) \right)}_{=0} \notag \\
& \quad + (1-{\bar q}) \sum_{j = 1 }^N p_{j} ({\bar q}-q_j) \left( \nabla F_{j} ({\bar {\mathbf{w}}}_{r-1}) - \nabla F ({\bar {\mathbf{w}}}_{r-1}) \right)
\underbrace{\sum_{j' = 1 }^N p_{j'}  \left( \nabla F_{j'} ({\bar {\mathbf{w}}}_{r-1}) - \sum_{i = 1}^N p_i \nabla F_{i} ({\bar {\mathbf{w}}}_{r-1})\right)}_{=0} \notag \\
& \quad + (1 -{\bar q}) \underbrace{\sum_{j = 1 }^N p_{j} \left( \nabla F_{j} ({\bar {\mathbf{w}}}_{r-1}) - \sum_{i = 1}^N p_i \nabla F_{i} ({\bar {\mathbf{w}}}_{r-1}) \right)}_{=0}
\sum_{j' = 1 }^N p_{j'} ({\bar q} - q_{j'})  \left( \nabla F_{j'} ({\bar {\mathbf{w}}}_{r-1}) - \sum_{i = 1}^N p_i \nabla F_{i} ({\bar {\mathbf{w}}}_{r-1})\right) \notag \\
& \quad +  \sum_{j = 1 }^N  \sum_{j' = 1 }^N   p_{j}p_{j'}({\bar q} - q_j)({\bar q}  -  q_{j'}) \left( \nabla  F_{j} ({\bar {\mathbf{w}}}_{r-1})  -  \nabla  F ({\bar {\mathbf{w}}}_{r-1}) \right)  \left( \nabla F_{j'} ({\bar {\mathbf{w}}}_{r-1}) - \nabla F ({\bar {\mathbf{w}}}_{r-1})\right) \bigg\} \bigg] \notag \\
= & \mathbb{E} \Bigg[ \sum_{v=2}^K \frac{(q_{\max})^{K-v}\mathbb{C}^v_K}{1-(q_{\max})^K}
\sum_{j = 1 }^N  \sum_{j' = 1 }^N  p_{j}p_{j'} ({\bar q} - q_j)({\bar q} - q_{j'}) ( \nabla F_{j} ({\bar {\mathbf{w}}}_{{r} - 1} ) - \nabla F ({\bar {\mathbf{w}}}_{r  -  1})  )  ( \nabla F_{{j'}} ({\bar {\mathbf{w}}}_{r  -  1}) - \nabla F ({\bar {\mathbf{w}}}_{r  -  1}) ) \Bigg] \notag \\
\overset{(a)}{\leq} & {\sum}_{v=2}^K  \frac{(q_{\max})^{K  -  v}\mathbb{C}^v_K}{1 -  (q_{\max})^K} {\sum}_{i = 1 }^N  p_{i} \|q_i  -  {\bar q} \|^2 \mathbb{E}  \left[ \| \nabla  F_{i} ({\bar {\mathbf{w}}}_{r  -  1})  -  \nabla  F ({\bar {\mathbf{w}}}_{r  -  1}) \|^2 \right] \notag \\
\overset{(b)}{\leq} & {\sum}_{v=2}^K  \frac{(q_{\max})^{K-v}\mathbb{C}^v_K}{1-(q_{\max})^K}
{\sum}_{i = 1 }^N   p_{i} \|q_i - {\bar q} \|^2 D_i^2
\, \text{,}
\end{align}}

\vspace{-0.3cm}
\noindent
where inequality (a) is due to the Young's inequality, i.e., $({\bar q}-q_j)({\bar q}-q_{j'})( \nabla F_{j} ({\bar {\mathbf{w}}}_{r-1}) - \nabla F ({\bar {\mathbf{w}}}_{r-1}) )$
$ ( \nabla F_{j'} ({\bar {\mathbf{w}}}_{r-1}) - \nabla F ({\bar {\mathbf{w}}}_{r-1}))$
$\leq  \frac{1}{2} \|{\bar q}-q_j \|^2 \| \nabla F_{j} ({\bar {\mathbf{w}}}_{r-1}) - \nabla F ({\bar {\mathbf{w}}}_{r-1}) \|^2 +  \frac{1}{2} \|{\bar q}-q_{j'} \|^2$ $\| \nabla F_{j'} ({\bar {\mathbf{w}}}_{r-1}) - \nabla F ({\bar {\mathbf{w}}}_{r-1}) \|^2$,
and inequality (b) is by Assumption 3.

Substituting \eqref{G_21}, \eqref{G_22}, \eqref{G_23}, and \eqref{G_24} into \eqref{G_2}, we have
{\small
\begin{align}\label{G_2_result}
G_2
\leq & 2 E L^2 {\sum}_{i = 1}^{N} {\bar \beta}_i {\sum}_{\ell = 1}^{E} \mathbb{E} \left[ \big \| \mathbf{w}^{r,\ell-1}_{i} - {\bar {\mathbf{w}}}_{r-1} \big\|^2  \right]
+  4 E^2 {\sum}_{v=2}^K   \frac{(q_{\max})^{K-v}\mathbb{C}^v_K}{1-(q_{\max})^K}
{\sum}_{i = 1 }^N   p_{i}  \left \|q_i  -  {\bar q}\right\|^2  D_i^2 \notag \\
& +  4 E^2 {\sum}_{i=1}^N {\bar \alpha}_i D_i^2
+ 4 E^2 \mathbb{E} \left[ \left\| \nabla F ({\bar {\mathbf{w}}}_{r-1}) \right\|^2 \right] \text{.}
\end{align}}

\vspace{-0.3cm}
Besides, for the term $G_3$ in \eqref{Proof_Thm1_lemma_2_step1}, we have
{\small
\begin{align}\label{G_3}
G_3
\overset{(a)}{=} \mathbb{E}\Bigg [  \frac{ {\sum}_{{i} \in \mathcal{S}_{r}} {\mathds{1}}^{r}_{i} \| \mathcal{Q}(\Delta \mathbf{w}^{r}_{i}) - \Delta \mathbf{w}^{r}_{i}  \|^2 }{( {\sum}_{{i} \in \mathcal{S}_{r}} {\mathds{1}}^{r}_i )^2 } \Bigg]
\overset{(b)}{=} {\sum}_{i = 1}^{N} {\bar \alpha}_i \mathbb{E}\left[  \| \mathcal{Q}(\Delta \mathbf{w}^{r}_{i}) - \Delta \mathbf{w}^{r}_{i} \|^2 \right]
\overset{(c)}{\leq} {\sum}_{i = 1}^{N} {\bar \alpha}_i J_{ir}^2
\, \text{,}
\end{align}}

\vspace{-0.3cm}
\noindent
where equality (a) is due to the unbiased quantization in \eqref{unbiased_estimation},
equality (b) is by \eqref{average_alpha_1} in Lemma \ref{beta_alpha_K},
and inequality (c) is due to the bounded QE in \eqref{quantization error_local_model}.

Finally,  by substituting (\ref{G_1}), (\ref{G_2_result}) and (\ref{G_3}) into (\ref{Proof_Thm1_lemma_2_step1}), we obtain Lemma \ref{Proof_Thm1_lemma_2}.
\hfill $\blacksquare$

\subsection{Proof of Lemma \ref{Proof_Thm1_lemma_3}}

According to \eqref{local model}, the local model in the $({r}+1)$-th communication round are updated by
{\small
\begin{align*}
\mathbf{w}_{i}^{r,\ell-1}
= \mathbf{\bar w}_{r-1} - \gamma {\sum}_{t = 1}^{\ell-1} \nabla F_i (\mathbf{w}_{i}^{r,t-1}, {\bm \xi}_{i}^{r,t}) \, \text{.}
\end{align*}}

\vspace{-0.3cm}
\noindent
Therefore,
{\small
\begin{align}\label{w_i-w_bar}
& \mathbb{E} \left[ \big\| \mathbf{w}^{r,\ell-1}_{i} - {\bar {\mathbf{w}}}_{r-1} \big\|^2  \right] \notag  \\
= & \mathbb{E} \left[ \Big \| \gamma {\sum}_{t = 1}^{\ell-1} \nabla F_i (\mathbf{w}_{i}^{r,t-1}, {\bm \xi}_{i}^{r,t}) \Big\|^2  \right] \notag  \\
\leq & \gamma^2 (\ell - 1) {\sum}_{t = 1}^{\ell-1} \mathbb{E} \left[ \left \| \nabla F_i (\mathbf{w}_{i}^{r,t-1}, {\bm \xi}_{i}^{r,t}) \right\|^2  \right] \notag  \\
\overset{(a)}{=} & \gamma^2 (\ell - 1) {\sum}_{t = 1}^{\ell-1} \mathbb{E} \left[ \left \| \nabla F_i (\mathbf{w}_{i}^{r,t-1}, {\bm \xi}_{i}^{r,t}) - \nabla F_i (\mathbf{w}_{i}^{r,t-1}) \right\|^2  \right]
+ \gamma^2 (\ell - 1) {\sum}_{t = 1}^{\ell-1} \mathbb{E} \left[ \left \| \nabla F_i (\mathbf{w}_{i}^{r,t-1}) \right\|^2  \right] \notag \\
\overset{(b)}{\leq} & \gamma^2 (\ell - 1)^2 \frac{\sigma^2}{b}
+ \gamma^2 (\ell - 1) {\sum}_{t = 1}^{\ell-1} \mathbb{E} \left[ \left\| \nabla F_i (\mathbf{w}_{i}^{r,t-1}) \right\|^2  \right]  \notag \\
\leq & \gamma^2 E^2 \frac{\sigma^2}{b}
+ \gamma^2 E {\sum}_{t = 1}^{\ell-1} \mathbb{E} \left[ \left\| \nabla F_i (\mathbf{w}_{i}^{r,t-1}) \right\|^2  \right] \notag \\
\leq & \gamma^2 E^2 \frac{\sigma^2}{b}
+  2 \gamma^2 E {\sum}_{t = 1}^{\ell-1} \mathbb{E} \left[ \left\| \nabla F_i (\mathbf{w}_{i}^{r,t-1})  -  \nabla F_i (\mathbf{\bar w}_{r-1}) \right\|^2  \right]
+ 2 \gamma^2 E {\sum}_{t = 1}^{\ell-1} \mathbb{E} \left[ \left \| \nabla F_i (\mathbf{\bar w}_{r-1}) \right\|^2  \right] \notag \\
\leq & \gamma^2 E^2 \frac{\sigma^2}{b}
+ 2 \gamma^2 E L^2 {\sum}_{t = 1}^{\ell-1} \mathbb{E} \left[ \left \| \mathbf{w}_{i}^{r,t-1} - \mathbf{\bar w}_{r-1} \right\|^2  \right]
+ 4 \gamma^2 E^2 \mathbb{E} \left[ \left \| \nabla F_i (\mathbf{\bar w}_{r-1}) -  \nabla F(\mathbf{\bar w}_{r-1}) \right\|^2
+ \left \| \nabla F(\mathbf{\bar w}_{r-1}) \right\|^2  \right] \notag \\
\overset{(c)}{\leq} & \gamma^2 E^2 \frac{\sigma^2}{b}
+ 2 \gamma^2 E L^2 {\sum}_{t = 1}^{\ell-1} \mathbb{E} \left[ \left\| \mathbf{w}_{i}^{r,t-1} - \mathbf{\bar w}_{r-1} \right\|^2 \right ]
+ 4 \gamma^2 E^2 D_i^2
+ 4 \gamma^2 E^2 \mathbb{E} \left[ \left\| \nabla F(\mathbf{\bar w}_{r-1}) \right\|^2  \right]
\text{,}
\end{align}}

\vspace{-0.3cm}
\noindent
where equality (a) is due to $\mathbb{E}[\| \mathbf{x} \|^2] = \mathbb{E}[\| \mathbf{x} - \mathbb{E}[\mathbf{x}] \|^2] + \| \mathbb{E} [\mathbf{x}] \|^2$ and $\mathbb{E} [ \nabla F_i (\mathbf{w}_{i}^{r,t-1}, {\bm \xi}_{i}^{r,t}) ] = \nabla F_i (\mathbf{w}_{i}^{r,t-1})$,
equality (b) is by Assumption 2 given the mini-batch size $b$,
and inequality (c) is by Assumption 3.
Then, summing both sides of \eqref{w_i-w_bar} from ${\ell} = 1$ to $E$ yields
{\small
\begin{align}\label{Proof_Thm1_lemma_3_step1}
& {\sum}_{\ell = 1}^{E} \mathbb{E} \left[ \big\| \mathbf{w}^{r,\ell-1}_{i} - {\bar {\mathbf{w}}}_{r-1} \big\|^2  \right] \notag \\
\leq & \gamma^2 E^3 \frac{\sigma^2}{b}
+ 2 \gamma^2 E L^2
\underbrace{{\sum}_{\ell = 1}^{E} {\sum}_{t = 1}^{\ell-1} \mathbb{E} \left[ \left\| \mathbf{w}_{i}^{r,t-1} - \mathbf{\bar w}_{r-1} \right\|^2  \right]}_{\rm (a)}
+ 4 \gamma^2 E^3 D_i^2
+ 4 \gamma^2 E^3 \mathbb{E} \left[ \left\| \nabla F(\mathbf{\bar w}_{r-1}) \right\|^2  \right] \notag \\
\overset{(b)}{\leq} & \gamma^2 E^3 \frac{\sigma^2}{b}
+ 2 \gamma^2 E^2 L^2 {\sum}_{\ell = 1}^{E} \mathbb{E} \left[ \big\| \mathbf{w}_{i}^{r,\ell-1} - \mathbf{\bar w}_{r-1} \big\|^2  \right]
+ 4 \gamma^2 E^3 D_i^2
+ 4 \gamma^2 E^3 \mathbb{E} \left[ \left\| \nabla F(\mathbf{\bar w}_{r-1}) \right\|^2  \right]
\text{,}
\end{align}}

\vspace{-0.3cm}
\noindent
where inequality (b) is because the occurrence number of $\mathbb{E} [  \| \mathbf{w}^{r,\ell-1}_{i} - {\bar {\mathbf{w}}}_{r-1} \|^2  ]$ for each $\ell \in  [1, E]$ in term (a) is less than the number of local updating steps $E$, and thus
${\rm (a)} \leq  E  \sum_{\ell = 1}^{E} \mathbb{E} [  \| \mathbf{w}_{i}^{r,\ell-1} - \mathbf{\bar w}_{r-1} \|^2  ]$.

Finally, rearranging the terms in (\ref{Proof_Thm1_lemma_3_step1}) yields Lemma \ref{Proof_Thm1_lemma_3}.
\hfill $\blacksquare$

\end{appendices}

%\bibliographystyle{IEEEtran}
%\footnotesize
%\bibliography{refs_journal}


\newpage

\setcounter{section}{0}

\renewcommand\thesection{\Alph{section}}

\section*{\huge Supplementary Material}

\section{ Proof of Lemma 1}

With $|w^{r,E}_{{i}j}| \in [{\underline{w} }^{r}_{{i}j}, {\bar w}^{r}_{{i}j}]$ and quantization level $B^{r}_{i}$, the quantized $w^{r,E}_{{i}j}$ is unbiasedly estimated since
{\small
\begin{align}\label{unbiased_estimation_each_parameter}
\mathbb{E}[ \mathcal{Q}(w^{r,E}_{{i}j}) ]
= & {\rm sign}(w^{r,E}_{{i}j}) \cdot c_u \cdot \Pr \left( \mathcal{Q}(w^{r,E}_{ij}) = {\rm sign}(w^{r,E}_{{i}j}) \cdot c_u \right) \notag \\
& + {\rm sign}(w^{r,E}_{{i}j}) \cdot c_{u+1} \cdot \Pr \left( \mathcal{Q}(w^{r,E}_{ij}) = {\rm sign}(w^{r,E}_{{i}j}) \cdot c_{u+1} \right) \notag \\
= & {\rm sign}(w^{r,E}_{{i}j}) \cdot \bigg( c_u \frac{c_{u+1}-|w^{r,E}_{{i}j}|}{c_{u+1}-c_u} + c_{u+1} \frac{|w^{r,E}_{{i}j}|-c_u}{c_{u+1}-c_u} \bigg)
= {\rm sign}(w^{r,E}_{{i}j}) \cdot |w^{r,E}_{{i}j}|
= w^{r,E}_{{i}j} \, \text{.}
\end{align}}

\vspace{-0.3cm}
Based on this, we have
{\small
\begin{align*}
\mathbb{E} [ \mathcal{Q}(\mathbf{w}^{r,E}_{i}) ]
= \left[ \mathbb{E}[ \mathcal{Q}(w^{r,E}_{{i}1}) ], \mathbb{E}[ \mathcal{Q}(w^{r,E}_{{i}2}) ], \cdots, \mathbb{E}[ \mathcal{Q}(w^{r,E}_{{i}{m}}) ] \right]
= \left[ w^{r,E}_{{i}1}, w^{r,E}_{{i}2}, \cdots, w^{r,E}_{{i}{m}} \right]
= \mathbf{w}^{r,E}_{i}
\, \text{.}
\end{align*}}

\vspace{-0.3cm}
With the stochastic quantization method in (\ref{quantization_method}), the quantization error is bounded by
{\small
\begin{align}\label{quantization_error_proof}
\mathbb{E} \left[ | \mathcal{Q}(w^{r,E}_{{i}j}) - w^{r,E}_{{i}j} |^2 \right]
= & (c_u  -  |w^{r,E}_{{i}j}| )^2  \cdot  \frac{c_{u+1}  -  |w^{r,E}_{{i}j}|}{c_{u+1}  -  c_u}
+  (c_{u+1}  -  |w^{r,E}_{{i}j}| )^2 \cdot \frac{|w^{r,E}_{{i}j}|   -  c_u}{c_{u+1}  -  c_u} \notag \\
= & \frac{ (|w^{r,E}_{{i}j}|-c_u ) (c_{u+1}-|w^{r,E}_{{i}j}| ) (|w^{r,E}_{{i}j}|-c_u+c_{u+1}-|w^{r,E}_{{i}j}| )}{c_{u+1}-c_u} \notag \\
= & (|w^{r,E}_{{i}j}| - c_u ) (c_{u+1}-|w^{r,E}_{{i}j}| ) \notag \\
= & - (|w^{r,E}_{{i}j}| )^2 + (c_u+c_{u+1} ) |w^{r,E}_{{i}j}| - c_u c_{u+1} \notag \\
= & - \left( |w^{r,E}_{{i}j}| - \frac{c_u+c_{u+1}}{2} \right)^2 + \left( \frac{c_u-c_{u+1}}{2} \right)^2
\leq \left( \frac{c_u - c_{u+1}}{2} \right)^2
\text{,}
\end{align}}

\vspace{-0.3cm}
\noindent
where with $c_u$ defined in (\ref{c_j}), the interval between neighboring knobs is given by
{\small
\begin{align}\label{c_j_interval}
| c_{u} - c_{u+1} |
= \frac{ | {\bar w}^{r}_{{i}j} - {\underline{w} }^{r}_{{i}j} |}{ 2^{B^{r}_{i}} - 1 }
\, \text{.}
\end{align}}

\vspace{-0.3cm}
Then, substituting (\ref{c_j_interval}) into (\ref{quantization_error_proof}), we have
{\small
\begin{align}\label{quantization error_parameter}
\mathbb{E} \left[ | \mathcal{Q}(w^{r,E}_{{i}j}) - w^{r,E}_{{i}j} |^2 \right]
\leq \frac{ ( {\bar w}^{r}_{{i}j} - {\underline{w} }^{r}_{{i}j} )^2 }{ 4 ( 2^{B^{r}_{i}} - 1 )^2 }
\, \text{,}
\end{align}}

\vspace{-0.3cm} \noindent
and the total QE of local model can be bounded by
{\small
\begin{align*}
\mathbb{E} \left[ | \mathcal{Q}(\mathbf{w}^{r,E}_{i}) - \mathbf{w}^{r,E}_{i} |^2 \right]
= \mathbb{E} \Bigg[ \Bigg| \sum_{j=1}^{m} \mathcal{Q}(w^{r,E}_{{i}j}) - w^{r,E}_{{i}j} \Bigg|^2 \Bigg]
\overset{(a)}{=} \sum_{j=1}^{m} \mathbb{E} \left[ | \mathcal{Q}(w^{r,E}_{{i}j}) - w^{r,E}_{{i}j} |^2 \right]
\overset{(b)}{\leq} \frac{ \sum_{j=1}^{m}  ( {\bar w}^{r}_{{i}j} - {\underline{w} }^{r}_{{i}j} )^2 }{ 4 ( 2^{B^{r}_{i}} - 1 )^2 }
\, \text{,}
\end{align*}}

\vspace{-0.3cm}
\noindent
where equality (a) is due to the unbiased quantization in (\ref{unbiased_estimation_each_parameter}), and inequality (b) is due to the error bound in (\ref{quantization error_parameter}).
\hfill $\blacksquare$

\section{Extended Discussion of Remark \ref{general_case_QE}}

\subsection{Performance analysis of general case}

For the general case, we consider the unfixed quantization level $B^{r}_i$ and the changed TO probabilities $q^{r}_i$ during the training process for different communication rounds.
Similar to Lemma \ref{beta_alpha_K}, we have some properties for the general case as shown in Lemma \ref{beta_alpha_K_general}.

\begin{lemma}\label{beta_alpha_K_general}
Considering FL algorithm in Algorithm 1, it holds true that
{\small
\begin{align}\label{avrage_beta_2}
\mathbb{E}
\left[  \left. \frac{ {\sum}_{{i} \in {\mathcal{S}}_{r}} \mathds{1}^{r}_{i} \Delta \mathbf{w}^{r}_i }{ {\sum}_{{i} \in {\mathcal{S}}_{r}} \mathds{1}^{r}_{i}} \right|
{\sum}_{{i} \in {\mathcal{S}}_{r}} \mathds{1}^{r}_{i} \neq 0  \right]
\overset{(a)}{=}
{\mathbb{E}}_{\mathcal{S}_{r}} \left[ {\sum}_{{i} \in {\mathcal{S}}_{r}} \beta^{r}_{i} \Delta \mathbf{w}^{r}_{i} \right]
\overset{(b)}{=}
{\sum}_{i=1}^N {\bar \beta}_i \Delta \mathbf{w}^{r}_{i}
\end{align}
}

\vspace{-0.3cm}
\noindent
for some $\beta^{r}_{i}, {\bar \beta}_i \in [0,1]$ with $\sum_{{i} \in {\mathcal{S}}_{r}} \beta^{r}_{i}  = 1$ and $\sum^N_{i=1} {\bar \beta}_i  = 1$,
where equality (a) is taken expected with respect to $\{ \mathds{1}^{r}_{i} \}$ while equality (b) is taken expected with respect to $\mathcal{S}_{r}$.

Moreover, we also have
\begin{small}
\begin{align}\label{avrage_alpha_2}
\mathbb{E}
\left[ \left. \frac{{\sum}_{{i} \in {\mathcal{S}}_{r}} \mathds{1}^{r}_{i}
\Delta \mathbf{w}^{r}_i }{ \left(\sum_{{i} \in {\mathcal{S}}_{r}} \mathds{1}^{r}_{i} \right)^2 } \right|
{\sum}_{{i} \in {\mathcal{S}}_{r}} \mathds{1}^{r}_{i} \neq 0 \right]
=
{\mathbb{E}}_{\mathcal{S}_{r}} \left[ {\sum}_{{i} \in {\mathcal{S}}_{r}} \alpha^{r}_{i} \Delta \mathbf{w}^{r}_{i} \right]
=
{\sum}_{i=1}^N {\bar \alpha}_i \Delta \mathbf{w}^{r}_{i}
\end{align}
\end{small}

\vspace{-0.3cm}
\noindent
for some $\alpha^{r}_{i},{\bar \alpha}_i\geq 0$ $\forall i =1,\cdots,N$ and $\forall r =1,\cdots,M$.

Finally, same with \eqref{bar_K_1}, we denote
\begin{small}
\begin{align*}
& {\mathbb{E}}
\! \left[ \left. \frac{1}{ \sum_{{i} \in {\mathcal{S}}_{r}} \mathds{1}^{r}_{i}} \right| {\sum}_{{i} \in {\mathcal{S}}_{r}} \mathds{1}^{r}_{i} \neq 0 \right]
= {\sum}_{i = 1}^N {\bar \alpha}_i
\triangleq \frac{1}{\bar K} \, \text{,}
\end{align*}
\end{small}

\vspace{-0.3cm} \noindent
where $\bar K$ represents the average effective number of active clients at each communication round.

If $q^{r}_i$ is uniform for all clients at all communication rounds,
i.e.,  $q^{r}_i = q$ $\forall i =1,\cdots,N$ and $\forall r =1,\cdots,M$,
then $\beta^{r}_i = 1/K$ and $\alpha^{r}_i = {1}/({K}{\bar K})$ $\forall i \in {\mathcal{S}}_{r}$,
${\bar \beta}_i = p_i$ and ${\bar \alpha}_i = p_i /{\bar K}$ $\forall i \in \{ 1,\cdots,N \}$,
and ${\bar K} = \frac{1 - (q)^K}{\sum_{v = 1}^K \frac{1}{v} \left(\mathbb{C}^v_K \left(1-q\right)^{v} (q)^{K-v}\right)}$ with $\mathbb{C}_K^v = \frac{K!}{v!(K-v)!}$.
In addition, if $q^{r}_i = 0$ $\forall i \in {\mathcal{S}}_{r}$ and $\forall r =1,\cdots,M$ (no TO), then ${\bar K} = K$.
\end{lemma}

From \eqref{avrage_beta_2}, one can see that $\{ \beta^{r}_i \}$ is the equivalent appearance probability of $\{ \Delta \mathbf{w}^{r}_i \}$ transmitted by each selected client $i \in \mathcal{S}_{r}$ in the global aggregation due to TO,
while $\beta_i$ is that of $\Delta \mathbf{w}^{r}_i$ transmitted by each client $i \in \{1,\cdots,N\}$ in the global aggregation due to client sampling and TO.
The main convergence result is stated below.

\begin{theorem}\label{Thm_diff_TO_general}
(General case) Let Assumptions 1 to 3 hold.
If one chooses $\gamma = {\bar K}^{\frac{1}{2}} / (8L{T}^{\frac{1}{2}}) $ and $E \leq T^{\frac{1}{4}}/{\bar K}^{\frac{3}{4}}$ where $T = ME \geq \max \{  {\bar K}^{3}, 1/{\bar K} \}$ is the total number of SGD updates per client,  we have
{\small
\begin{align}\label{theorem_2}
& \frac{1}{M} {\sum}_{r = 1}^{M} \mathbb{E}\left[ \left\| \nabla  F({\bar {\mathbf{w}}}_{r-1}) \right\|^2 \left| {\sum}_{{i} \in {\mathcal{S}}_{r}} \mathds{1}^{r}_{i} \neq 0 \right. \right] \notag \\
\leq & \frac{496 L  \left( \mathbb{E}[F({\bar {\mathbf{w}}}_{0})] - \underline{F} \right)}{11  \left( T  {\bar K} \right)^{\frac{1}{2}} }
+  \Bigg(  \frac{ 39 }{ 88  \left( T {\bar K} \right)^{\frac{1}{2}} }
+  \frac{1}{ 88  \left( T  {\bar K} \right)^{\frac{3}{4}} }  \Bigg)  \frac{\sigma^2}{b}
+  \underbrace{ \frac{31 {\bar K}^{\frac{1}{2}}}{ 88 T^{\frac{3}{2}} } {\sum}_{r = 1}^{M} \mathbb{E}_{\mathcal{S}_{r}} \Bigg[ {\sum}_{{i} \in \mathcal{S}_{r}} \alpha^{r}_{i} J_{ir}^2 \Bigg]}_{ {\rm (a) (caused \ by \ QE)} }
\notag \\
& +  \underbrace{ \frac{31 }{ 22 \left( T {\bar K} \right)^{\frac{1}{4}} } {\sum}_{i=1}^N {\bar \alpha}_i D_i^2 }_{ {\rm (b) (caused \ by \ partial \ participation} \atop {\rm  and \ data \ variance)} }
+  \underbrace{ \Bigg(  \frac{4 }{ 11 \left( T {\bar K} \right)^{\frac{1}{2}} }
+  \frac{1}{ 22 \left( T {\bar K} \right)^{\frac{3}{4}} }   \Bigg) {\sum}_{i = 1}^{N}  {\bar \beta}_i D_i^2 }_{ {\rm (c) (caused \ by \ data \ variance)} }
+  \underbrace{ \frac{62}{11} \chi^2_{\bm{\beta}\|\mathbf{p}}  {\sum}_{i = 1}^N p_i D_i^2 }_{ {\rm (d) (caused \ by \ TO \ and} \atop {\rm data \ variance)} }
\notag \\
& + \underbrace{ \frac{31 }{ 22 T {\bar K} } {\sum}_{v=2}^K  \frac{(q_{\max})^{K-v}\mathbb{C}^v_K}{1-(q_{\max})^K}
{\sum}_{r = 1}^M \mathbb{E}_{\mathcal{S}_{r}} \left[ \frac{1}{K} {\sum}_{{i} \in \mathcal{S}_{r}} \left \|q^{r}_{i} - {\bar q}\right\|^2 D^2_{i} \right]
}_{ {\rm (e) (caused \ by \ TO \ and \ data \ variance)} }
\text{,}
\end{align}}

\vspace{-0.3cm}
\noindent where $\chi^2_{\bm{\beta}\|\mathbf{p}} \triangleq  \sum_{i = 1}^N {({\bar \beta}_i - p_i)^2}/{ p_i}$ is the chi-square divergence \cite{wang2020tackling},
$q_{\max} = \max_{i\in \mathcal{S}_{r}, \forall \mathcal{S}_{r}} \{q^{r}_i\}$ and $\bar q = \mathbb{E}_{\mathcal{S}_{r}} \left[ \frac{1}{K} \sum_{i\in \mathcal{S}_{r}} q^{r}_i \right]$ are the maximum and average TO probabilities, respectively.
\end{theorem}

\emph{Proof:} See the subsequent Subsection \ref{sub_Thm_2}. \hfill $\blacksquare$

The upper bound in \eqref{theorem_2} reveals similar insights as discussed in Theorem \ref{Thm_diff_TO_fixed}.
Also, when the clients have a uniform TO probability, the terms (d) and (e) would vanish.
Then, combining with Lemma \ref{beta_alpha_K_general}, we can derive the following Corollary \ref{uniform_TO_general} for the uniform-TO case with unfixed quantization level $B^{r}_i$.
As shown in \eqref{corollary_2}, the FL algorithm can also achieve a linear speed-up with respect to ${\bar K}$ even when both TO and QE are present.

\begin{corollary}\label{uniform_TO_general}
Under the same conditions as Theorem \ref{Thm_diff_TO_general}, if all clients have a uniform TO probability $q$, we have
\begin{small}
\begin{align}\label{corollary_2}
& \frac{1}{M} {\sum}_{r = 1}^{M} \mathbb{E}\left[ \| \nabla  F({\bar {\mathbf{w}}}_{r-1}) \|^2 \left| \sum_{{i} \in {\mathcal{S}}_{r}} \mathds{1}^{r}_{i} \neq 0 \right. \right] \notag \\
\leq & \frac{496 L }{11 ( T {\bar K} )^{\frac{1}{2}} }  \left( \mathbb{E}[F({\bar {\mathbf{w}}}_{0})] - \underline{F} \right)
+ \left(  \frac{ 39 }{ 88 ( T {\bar K} )^{\frac{1}{2}} }
+  \frac{1}{ 88 ( T {\bar K} )^{\frac{3}{4}} }  \right)  \frac{\sigma^2}{b}
+ \frac{31}{ 88 T^{\frac{3}{2}} {\bar K}^{\frac{1}{2}}} {\sum}_{r = 1}^{M} {\mathbb{E}}_{\mathcal{S}_{r}} \left[ \frac{1}{K} {\sum}_{{i} \in {\mathcal{S}}_{r}} J_{ir}^2 \right] \notag \\
& + \left( \frac{4 }{ 11 ( T {\bar K} )^{\frac{1}{2}} }
+ \frac{1}{ 22 ( T {\bar K} )^{\frac{3}{4}} } + \frac{31 }{ 22 T^{\frac{1}{4}} {\bar K}^{\frac{5}{4}} } \right) {\sum}_{i = 1}^{N} p_i D_i^2 \, \text{.}
\end{align}
\end{small}
\end{corollary}
\hfill $\blacksquare$

\subsection{Proof of Theorem \ref{Thm_diff_TO_general}}\label{sub_Thm_2}

In the general case, for the same client $i$, its TO probability $q^{r}_i$ and quantization level $B^{r}_i$ would vary with the selected client set $\mathcal{S}_{r}$.
For example, the TO probability and quantization level of client 1 in ${\mathcal{S}}_r^g = \{1,2,3,\cdots,K\}$ and those in ${\mathcal{S}}_r^g = \{1,3,4,\cdots,K+1\}$ are different.
Based on this, since different communication rounds correspond to different $\mathcal{S}_{r}$, the TO probability $q^{r}_i$ and quantization level $B^{r}_i$ of the same selected client $i$ vary with the communication round.

For simplicity, we assume that for each possible set ${\mathcal{S}}_r^g$, both the wireless resource (including bandwidth and transmit power) and quantization level follow a fixed allocation scheme whenever ${\mathcal{S}}_r^g$ appears.
In this way, for each possible set ${\mathcal{S}}_r^g$, there is a unique set of the TO probabilities and quantization levels for the clients in ${\mathcal{S}}_r^g$.
Then, with denoting $q_{gi}$ and $ B_{gi}$ as the TO probability and the quantization level of the client $i \in {\mathcal{S}}_r^g$, we have $q^{r}_{i} = q_{gi}$ and $B^{r}_{i} = B_{gi}$ if ${\mathcal{S}}_r = {\mathcal{S}}_r^g$.

The proof of Theorem 2 is similar to that of Theorem 1 (Appendix B) except for the following differences.

\subsubsection{Difference 1}

The formulation \eqref{G_3} in Appendix B becomes
{\small
\begin{align}\label{G_3_general}
G_3
\overset{(a)}{=} & \mathbb{E}\Bigg [  \frac{ {\sum}_{{i} \in \mathcal{S}_{r}} {\mathds{1}}^{r}_{i} \| \mathcal{Q}(\Delta \mathbf{w}^{r}_{i}) - \Delta \mathbf{w}^{r}_{i}  \|^2 }{( {\sum}_{{i} \in \mathcal{S}_{r}} {\mathds{1}}^{r}_i )^2 } \Bigg]
\overset{(b)}{=} \mathbb{E}_{\mathcal{S}_{r}} \left[ {\sum}_{{i} \in \mathcal{S}_{r}} \alpha^{r}_{i} \mathbb{E} \left[ \| \mathcal{Q}(\Delta \mathbf{w}^{r}_{i}) - \Delta \mathbf{w}^{r}_{i} \|^2 \right] \right]
\overset{(c)}{\leq} \mathbb{E}_{\mathcal{S}_{r}} \left[ {\sum}_{{i} \in \mathcal{S}_{r}} \alpha^{r}_{i} J_{ir}^2 \right]
\, \text{,}
\end{align}}

\vspace{-0.3cm}
\noindent
where equality (a) is due to the unbiased quantization in (\ref{unbiased_estimation}),
equality (b) is caused by \eqref{avrage_alpha_2} in Lemma \ref{beta_alpha_K_general},
and inequality (c) is due to the bounded QE in \eqref{quantization error_local_model}.
Based on \eqref{G_3_general}, the term (a) in Theorem 1 (i.e., $\frac{31 {\bar K}^{1/2}}{ 88 T^{3/2} } \sum_{r = 1}^{M} \sum_{i = 1}^{N} {\bar \alpha}_i J_{ir}^2$) turns into $\frac{31 {\bar K}^{1/2}}{ 88 T^{3/2} } \sum_{r = 1}^{M} \mathbb{E}_{\mathcal{S}_{r}} \left[ \sum_{{i} \in \mathcal{S}_{r}} \alpha^{r}_{i} J_{ir}^2 \right]$ in Theorem 2.

\subsubsection{Difference 2}

With the maximum TO probability $q_{\max} = \max \limits_{i\in \mathcal{S}_{r}, \forall \mathcal{S}_{r}} \{q^{r}_i\}
= \max \limits_{g \in \{1,\cdots,N^K\}} \big\{ \max \limits_{{i} \in {\mathcal{S}}_r^g} q_{g{i}} \big\}$, \eqref{Pr_1_fixed} in Appendix B becomes
{\small
\begin{align*}
\Pr\left[{\mathds{1}}^{r}_{k}=1, {\mathds{1}}^{r}_{{k'}}=1, {\sum}_{{i} \in \mathcal{S}_{r}}{\mathds{1}}^{r}_{i}=v \right]
\leq \frac{(1-q^{r}_{k})(1-q^{r}_{{k'}}) (q_{\max})^{K-v} \mathbb{C}_{K-2}^{v-2} }{1-(q_{\max})^K}
\, \text{.}
\end{align*}}

\vspace{-0.3cm} \noindent
Then, with the average TO probability
$\bar q
= \mathbb{E}_{\mathcal{S}_{r}} \big[ \frac{1}{K} \sum_{i\in \mathcal{S}_{r}} q^{r}_i \big]
=\sum_{g=1}^{N^K} \big( \prod_{{i} \in {\mathcal{S}}_r^g} p_{i} \cdot \frac{1}{K} \sum_{{i} \in {\mathcal{S}}_r^g} q_{g{i}} \big)$, the formulation \eqref{G_24_step1} in Appendix B turns into
{\small
\begin{align}\label{G_24_general_1}
G_{24}
= & {\sum}_{v=2}^K  \frac{(q_{\max})^{K-v}\mathbb{C}_{K-2}^{v-2}}{(1  -  (q_{\max})^K)v^2} \notag \\
& \cdot
\mathbb{E} \left[ {\sum}_{k \in \mathcal{S}_{r}} {\sum}_{k' \in \mathcal{S}_{r} \atop k' \neq k}  \Big( (1  -  q^{r}_{k})(1  -  q^{r}_{{k'}})( \nabla  F_{k} ({\bar {\mathbf{w}}}_{{r} - 1} )  -  \nabla F ({\bar {\mathbf{w}}}_{{r} - 1})  )  ( \nabla  F_{{k'}} ({\bar {\mathbf{w}}}_{{r} - 1})  -  \nabla  F ({\bar {\mathbf{w}}}_{{r} - 1}) )  \Big)   \right] \notag \\
\overset{(a)}{=} & {\sum}_{v=2}^K  \frac{(q_{\max})^{K-v}\mathbb{C}_{K-2}^{v-2}}{(1  -  (q_{\max})^K)v^2} \notag \\
& \cdot
\mathbb{E} \left[
(1-{\bar q})^2 {\sum}_{k \in \mathcal{S}_{r}} {\sum}_{k' \in \mathcal{S}_{r} \atop k' \neq k} \left( \nabla F_{k} ({\bar {\mathbf{w}}}_{r-1}) - \nabla F ({\bar {\mathbf{w}}}_{r-1}) \right)  \left( \nabla F_{k'} ({\bar {\mathbf{w}}}_{r-1}) - \nabla F ({\bar {\mathbf{w}}}_{r-1}) \right) \right.  \notag \\
& \qquad + (1-{\bar q}) {\sum}_{k \in \mathcal{S}_{r}} {\sum}_{k' \in \mathcal{S}_{r} \atop k' \neq k} ({\bar q}-q^{r}_{k}) \left( \nabla F_{k} ({\bar {\mathbf{w}}}_{r-1}) - \nabla F ({\bar {\mathbf{w}}}_{r-1}) \right)
\left( \nabla F_{k'} ({\bar {\mathbf{w}}}_{r-1}) - \nabla F ({\bar {\mathbf{w}}}_{r-1}) \right) \notag \\
& \qquad + (1-{\bar q}) {\sum}_{k \in \mathcal{S}_{r}} {\sum}_{k' \in \mathcal{S}_{r} \atop k' \neq k} ({\bar q}-q^{r}_{k'}) \left( \nabla F_{k} ({\bar {\mathbf{w}}}_{r-1}) - \nabla F ({\bar {\mathbf{w}}}_{r-1}) \right)
\left( \nabla F_{k'} ({\bar {\mathbf{w}}}_{r-1}) - \nabla F ({\bar {\mathbf{w}}}_{r-1}) \right) \notag \\
& \qquad \left.  + {\sum}_{k \in \mathcal{S}_{r}} {\sum}_{k' \in \mathcal{S}_{r} \atop k' \neq k} ({\bar q} -  q^{r}_{k})({\bar q} - q^{r}_{k'}) \left( \nabla F_{k} ({\bar {\mathbf{w}}}_{r-1})  -  \nabla F ({\bar {\mathbf{w}}}_{r-1}) \right) \left( \nabla F_{k'} ({\bar {\mathbf{w}}}_{r-1}) - \nabla F ({\bar {\mathbf{w}}}_{r-1}) \right) \right]\notag \\
= & {\sum}_{v=2}^K  \frac{(q_{\max})^{K-v}\mathbb{C}_{K-2}^{v-2}}{(1  -  (q_{\max})^K)v^2} \notag \\
& \cdot
\mathbb{E} \left[
(1-{\bar q})^2 {\sum}_{k \in \mathcal{S}_{r}} \left( \nabla F_{k} ({\bar {\mathbf{w}}}_{r-1}) - \nabla F ({\bar {\mathbf{w}}}_{r-1}) \right)
{\sum}_{k' \in \mathcal{S}_{r} \atop k' \neq k}  \left( \nabla F_{k'} ({\bar {\mathbf{w}}}_{r-1}) - \nabla F ({\bar {\mathbf{w}}}_{r-1}) \right) \right.  \notag \\
& \qquad + (1-{\bar q}) {\sum}_{k \in \mathcal{S}_{r}} ({\bar q}-q^{r}_{k}) \left( \nabla F_{k} ({\bar {\mathbf{w}}}_{r-1}) - \nabla F ({\bar {\mathbf{w}}}_{r-1}) \right)
{\sum}_{k' \in \mathcal{S}_{r} \atop k' \neq k} \notag \left( \nabla F_{k'} ({\bar {\mathbf{w}}}_{r-1}) - \nabla F ({\bar {\mathbf{w}}}_{r-1}) \right) \\
& \qquad  + (1-{\bar q})
{\sum}_{k' \in \mathcal{S}_{r} } ({\bar q}-q^{r}_{k'}) \left( \nabla F_{k'} ({\bar {\mathbf{w}}}_{r-1}) - \nabla F ({\bar {\mathbf{w}}}_{r-1}) \right)
{\sum}_{k \in \mathcal{S}_{r} \atop k \neq k'} \left( \nabla F_{k} ({\bar {\mathbf{w}}}_{r-1}) - \nabla F ({\bar {\mathbf{w}}}_{r-1}) \right)
\notag \\
& \qquad \left.  + {\sum}_{k \in \mathcal{S}_{r}} ({\bar q} -  q^{r}_{k}) \left( \nabla F_{k} ({\bar {\mathbf{w}}}_{r-1})  -  \nabla F ({\bar {\mathbf{w}}}_{r-1}) \right)
{\sum}_{k' \in \mathcal{S}_{r} \atop k' \neq k} ({\bar q} - q^{r}_{k'})  \left( \nabla F_{k'} ({\bar {\mathbf{w}}}_{r-1}) - \nabla F ({\bar {\mathbf{w}}}_{r-1}) \right) \right]
\text{,}
\end{align}}

\vspace{-0.3cm} \noindent
where equality (a) follows from
$ (1-q^{r}_{k})(1-q^{r}_{k'})
= (1-{\bar q}+{\bar q}-q^{r}_{k})(1-{\bar q}+{\bar q}-q^{r}_{k'})
= (1-{\bar q})^2 + (1-{\bar q})({\bar q}-q^{r}_{k}) + (1-{\bar q})({\bar q}-q^{r}_{k'})
+ ({\bar q}-q^{r}_{k})({\bar q}-q^{r}_{k'})$.

Next, since the clients $k,k' \in \mathcal{S}_{r}$ are selected independently and with replacement, then based on $\nabla F ({\bar {\mathbf{w}}}_{r - 1}) = {\sum}_{i = 1}^N p_i \nabla  F_{i} ({\bar {\mathbf{w}}}_{r-1})$ and the same reason as obtaining (c) in \eqref{same_outage probability_beta_proof_1}, \eqref{G_24_general_1} becomes
{\small
\begin{align}\label{G_24_general}
G_{24}
\leq & {\sum}_{v=2}^K  \frac{(q_{\max})^{K-v}\mathbb{C}_{K-2}^{v-2}}{(1  -  (q_{\max})^K)v^2}  \notag \\
& \cdot \mathbb{E} \Bigg[ (1-{\bar q})^2 K(K \! - \! 1)
\underbrace{ \sum_{j = 1 }^N p_{j} \bigg( \nabla \! F_{j} ({\bar {\mathbf{w}}}_{r-1}) \! - \! \sum_{i = 1}^N p_i \nabla \! F_i ({\bar {\mathbf{w}}}_{r-1}) \bigg)}_{=0}
\underbrace{ \sum_{j' = 1 }^N p_{j'}
\bigg( \nabla \! F_{j'} ({\bar {\mathbf{w}}}_{r-1}) \! - \! \sum_{i = 1}^N p_i \nabla \! F_i ({\bar {\mathbf{w}}}_{r-1}) \bigg)}_{=0} \notag \\
& \qquad + (1 \! - \!  {\bar q})(K \! - \! 1)
\sum_{k \in \mathcal{S}_{r}} ({\bar q} \! - \! q^{r}_{k})\left( \nabla F_{k} ({\bar {\mathbf{w}}}_{r-1}) \! - \! \nabla F ({\bar {\mathbf{w}}}_{r-1}) \right)
\underbrace{\sum_{j' = 1 }^N p_{j'} \bigg( \nabla \! F_{j'} ({\bar {\mathbf{w}}}_{r-1}) \! - \! \sum_{i = 1}^N p_i \nabla \! F_i ({\bar {\mathbf{w}}}_{r-1}) \bigg)}_{=0} \notag \\
& \qquad + (1 \! - \! {\bar q}) (K \! - \! 1)
\sum_{k' \in \mathcal{S}_{r}} ({\bar q} \! - \! q^{r}_{k'})\left( \nabla F_{k'} ({\bar {\mathbf{w}}}_{r-1}) \! - \! \nabla F ({\bar {\mathbf{w}}}_{r-1}) \right)
\underbrace{\sum_{j = 1 }^N p_{j} \bigg( \nabla F_{j} ({\bar {\mathbf{w}}}_{r-1}) \! - \! \sum_{i = 1}^N p_i \nabla F_i ({\bar {\mathbf{w}}}_{r-1}) \bigg)}_{=0}
  \notag \\
& \qquad + {\sum}_{k \in \mathcal{S}_{r}}  {\sum}_{k' \in \mathcal{S}_{r} \atop k' \neq k} ({\bar q}-q^{r}_{k})({\bar q}-q^{r}_{k'})
\left( \nabla  F_{k} ({\bar {\mathbf{w}}}_{r-1})  - \nabla  F ({\bar {\mathbf{w}}}_{r-1}) \right) \left( \nabla  F_{k'} ({\bar {\mathbf{w}}}_{r-1})  -  \nabla F  ({\bar {\mathbf{w}}}_{r-1}) \right)  \Bigg] \notag \\
= & \sum_{v=2}^K \! \frac{(q_{\max})^{K-v}\mathbb{C}_{K-2}^{v-2}}{(1  -  (q_{\max})^K)v^2}
\cdot \mathbb{E} \Bigg[ \sum_{k \in \mathcal{S}_{r}} \!  \sum_{k' \in \mathcal{S}_{r} \atop k' \neq k } \! \! ({\bar q} \! - \! q^{r}_{k})({\bar q} \! - \! q^{r}_{k'})
\left( \nabla \!  F_{k} ({\bar {\mathbf{w}}}_{r-1}) \!  - \! \nabla  \! F ({\bar {\mathbf{w}}}_{r-1}) \right) \left( \nabla \!  F_{k'} ({\bar {\mathbf{w}}}_{r-1}) \!  -  \! \nabla \!  F ({\bar {\mathbf{w}}}_{r-1}) \right)  \! \! \Bigg] \notag \\
\overset{(a)}{\leq} & {\sum}_{v=2}^K  \frac{(q_{\max})^{K-v}\mathbb{C}_{K-2}^{v-2}}{(1 \! - \! (q_{\max})^K)v^2} \cdot \mathbb{E}  \bigg[ {\sum}_{k\in \mathcal{S}_{r}} {\sum}_{k' \in \mathcal{S}_{r} \atop k' \neq k}  \frac{1}{2}
\Big( \left\|q^{r}_{k} \! - \! {\bar q} \right\|^2 \left\| \nabla  F_{k} ({\bar {\mathbf{w}}}_{r-1}) \! - \! \nabla  F ({\bar {\mathbf{w}}}_{r-1}) \right\|^2 \notag \\
& \qquad\qquad\qquad\qquad\qquad\qquad  + \left \|q^{r}_{k'}-{\bar q} \right\|^2 \left\| \nabla F_{k'} ({\bar {\mathbf{w}}}_{r-1}) \! - \! \nabla F ({\bar {\mathbf{w}}}_{r-1}) \right\|^2 \Big) \! \bigg] \notag \\
= & {\sum}_{v=2}^K  \frac{(q_{\max})^{K-v}\mathbb{C}_{K-2}^{v-2}}{(1  -  (q_{\max})^K)v^2}
\cdot \frac{K  -  1}{2} \cdot
\mathbb{E} \bigg[ {\sum}_{{k} \in \mathcal{S}_{r}}  \left \|q^{r}_{k}  -  {\bar q} \right\|^2 \left\| \nabla F_{k} ({\bar {\mathbf{w}}}_{r-1})  -  \nabla F ({\bar {\mathbf{w}}}_{r-1}) \right\|^2 \notag \\
& \qquad\qquad\qquad\qquad\qquad\qquad\qquad\qquad + {\sum}_{k' \in \mathcal{S}_{r}}  \left\|q^{r}_{k'}  -  {\bar q} \right\|^2 \left\| \nabla F_{k'} ({\bar {\mathbf{w}}}_{r-1})   -  \nabla F ({\bar {\mathbf{w}}}_{r-1}) \right\|^2 \bigg] \notag \\
= & {\sum}_{v=2}^K  \frac{(q_{\max})^{K-v} K(K  -  1) \mathbb{C}_{K-2}^{v-2}}{(1  -  (q_{\max})^K)v^2} \cdot \mathbb{E} \left[ \frac{1}{K} {\sum}_{{i} \in \mathcal{S}_{r}} \left \|q^{r}_{i}  -  {\bar q} \right\|^2 \left\| \nabla F_{i} ({\bar {\mathbf{w}}}_{r-1})  -  \nabla F ({\bar {\mathbf{w}}}_{r-1}) \right\|^2 \right] \notag \\
\overset{(b)}{\leq} & {\sum}_{v=2}^K \frac{ (q_{\max})^{K-v} \mathbb{C}^v_K}{ 1-(q_{\max})^K} \mathbb{E}_{\mathcal{S}_{r}} \left[ \frac{1}{K} {\sum}_{{i} \in \mathcal{S}_{r}} \left \|q^{r}_{i}-{\bar q} \right\|^2 D^2_{i} \right]
\text{,}
\end{align}}

\vspace{-0.3cm}
\noindent
where inequality (a) is due to Young's Inequality,
and inequality (b) is obtained by $\frac{K(K-1)\mathbb{C}_{K-2}^{v-2}}{v^2} \leq \frac{K(K-1)}{v(v-1)} \mathbb{C}_{K-2}^{v-2} = \mathbb{C}^v_K$ and Assumption 3.

Based on \eqref{G_24_general}, the last term of \eqref{Proof_Thm1_lemma_4_formulation} in Appendix B becomes
{\small
\begin{align*}\frac{2 \gamma E L}{M} \cdot \sum_{v=2}^K \frac{(q_{\max})^{K-v}\mathbb{C}^v_K}{1-(q_{\max})^K}
{\sum}_{r = 1 }^M \mathbb{E}_{\mathcal{S}_{r}} \left[ \frac{1}{K} {\sum}_{{i} \in \mathcal{S}_{r}} \left \|q^{r}_{i}-{\bar q} \right\|^2 D^2_{i} \right]\, \text{.}
\end{align*}}

\vspace{-0.3cm}\noindent
and the coefficient $H_6$ in \eqref{Proof_Thm1_lemma_4_formulation} is redefined as $H_6 \triangleq \frac{2 \gamma E L}{M} = \frac{2 \gamma E^2 L}{T}$.
If one chooses $\gamma = {\bar K}^{\frac{1}{2}} / (8L{T}^{\frac{1}{2}}) $ and $E \leq T^{\frac{1}{4}}/{\bar K}^{\frac{3}{4}}$, we have
{\small
\begin{align*}
H_6 \leq \frac{2}{8L} \sqrt{\frac{\bar K}{T}}
\cdot \left( \frac{T^{\frac{1}{4}}}{{\bar K}^{\frac{3}{4}}} \right)^2 \cdot \frac{L}{T} = \frac{1}{4T{\bar K}}
\, \text{.}
\end{align*}}

\vspace{-0.1cm} \noindent
Therefore, the term (e) (i.e., $\frac{31 }{ 22 ( T {\bar K} )^{{1}/{4}} } \sum_{v=2}^K  \frac{(q_{\max})^{K-v}\mathbb{C}^v_K}{1-(q_{\max})^K}
\sum_{i = 1 }^N   p_{i}  \left \|q_i - {\bar q}\right\|^2 D_i^2$) in Theorem 1 becomes
$\frac{31 }{ 22 T {\bar K} } \sum_{v=2}^K  \frac{(q_{\max})^{K-v}\mathbb{C}^v_K}{1-(q_{\max})^K}
\sum_{r = 1}^M \mathbb{E}_{\mathcal{S}_{r}} \left[ \frac{1}{K} \sum_{{i} \in \mathcal{S}_{r}} \left \|q^{r}_{i} - {\bar q}\right\|^2 D^2_{i} \right]$ in Theorem 2.
\hfill $\blacksquare$

\section{Average uplink transmission delay in \eqref{average_transmission_delay}}

\subsection{Derivation process of ${\bar \tau}^{r}_i$}

If the TO probabilities of the selected clients in $\mathcal{S}_{r}$ all equal to 1, the probability that all selected clients fail to transmit data without TO is $\Pr ( \sum_{{i} \in {\mathcal{S}}_{r}} \mathds{1}^{r}_{i} ) = {\prod}_{{i} \in \mathcal{S}_{r}} q_{i} = 1$. In such case, the retransmission process will be repeated infinitely,
and the transmission delay will become infinite. However, this extreme situation can be easily avoided in the wireless system if the conditions in Lemma \ref{uplink_delay_lemma} are satisfied.

\begin{lemma}\label{uplink_delay_lemma}
With the definition of TO probability in (\ref{outage_probability_shadow}), if the uplink transmission rate $R_i < + \infty$, the transmit power $P_i > 0$ (in Watt) and the allocated bandwidth $W_i > 0$ for each client ${i}$ are satisfied, then the outage probability of each client $q_i < 1$.
\end{lemma}

\emph{Proof:} See the subsequent Subsection \ref{proof_Lemma_q_larger_0}.
\hfill $\blacksquare$

Actually, as shown in Proposition \ref{optimal_condition}, the above conditions are satisfied in the optimal condition of problem (\ref{obj_func_A1}).

Then, since retransmission is performed if all selected clients experience outage in the uplink transmission (i.e., $\sum_{{j} \in {\mathcal{S}}_{r}} \mathds{1}^{r}_{j} = 0$), the average transmission delay of the client ${i} \in \mathcal{S}_{r}$ is computed by
{\small \begin{align}\label{average_transmission_delay_formulation_1}
{\bar \tau}^{r}_i
= \sum \limits_{k=1}^{\infty}
\underbrace{ \left( {\prod}_{{j} \in \mathcal{S}_{r}} q_{j} \right)^{k-1}
\left( 1- {\prod}_{{j} \in \mathcal{S}_{r}} q_{j} \right)}_{\rm (a)}
\underbrace{k \cdot \max \limits_{{j} \in \mathcal{S}_{r}} \frac{{\hat B}_{j}}{R_{j}}}_{\rm (b)}
= \underbrace{\left( 1- {\prod}_{{j} \in \mathcal{S}_{r}} q_{j} \right) \sum \limits_{k=1}^{\infty} k
\left( {\prod}_{{j} \in \mathcal{S}_{r}} q_{j} \right)^{k-1}}_{(c)}
\cdot \max \limits_{{j} \in \mathcal{S}_{r}} \frac{{\hat B}_{j}}{R_{j}}
\, \text{,}
\end{align}}

\vspace{-0.3cm} \noindent
where (a) denotes the probability that there isn't any client successfully uploading its model until the $k$-th transmission round, and (b) is the uplink delay of $k$ successive transmissions.

Next, with
{\small
\begin{align*}
& \left( 1 - {\prod}_{{j} \in \mathcal{S}_{r}} q_{j} \right) \sum \limits_{k=1}^{N} k \left( {\prod}_{{j} \in \mathcal{S}_{r}} q_{j} \right)^{k-1} \notag \\
= & \sum \limits_{k=1}^{N} k \left( {\prod}_{{j} \in \mathcal{S}_{r}} q_{j} \right)^{k-1}
- \sum \limits_{k=1}^{N} k \left( {\prod}_{{j} \in \mathcal{S}_{r}} q_{j} \right)^{k}
= \sum \limits_{k=0}^{N-1} \left( {\prod}_{{j} \in \mathcal{S}_{r}} q_{j} \right)^{k}
-  N  \left( {\prod}_{{j} \in \mathcal{S}_{r}} q_{j} \right)^{N} \notag \\
= & \frac{1 - \left( {\prod}_{{j} \in \mathcal{S}_{r}} q_{j} \right)^{N}}{1 - {\prod}_{{j} \in \mathcal{S}_{r}} q_{j}} - N \left( {\prod}_{{j} \in \mathcal{S}_{r}} q_{j} \right)^{N}
= \frac{1 - (1+N)\left( {\prod}_{{j} \in \mathcal{S}_{r}} q_{j} \right)^{N} + N\left( {\prod}_{{j} \in \mathcal{S}_{r}} q_{j} \right)^{N+1}}{1 - {\prod}_{{j} \in \mathcal{S}_{r}} q_{j}}
\, \text{,}
\end{align*}}

\vspace{-0.15cm}
\noindent
and $ \prod_{{j} \in \mathcal{S}_{r}} q_{j} < 1 $, the term (c) in \eqref{average_transmission_delay_formulation_1} is given by
{\small
\begin{align}\label{average_transmission_delay_formulation_2}
(c)
= \lim \limits_{N \rightarrow \infty} \left( 1- {\prod}_{{j} \in \mathcal{S}_{r}} q_{j} \right) \sum \limits_{k=1}^{N} k \left( {\prod}_{{j} \in \mathcal{S}_{r}} q_{j} \right)^{k-1}
= \frac{1}{1 - {\prod}_{{j} \in \mathcal{S}_{r}} q_{j}}
\, \text{.}
\end{align}}

\vspace{-0.15cm}
Finally, combining (\ref{average_transmission_delay_formulation_1}) and (\ref{average_transmission_delay_formulation_2}), we can obtain
{\small
\begin{align*}
{\bar \tau}^{r}_i
= \frac{1}{1- \prod_{{j} \in \mathcal{S}_{r}} q_{j} } \max \limits_{{j} \in \mathcal{S}_{r}} \frac{{\hat B}_{j}}{R_{j}}
\, \text{.}
\end{align*}}
\hfill $\blacksquare$

\subsection{Proof of Lemma \ref{uplink_delay_lemma}}\label{proof_Lemma_q_larger_0}

According to (\ref{outage_probability_shadow}), if $\rho_i < + \infty$, we have $Q( \rho_i/\sigma_{\rm dB} ) > Q( + \infty ) = 0 $ and then $q_{i} < 1$.
Therefore, if we want $q_{i} < 1$, the following conditions need to be satisfied to make $\rho_i < + \infty$.
\begin{enumerate}[(i)]
\item The uplink transmission rate $R_{i} < + \infty$.
Otherwise, according to the definition of $\rho_i$ in (\ref{outage_probability_shadow}), i.e., $\rho_i \triangleq [ (2^{R_{i}/W_{i}} -1)W_{i} N_0 ]_{\rm dB} - [ P_{i} ]_{\rm dB} - [ {\mathcal{K}} ]_{\rm dB} + \lambda [ d_{i} ]_{\rm dB}$, if $R_{i} = + \infty$, we have $\rho_i = + \infty$.
\item The transmit power $P_{i} > 0$ (Watt). Otherwise, if $P_{i} = 0$ (Watt), we have $[ P_{i} ]_{\rm dB} = - \infty$ and then $\rho_i = + \infty$.
\item The allocated bandwidth $W_i > 0$. Otherwise, if $W_i = 0$, then $\rho_i = + \infty$ since
{\small
\begin{align}
\lim \limits_{W_i \rightarrow 0} (2^{\frac{R_{i}}{W_{i}}} -1)W_{i}
= \lim \limits_{W_i \rightarrow 0} \frac{2^{\frac{R_{i}}{W_{i}}} -1}{ \frac{1}{W_{i}} }
\overset{(a)}{=} & \lim \limits_{W_i \rightarrow 0} \frac{- 2^{\frac{R_{i}}{W_{i}}} \cdot \ln 2 \cdot \frac{R_i}{W_{i}^2} }{ - \frac{1}{W_{i}^2} }
= \lim \limits_{W_i \rightarrow 0} 2^{\frac{R_{i}}{W_{i}}} R_{i} \ln 2
= + \infty
\end{align}}

\vspace{-0.3cm}\noindent
where (a) is due to the L'Hospital's Rule.
\end{enumerate}

Therefore, with $R_{i} < + \infty$, $P_{i} > 0$ (in Watt), and $W_i > 0$, we have $q_i < 1$.
\hfill $\blacksquare$

\section{Monotonically increasing property of $\overline{W}_i(B_i)$}

According to (\ref{R_i_optimal}) and (\ref{quantization_level}), we have the quantization level satisfies
{\small
\begin{align}\label{B_i_full_expression}
& B_i
= {\bar B}_i(W_i)
= \frac{\tau_{\max}}{m} W_i \log_2 \left( 1 + \frac{\theta_i P_{\max}}{W_i N_0} \right) - \frac{\mu}{m} \, \text{.}
\end{align}}

\vspace{-0.3cm} \noindent
Based on this, the first-order derivative of ${\bar B}_i(W_i)$ with respect to the allocated bandwidth $W_i$ is
{\small
\begin{align}\label{B_i_W_i_gradient}
\frac{\partial {\bar B}_i(W_i)}{\partial W_i}
= & \frac{\tau_{\max}}{m} \log_2 \left( 1 + \frac{\theta_i P_{\max}}{W_i N_0} \right)
+ \frac{\tau_{\max}}{m} \frac{W_i}{\left( 1 + \frac{\theta_i P_{\max}}{W_i N_0} \right) \ln 2} \cdot \left( - \frac{\theta_i P_{\max}}{W_i^2 N_0} \right) \notag \\
= & \frac{\tau_{\max}}{m} \log_2  \left(  1  +  \frac{\theta_i P_{\max}}{W_i N_0}  \right)
-  \frac{\tau_{\max}\theta_i P_{\max}}{m\left( W_i N_0  +  \theta_i P_{\max} \right) \ln 2}
\: \text{,}
\end{align}}

\vspace{-0.3cm} \noindent
and then the associated second-order derivative is
{\small
\begin{align*}
\frac{\partial^2 {\bar B}_i(W_i)}{\partial W^2_i}
= & \frac{\tau_{\max}}{m \left( 1 + \frac{\theta_i P_{\max}}{W_i N_0} \right) \ln 2} \cdot \left( - \frac{\theta_i P_{\max}}{W_i^2 N_0} \right)
+ \frac{\tau_{\max}\theta_i P_{\max}N_0}{m\left( W_i N_0 +  \theta_i P_{\max} \right)^2 \ln 2} \notag \\
= &  - \frac{\tau_{\max}\theta_i P_{\max}}{m\left( W_i N_0  +  \theta_i P_{\max} \right) W_i\ln 2}
+ \frac{\tau_{\max}\theta_i P_{\max} N_0 }{m\left( W_i N_0  +  \theta_i P_{\max} \right)^2 \ln 2}
= - \frac{\tau_{\max}\theta_i^2 P_{\max}^2}{m\left( W_i N_0  +  \theta_i P_{\max} \right)^2 W_i\ln 2}
\, \text{.}
\end{align*}}

\vspace{-0.3cm}
In the practical wireless environment, the shadowing variance $\sigma_{\rm dB} > 0$, the constant $[ \mathcal{K} ]_{\rm dB} > - \infty$, the distance $d_i < + \infty$ (in meter), and it is reasonable to set the TO probability constraint $q_{\max} \in (0,1]$.
Thus, the parameter $\theta_i \triangleq 10^{\frac{1}{10}\left(\sigma_{\rm dB} \cdot Q^{-1}\left(1 - q_{\max}\right) + [ \mathcal{K} ]_{\rm dB} - \lambda [d_{i}]_{\rm dB} \right)}$ defined in (\ref{R_i_optimal}) satisfies $\theta_i \in (0, + \infty)$.
Meanwhile, in the real communication systems, the number of parameters ${m} \in (0,+ \infty)$, the delay constraint $\tau_{\max} \in (0,+ \infty)$, and the transmit power constraint $P_{\max} \in (0,+ \infty)$ (in Watt).
Therefore, $\frac{\partial^2 {\bar B}_i(W_i)}{\partial W^2_i} < 0$ with the allocated bandwidth $W_i \in [0,+ \infty)$, which means that $\frac{\partial {\bar B}_i(W_i)}{\partial W_i} $ monotonically decreases with the increasing $W_i \in [0,+ \infty)$.
Then, combining with $\lim_{W_i \rightarrow \infty} \frac{\partial {\bar B}_i(W_i)}{\partial W_i} = 0$ in (\ref{B_i_W_i_gradient}), we have
{\small
\begin{align}\label{Bi_increase}
\frac{\partial {\bar B}_i(W_i)}{\partial W_i} > 0
\end{align}}

\vspace{-0.3cm} \noindent
for $W_i \in [0,+ \infty)$, which means that $B_i$ in \eqref{B_i_full_expression} monotonically increases with $W_i \in [0,+ \infty)$.

Next, based on \eqref{B_i_full_expression} and the implicit function theorem \cite{Krantz_Parks02}, we can define a function $\Psi_i(W_i, B_i)$ to describe the relation between $W_i$ and $B_i$ as
\begin{align}\label{Psi_Wi_Bi}
& \Psi_i(W_i, B_i)
= \Psi_i({\bar W}_i(B_i), B_i)
= {\bar B}_i(W_i) - B_i
= 0
\, \text{.}
\end{align}
Then, taking the derivatives of both sides in \eqref{Psi_Wi_Bi} with respect to $B_i$, we have
{\small
\begin{align*}
& \frac{\partial \Psi_i(W_i, B_i)}{\partial B_i}
+ \frac{\partial \Psi_i(W_i, B_i)}{\partial W_i}
\cdot \frac{\partial {\bar W}_i(B_i)}{\partial B_i}
= 0
\, \text{.}
\end{align*}}

\vspace{-0.3cm} \noindent
Thus, combining with $ \frac{\partial \Psi_i(W_i, B_i)}{\partial W_i} = \frac{\partial {\bar B}_i(W_i)}{\partial W_i} $ and $ \frac{\partial \Psi_i(W_i, B_i)}{\partial B_i} = -1$, we can obtain that
{\small
\begin{align*}
& \frac{\partial {\bar W}_i(B_i)}{\partial B_i}
= - \frac{ \frac{\partial \Psi_i(W_i, B_i)}{\partial B_i} }{ \frac{\partial \Psi_i(W_i, B_i)}{\partial W_i} }
= \frac{1}{\frac{\partial {\bar B}_i(W_i)}{\partial W_i}}
\overset{(a)}{>} 0
\end{align*}}

\vspace{-0.3cm} \noindent
where (a) is due to \eqref{Bi_increase}.
Therefore, ${\bar W}_i(B_i)$ monotonically increases with $B_i$.
\hfill $\blacksquare$

\section{Proof of Proposition \ref{convexity}}

Based on \eqref{R_i_optimal} and \eqref{obj_func_A2_a}, we can denote $\phi_i \triangleq \frac{1}{\left( 2^{\frac{\tau_{\max}}{m} {\bar R}_i(W_i)-  \frac{\mu}{m} } - 1 \right)^2 }
= \frac{1}{\left( 2^{\frac{\tau_{\max}}{m} W_i \log_2 \left( 1 + \frac{\theta_i P_{\max}}{W_i N_0} \right)-  \frac{\mu}{m} } - 1 \right)^2 } $. Then, we have
{\small
\begin{align*}
\frac{\partial \phi_i}{\partial W_i}
= & - \frac{2}{\left( 2^{\frac{\tau_{\max}}{m} {\bar R}_i(W_i)-  \frac{\mu}{m} } - 1 \right)^3}
\cdot \frac{\partial \left( 2^{\frac{\tau_{\max}}{m} {\bar R}_i(W_i)-  \frac{\mu}{m} } - 1 \right)}{\partial W_i} \notag \\
= & - \frac{2}{\left( 2^{\frac{\tau_{\max}}{m} {\bar R}_i(W_i)-  \frac{\mu}{m} }  -  1 \right)^3}
\cdot 2^{\frac{\tau_{\max}}{m} {\bar R}_i(W_i)-  \frac{\mu}{m} } \ln 2
\cdot \frac{\tau_{\max}}{m}  \left( \log_2  \left(  1  +  \frac{\theta_i P_{\max}}{W_i N_0}  \right)  -  \frac{\theta_i P_{\max}}{( W_i N_0  +  \theta_iP_{\max} ) \ln2} \right) \notag \\
= & - \frac{2}{\left( 2^{\frac{\tau_{\max}}{m} {\bar R}_i(W_i)-  \frac{\mu}{m} } - 1 \right)^3}
\cdot 2^{\frac{\tau_{\max}}{m} {\bar R}_i(W_i)-  \frac{\mu}{m} }
\cdot \frac{\tau_{\max}}{m}  \left( \ln  \left(  1  +  \frac{\theta_i P_{\max}}{W_i N_0}  \right)  -  \frac{\theta_i P_{\max}}{W_i N_0  +  \theta_iP_{\max} } \right) \notag \\
= & \frac{2\tau_{\max}}{m} \cdot
\frac{
\overbrace{
2^{\frac{\tau_{\max}}{m} {\bar R}_i(W_i)-  \frac{\mu}{m} }
\cdot \left( \frac{\theta_i P_{\max}}{W_i N_0  +  \theta_iP_{\max} } - \ln  \left(  1  +  \frac{\theta_i P_{\max}}{W_i N_0}  \right) \right)
}^{\triangleq \varphi_i}
}{ \underbrace{ \left( 2^{\frac{\tau_{\max}}{m} {\bar R}_i(W_i)-  \frac{\mu}{m} } - 1 \right)^3 }_{\triangleq \rho_i} }
\, \text{.}
\end{align*}}

\vspace{-0.3cm}
\noindent
Based on this, we have
{\small
\begin{align*}
\frac{\partial^2 \phi_i}{\partial W_i^2}
= & \frac{2\tau_{\max}}{m} \cdot \frac{ \frac{\partial \varphi_i}{\partial W_i}\rho_i - \frac{\partial \rho_i}{\partial W_i}\varphi_i }{\rho_i^2}
\, \text{,}
\end{align*}}

\vspace{-0.3cm} \noindent
where $\rho_i^2 = \left( 2^{\frac{\tau_{\max}}{m} {\bar R}_i(W_i)-  \frac{\mu}{m} } - 1 \right)^6 \geq 1$ since the quantization level $B_i = \frac{\tau_{\max}}{m} {\bar R}_i(W_i)-  \frac{\mu}{m} \geq 1$,
{\small
\begin{align*}
\frac{\partial \rho_i}{\partial W_i}
= & \frac{\partial \left( 2^{\frac{\tau_{\max}}{m} {\bar R}_i(W_i)-  \frac{\mu}{m} } - 1 \right)^3}{\partial W_i} \notag \\
= & 3 \left( 2^{\frac{\tau_{\max}}{m} {\bar R}_i(W_i)-  \frac{\mu}{m} } - 1 \right)^2
\cdot 2^{\frac{\tau_{\max}}{m} {\bar R}_i(W_i)-  \frac{\mu}{m} }
\cdot \frac{\tau_{\max}}{m}  \left( \ln  \left(  1  +  \frac{\theta_i P_{\max}}{W_i N_0}  \right)  -  \frac{\theta_i P_{\max}}{W_i N_0  +  \theta_iP_{\max} } \right) \notag \\
= & 3 \left( 2^{\frac{\tau_{\max}}{m} {\bar R}_i(W_i)-  \frac{\mu}{m} }  - 1 \right)^2
\cdot \left( 2^{\frac{\tau_{\max}}{m} {\bar R}_i(W_i)-  \frac{\mu}{m} }  -  1  +  1 \right)
\cdot \frac{\tau_{\max}}{m}  \left( \ln  \left(  1  +  \frac{\theta_i P_{\max}}{W_i N_0}  \right)  -  \frac{\theta_i P_{\max}}{W_i N_0  +  \theta_iP_{\max} } \right) \notag \\
= & \frac{3\tau_{\max}}{m}  \left( \ln  \left(  1  +  \frac{\theta_i P_{\max}}{W_i N_0}  \right)  -  \frac{\theta_i P_{\max}}{W_i N_0  +  \theta_iP_{\max} } \right)
\cdot \bigg(  \left( 2^{\frac{\tau_{\max}}{m} {\bar R}_i(W_i)-  \frac{\mu}{m} }  -  1 \right)^3
+  \left( 2^{\frac{\tau_{\max}}{m} {\bar R}_i(W_i) - \frac{\mu}{m} } -   1 \right)^2 \bigg)
\text{,}
\end{align*}}

\vspace{-0.3cm} \noindent
and
{\small
\begin{align*}
\frac{\partial \varphi_i}{\partial W_i}
= &  \frac{ \partial \left( 2^{\frac{\tau_{\max}}{m} {\bar R}_i(W_i) - \frac{\mu}{m} }
\cdot \left(  \frac{\theta_i P_{\max}}{ W_i N_0 + \theta_iP_{\max} }  -  \ln  \left( 1  +  \frac{\theta_i P_{\max}}{W_i N_0} \right)  \right) \right) }
 {\partial W_i} \notag \\
= & - 2^{\frac{\tau_{\max}}{m} {\bar R}_i(W_i)  -  \frac{\mu}{m} } \cdot \frac{\tau_{\max}}{m}  \left( \ln  \left(  1  +  \frac{\theta_i P_{\max}}{W_i N_0}  \right)  -  \frac{\theta_i P_{\max}}{W_i N_0  +  \theta_iP_{\max} } \right)^2 \notag \\
& + 2^{\frac{\tau_{\max}}{m} {\bar R}_i(W_i)  -  \frac{\mu}{m} }
\frac{\theta_i^2 P_{\max}^2}{(W_i N_0 + \theta_i P_{\max})^2 W_i}
\, \text{.}
\end{align*}}

\vspace{-0.3cm}
Thus,
{\small
\begin{align*}
& \frac{\partial \varphi_i}{\partial W_i}\rho_i - \frac{\partial \rho_i}{\partial W_i}\varphi_i
\notag \\
= & - \frac{\tau_{\max}}{m} 2^{\frac{\tau_{\max}}{m} {\bar R}_i(W_i)  -  \frac{\mu}{m} }
\cdot \left( \ln  \left(  1  +  \frac{\theta_i P_{\max}}{W_i N_0}  \right)  -  \frac{\theta_i P_{\max}}{W_i N_0  +  \theta_iP_{\max} } \right)^2
\cdot \left( 2^{\frac{\tau_{\max}}{m} {\bar R}_i(W_i)-  \frac{\mu}{m} } - 1 \right)^3 \notag \\
& + 2^{\frac{\tau_{\max}}{m} {\bar R}_i(W_i)  -  \frac{\mu}{m} }
\frac{\theta_i^2 P_{\max}^2}{(W_i N_0 + \theta_i P_{\max})^2 W_i}
\cdot \left( 2^{\frac{\tau_{\max}}{m} {\bar R}_i(W_i)-  \frac{\mu}{m} } - 1 \right)^3 \notag \\
& + \frac{3\tau_{\max}}{m} 2^{\frac{\tau_{\max}}{m} {\bar R}_i(W_i)  -  \frac{\mu}{m} }
\cdot \left( \ln  \left(  1  +  \frac{\theta_i P_{\max}}{W_i N_0}  \right)  -  \frac{\theta_i P_{\max}}{W_i N_0  +  \theta_iP_{\max} } \right)^2
\cdot \left( 2^{\frac{\tau_{\max}}{m} {\bar R}_i(W_i)-  \frac{\mu}{m} } - 1 \right)^3 \notag \\
& + \frac{3\tau_{\max}}{m} 2^{\frac{\tau_{\max}}{m} {\bar R}_i(W_i)  -  \frac{\mu}{m} }
\cdot \left( \ln  \left(  1  +  \frac{\theta_i P_{\max}}{W_i N_0}  \right)  -  \frac{\theta_i P_{\max}}{W_i N_0  +  \theta_iP_{\max} } \right)^2
\cdot \left( 2^{\frac{\tau_{\max}}{m} {\bar R}_i(W_i)-  \frac{\mu}{m} } - 1 \right)^2 \notag \\
= & \frac{2 \tau_{\max}}{m} 2^{\frac{\tau_{\max}}{m} {\bar R}_i(W_i)  -  \frac{\mu}{m} }
\cdot \left( \ln  \left(  1  +  \frac{\theta_i P_{\max}}{W_i N_0}  \right)  -  \frac{\theta_i P_{\max}}{W_i N_0  +  \theta_iP_{\max} } \right)^2
\cdot \left( 2^{\frac{\tau_{\max}}{m} {\bar R}_i(W_i)-  \frac{\mu}{m} } - 1 \right)^3 \notag \\
& + 2^{\frac{\tau_{\max}}{m} {\bar R}_i(W_i)  -  \frac{\mu}{m} }
\frac{\theta_i^2 P_{\max}^2}{(W_i N_0 + \theta_i P_{\max})^2 W_i}
\cdot \left( 2^{\frac{\tau_{\max}}{m} {\bar R}_i(W_i)-  \frac{\mu}{m} } - 1 \right)^3 \notag \\
& + \frac{3\tau_{\max}}{m} 2^{\frac{\tau_{\max}}{m} {\bar R}_i(W_i)  -  \frac{\mu}{m} }
\cdot \left( \ln  \left(  1  +  \frac{\theta_i P_{\max}}{W_i N_0}  \right)  -  \frac{\theta_i P_{\max}}{W_i N_0  +  \theta_iP_{\max} } \right)^2
\cdot \left( 2^{\frac{\tau_{\max}}{m} {\bar R}_i(W_i)-  \frac{\mu}{m} } - 1 \right)^2
\text{.}
\end{align*}}

\vspace{-0.3cm}
Since the quantization level $B_i = \frac{\tau_{\max}}{m} {\bar R}_i(W_i)-  \frac{\mu}{m} \geq 1$, we have $2^{\frac{\tau_{\max}}{m} {\bar R}_i(W_i)-  \frac{\mu}{m} } - 1 \geq 1$.
Besides, with the allocated bandwidth $W_i \in [\overline{W}_i(1),+ \infty)$ in the constraint \eqref{obj_func_A2_b}, as well as the number of parameters ${m} \in (0,+ \infty)$ and the delay constraint $\tau_{\max} \in (0,+ \infty)$ in the practical communication systems, we have $\frac{\partial^2 \phi_i}{\partial W_i^2} \geq 0$, which means $\phi_i$ is convex with respect to $W_i$ in the feasible region of \eqref{obj_func_A2_b}.
Therefore, the objective function (\ref{obj_func_A2}) is convex.
\hfill $\blacksquare$

\end{document}